\documentclass[11pt]{article}
\usepackage{amsmath,amssymb,amsfonts,latexsym,graphicx,amsthm}
\usepackage{thmtools}
\makeatletter
\@ifundefined{newcounteralias}{}{%
  \renewcommand\thmt@autorefsetup{%
    \@xa\def\csname\thmt@envname autorefname\@xa\endcsname\@xa{\thmt@thmname}%
  }%
}
\makeatother
\usepackage{color}
\usepackage{url,hyperref}
\usepackage{comment}

\usepackage[linesnumbered,ruled,vlined]{algorithm2e}

\usepackage{xcolor} \usepackage{hyperref} \hypersetup{
    colorlinks = true, allcolors = black, citecolor = {blue!0!black}, linkcolor = {blue!0!black}
}
\usepackage[shortlabels]{enumitem}
\usepackage{cleveref}

\usepackage[margin=1in]{geometry}
\newsavebox{\openproblemcontent}
\newenvironment{openproblem}{%
    \par\addvspace{8pt}\begingroup
    \setlength{\fboxsep}{8pt}%
    \setlength{\fboxrule}{0.6pt}%
    \begin{lrbox}{\openproblemcontent}%
    \begin{minipage}{\dimexpr\linewidth-2\fboxsep-2\fboxrule\relax}%
    \setlength{\parindent}{0pt}\raggedright
    {\color{blue!50!black}\bfseries Open Problem}\par\smallskip
}{%
    \end{minipage}%
    \end{lrbox}%
    \noindent\fcolorbox{blue!50!black}{blue!2!white}{\usebox{\openproblemcontent}}%
    \par\endgroup\addvspace{8pt}%
}

\newtheorem{theorem}{Theorem}[section]
\newtheorem{lemma}[theorem]{Lemma}

\theoremstyle{definition}
\newtheorem{definition}[theorem]{Definition}

        {\medskip}

\newcommand{\Otilde}{\widetilde{O}}

\newcommand{\lo}{{\text{lo}}}
\newcommand{\hi}{{\text{hi}}}
\newcommand{\mi}{{\text{mid}}}
\newcommand{\mis}{\mathcal{M}}

\newcounter{paragraphCounter} % Create a new counter

\usepackage{tikz}
\usetikzlibrary{trees,arrows.meta}
\usetikzlibrary{shapes.geometric}

\definecolor{dsLayerHigh}{RGB}{242,244,249}
\definecolor{dsLayerMid}{RGB}{239,246,248}
\definecolor{dsLayerLow}{RGB}{247,248,244}
\definecolor{dsHighSet}{RGB}{91,75,145}
\definecolor{dsCOne}{RGB}{196,121,25}
\definecolor{dsCTwo}{RGB}{187,69,96}
\definecolor{dsFocus}{RGB}{226,153,20}
\definecolor{dsInk}{RGB}{44,49,57}

\usepackage{booktabs,array}
\usetikzlibrary{calc,positioning,fit}

\definecolor{dsMidSet}{RGB}{38,124,151}
\definecolor{dsLowSet}{RGB}{70,133,96}
\hypersetup{hypertexnames=false}

\title{Faster Dynamic Maximal Independent Set\\
against Adaptive Adversaries}
\author{
	Hongyi Chen\thanks{State Key Laboratory for Novel Software Technology, Nanjing University, \href{mailto:hychener01@gmail.com}{hychener01@gmail.com}}\and
	Shuyan Fan \thanks{State Key Laboratory for Novel Software Technology, Nanjing University, \href{mailto:231220005@smail.nju.edu.cn}{231220005@smail.nju.edu.cn}} \and
	Tianyi Zhang\thanks{State Key Laboratory for Novel Software Technology, Nanjing University, \href{mailto:tianyiz25@nju.edu.cn}{tianyiz25@nju.edu.cn}}
}
\date{}

\hypersetup{pdftitle={Faster Dynamic Maximal Independent Set against Adaptive Adversaries},
 pdfauthor={Hongyi Chen, Shuyan Fan, Tianyi Zhang},pageanchor=false}
\begin{document}
\pagestyle{plain}

\maketitle

\begin{NoHyper}
\begin{abstract}
We study the fully dynamic maximal independent set (MIS) problem: maintaining an MIS in an undirected graph on \(n\) vertices with at most \(m\) edges under edge insertions and deletions.

Against an oblivious adversary, which fixes the edge update sequence in advance,
the randomized algorithms of Behnezhad et al. (2019) and Chechik and Zhang
(2019) achieve polylogarithmic update
time.
Against an adaptive adversary, which may choose each update after observing
the previously reported MISs, the best previously known update bound was
\(O(m^{2/3})\), given by the deterministic algorithm of Gupta and Khan
(2018, 2021). Assuming the combinatorial Boolean
matrix multiplication hypothesis, Bernstein et al. (2026) proved that any
combinatorial algorithm against an adaptive adversary requires at least
\(m^{1/2-o(1)}\) update time even under edge insertions
alone, leaving a polynomial gap between upper and lower bounds.

We present a randomized fully dynamic MIS algorithm with
\(\widetilde O(m^{5/8})=\widetilde O(m^{2/3-1/24})\) update time
against adaptive adversaries. This gives the first polynomial improvement
over the \(m^{2/3}\) bound for adaptive fully dynamic MIS.
\end{abstract}
\end{NoHyper}

\thispagestyle{empty}
\clearpage

\thispagestyle{empty}
\tableofcontents
\clearpage

\setcounter{page}{1}
\pagestyle{plain}
\hypersetup{pageanchor=true}

\section{Introduction}

Dynamic graph algorithms maintain information about a graph as its edges are
inserted and deleted, aiming to process each update much faster than
recomputing a solution from scratch.  Fundamental problems studied in this
setting include connectivity and related connectivity
notions~\cite{DBLP:conf/stoc/HolmLT98,DBLP:conf/focs/0001S21,
DBLP:conf/stoc/HolmNR025}, minimum spanning
forest~\cite{DBLP:conf/stoc/HolmLT98,DBLP:conf/focs/NanongkaiSW17,
DBLP:journals/iandc/HenzingerP21}, reachability and transitive
closure~\cite{DBLP:conf/focs/Sankowski04,DBLP:conf/wea/HanauerH020,
DBLP:conf/focs/BrandK23}, shortest
paths~\cite{DBLP:journals/jacm/DemetrescuI04,DBLP:conf/stoc/ChuzhoyZ23,
DBLP:conf/focs/BrandK23},
matching~\cite{DBLP:conf/focs/BaswanaGS11,DBLP:conf/stoc/Wajc20,
DBLP:conf/focs/Liu24,DBLP:conf/focs/BehnezhadG24}, and network
flow~\cite{DBLP:conf/focs/GaoLP21,DBLP:conf/stoc/BrandLS23}, among many
others.  

We study the fully dynamic maximal independent set (MIS) problem.
We consider an undirected graph \(G=(V,E)\) with \(n\) vertices that undergoes a sequence of edge insertions and deletions. Throughout,
\(m\) denotes an upper bound on the number of edges \(|E|\) present at any time.
Updates are revealed online, and the algorithm receives no
predictions of future updates.
A set
\(\mis\subseteq V\) is an \emph{independent set} if no two vertices in \(\mis\) are
adjacent. It is a \emph{maximal independent set} (MIS) if no proper superset of
\(\mis\) is independent; equivalently, every vertex in \(V\setminus \mis\) has a
neighbor in \(\mis\). A vertex outside \(\mis\) is \emph{dominated} if it
has a neighbor in \(\mis\), and \emph{undominated} otherwise.
MIS is a fundamental symmetry-breaking problem in parallel and distributed computation, with applications to coloring, matching, and network coordination. Although a static MIS can be
computed by a single greedy scan, the fully dynamic MIS problem requires
maintaining such a set after every edge update. This is substantially more
challenging: if an inserted edge joins two vertices of the current MIS, one
of these vertices must leave, potentially leaving many of its neighbors undominated
and triggering a cascade of further repairs.

An elementary deterministic algorithm maintains an MIS in \(O(\max_{v\in V}\{\deg_G(v)\}) = O(n)\)
amortized update time. Assadi, Onak, Schieber, and Solomon gave the first nontrivial fully dynamic MIS algorithm
with update time sublinear in the number of edges, obtaining a deterministic
\(O(m^{3/4})\) amortized bound~\cite{DBLP:conf/stoc/AssadiOSS18}.  Gupta and
Khan subsequently improved this to \(O(m^{2/3})\) with a particularly simple
deterministic algorithm, first announced in a 2018
preprint~\cite{DBLP:journals/corr/abs-1804-01823} and published in
2021~\cite{DBLP:conf/sosa/GuptaK21}; Du and Zhang
independently obtained \(O(m^{2/3}\sqrt{\log m})\) update
time~\cite{DBLP:journals/corr/abs-1804-08908}. 
The resulting \(O(\min\{n,m^{2/3}\})\) amortized update time remains
the best known deterministic bound.

For randomized algorithms, the adversarial model is important. An \emph{oblivious adversary} fixes the entire update sequence in advance without revealing it to the algorithm. In this model, Assadi, Onak,
Schieber, and Solomon achieved
\(\widetilde O(\sqrt n)\)\footnote{We use $\widetilde O(\cdot)$ to suppress factors polynomial in $\log n$. } expected amortized update
time~\cite{DBLP:conf/soda/AssadiOSS19}, and two independent works subsequently
obtained polylogarithmic update time~\cite{DBLP:conf/focs/BehnezhadDHSS19,
DBLP:conf/focs/ChechikZ19}. These guarantees rely on the update sequence being
fixed before the execution begins and, in particular, being independent of the
algorithm's randomness.

An \emph{adaptive adversary}, in contrast, may choose each update after
observing the solutions reported by the algorithm so far. Since a deterministic
algorithm uses no random choices, its guarantee is unaffected by such
adaptivity. In particular, the deterministic \(O(m^{2/3})\) amortized update
bound of Gupta and Khan~\cite{DBLP:conf/sosa/GuptaK21} is automatically robust
against adaptive adversaries. For algorithms that explicitly maintain the MIS, an adaptive adversary can
force \(\Omega(n)\) membership changes in a single edge
update~\cite[Appendix~A]{DBLP:conf/stoc/AssadiOSS18}.
Consequently, \(o(n)\) worst-case update time is impossible in general,
necessitating amortization for sublinear update bounds.

Beimel et al.~\cite{DBLP:conf/stoc/BeimelKMNSS22} explain why committing to
random choices, as in fast oblivious MIS algorithms, does not directly extend
to this setting: an adaptive adversary can target the revealed choices to
force costly recomputation. Recent work establishes a conditional lower bound for dynamic MIS against adaptive adversaries. Assuming
the combinatorial Boolean matrix multiplication hypothesis, Bernstein,
Bhattacharya, Fischer, Kiss, and Saranurak proved an
\(n^{1-o(1)}\) amortized update lower bound for combinatorial algorithms against
adaptive adversaries, even under edge insertions
alone~\cite{DBLP:conf/soda/BernsteinBFKS26}.
By adding isolated vertices to the graphs in the dense hard instances, this lower bound
extends to $m^{1/2-o(1)}$ amortized update time for every graph sparsity
from $m=\Theta(n)$ to $m=\Theta(n^2)$. This leaves a gap between the $m^{1/2-o(1)}$
lower bound and the $O(m^{2/3})$ upper bound.
Gupta, Khan, and Surendra recently obtained an alternative deterministic
\(O(m^{2/3})\) fully dynamic algorithm and highlighted the goal of improving
this bound to match their $O(\sqrt m)$ incremental
bound~\cite{DBLP:conf/swat/GuptaKS26}. These results motivate the following question.

\begin{openproblem}
Can a maximal independent set be maintained in a fully dynamic graph against
an adaptive adversary in \(O(m^{2/3-\epsilon})\) amortized update time for some constant $\epsilon > 0$?
\end{openproblem}

\subsection{Our Result}

We answer this question affirmatively by giving the first polynomial improvement over the \(m^{2/3}\) update-time bound for fully dynamic MIS against adaptive adversaries.

\begin{theorem}\label{thm:main}
	Let $G=(V,E)$ be an initially empty undirected graph on $n$ vertices
	that undergoes a sequence of edge insertions and deletions, with at most
	$m$ edges at any time. There is a randomized algorithm that maintains
	a maximal independent set of $G$ against adaptive adversaries with
	amortized update time
	$\Otilde(m^{5/8})=\Otilde(m^{2/3-1/24})$.
\end{theorem}

\paragraph{The basic \(O(m^{2/3})\) algorithm.}
Our starting point is the simple deterministic algorithm of Gupta and
Khan~\cite{DBLP:conf/sosa/GuptaK21}. Partition vertices into low and high degree classes at a threshold \(D\). Maintain an MIS of the low degree
subgraph by tracking each vertex's number of low degree MIS neighbors and
greedily repairing maximality after each update, at an amortized cost of
\(O(D)\). Then extend it to an MIS of the whole graph by rebuilding an MIS
on the high degree vertices with no low degree MIS neighbor. There are
\(O(m/D)\) high degree vertices, so rebuilding costs \(O((m/D)^2)\).
Balancing the two costs with \(D=m^{2/3}\) gives the \(O(m^{2/3})\) bound.

\paragraph{Our approach.}
The main bottleneck is rebuilding the MIS on the high degree subgraph: lowering \(D\) reduces the cost of low degree repairs but increases the rebuilding cost.
Our approach avoids rebuilding the high layer after every update by
maintaining a \emph{candidate set} of low degree vertices outside the MIS
that are dominated solely by high degree MIS vertices. We allow these
candidates to remain outside the MIS, and include a candidate in the MIS only when it has few
high degree MIS neighbors. This limits the conflicts caused when it joins the MIS
and allows us to repair the high layer locally. This new candidate-set design introduces two main technical challenges.

\paragraph{Structural challenges.}
Maintaining the candidate set in less than \(m^{2/3}\) amortized time creates
further bottlenecks: a high degree MIS change may affect many candidates,
and candidates may accumulate so quickly that periodic rebuilding becomes
too frequent to achieve the desired amortized update time.
To resolve these bottlenecks, we develop additional data structures for maintaining the candidate set. These structures may be of independent interest; simplifying or extending them could yield further improvements for dynamic MIS.

\paragraph{Adaptivity challenges.}
Our design maintains randomly sampled high degree MIS neighbors to verify
that candidates are dominated. The reported MIS can therefore leak information
about these private samples to the adversary, whose future updates may
exploit that information. To analyze this dependence, we develop a new
tool for adaptivity analysis: \emph{probabilistic games}. These games characterize the information revealed by the data structure and establish the sampling guarantees needed under adaptive updates. This adaptivity argument may also be of independent interest, and the games may provide tools for analyzing other randomized data structures against adaptive adversaries.

\subsection{Further Related Work}

\subsubsection{Other Settings of Dynamic MIS}
\paragraph{Partially dynamic MIS.}
In the decremental setting, where edges are only deleted, a simple monotone
algorithm is optimal: include a vertex in the MIS whenever it ceases to have a neighbor
in the MIS. No vertex ever needs to leave the MIS, and each vertex joins the MIS at
most once, giving optimal total time linear in the size of the initial
graph~\cite{DBLP:conf/sosa/GuptaK21}.
In the incremental setting, where edges are only inserted, Gupta, Khan, and
Surendra gave a deterministic algorithm with \(O(\sqrt m)\) amortized update
time~\cite{DBLP:conf/swat/GuptaKS26}.
Moreover, the conditional \(m^{1/2-o(1)}\) lower bound against adaptive
adversaries also holds for incremental
MIS~\cite{DBLP:conf/soda/BernsteinBFKS26}.

\paragraph{Bounded arboricity.}
For graphs whose arboricity---the minimum number of forests into which the edge set can be partitioned---remains at most \(\alpha\), Onak, Schieber,
Solomon, and Wein gave a deterministic fully dynamic MIS algorithm with
\(O(\alpha^2\log^2 n)\) amortized update
time~\cite{DBLP:journals/talg/OnakSSW20}. In particular, this gives
polylogarithmic update time for graph families of bounded arboricity.

\paragraph{Dynamic MIS with predictions.}
Khan and Kumar~\mbox{\cite{KhanKumar26Predictions}} study dynamic MIS
when a predicted update sequence is available for offline
preprocessing.
For a sequence of $u$ fully dynamic updates,
their algorithm uses $O(un)$ preprocessing time and space and
achieves $O(\eta_e+\eta_v^*)$ worst-case time per update after
the first prediction error.
In their basic error model, $\eta_e$ counts the distinct edges
updated since the first mismatch.
The parameter $\eta_v^*$ measures the sum of degrees of vertices
whose MIS membership differs from that at the end of the correctly
predicted prefix, maximized over subsequent updates processed so
far; degrees are maximized over the relevant graphs.

\subsubsection{Adaptive Algorithms for Symmetry-Breaking Problems}

MIS, maximal matching, vertex coloring, and edge coloring are fundamental
symmetry-breaking problems. Recent work has established improved
update-time bounds for their dynamic variants against adaptive adversaries.

\paragraph{Maximal matching.}
Against oblivious adversaries, Baswana, Gupta, and Sen obtained
\(O(\log n)\) expected amortized update
time~\cite{DBLP:conf/focs/BaswanaGS11}, which Solomon improved to
\(O(1)\) expected amortized update
time~\cite{DBLP:conf/focs/Solomon16}.
In the parallel setting, Ghaffari and Trygub gave a randomized dynamic
maximal matching algorithm that processes batches of edge insertions and
deletions with polylogarithmic amortized work per update and
polylogarithmic depth per batch, with high probability over
polynomial-length update sequences against oblivious
adversaries~\cite{DBLP:conf/spaa/0001T24}.
Against adaptive adversaries, Neiman and Solomon gave a deterministic
\(O(\sqrt m)\) worst-case update
bound~\cite{DBLP:conf/stoc/NeimanS13}.
This bound is \(o(n)\) when \(m=o(n^2)\), but remains \(O(n)\)
when \(m=\Theta(n^2)\).
In 2025, Bernstein, Bhattacharya, Kiss, and Saranurak
gave a deterministic algorithm with \(\widetilde O(n^{8/9})\) amortized
update time and a randomized algorithm with
\(\widetilde O(n^{3/4})\) amortized update
time against adaptive adversaries~\cite{DBLP:conf/stoc/BernsteinBKS25}.
In 2026, Chuzhoy, Khanna, and Song improved the deterministic bound
to \(n^{1/2+o(1)}\) amortized update
time~\cite{DBLP:conf/stoc/ChuzhoyKS26}.

\paragraph{Vertex coloring.}
For \((\Delta+1)\)-coloring, the goal is to maintain a proper vertex
coloring using one more color than a known upper bound \(\Delta\) on the
maximum degree over the update sequence.
Bhattacharya, Grandoni, Kulkarni, Liu, and
Solomon~\cite{DBLP:journals/talg/BhattacharyaGKL22} and, independently,
Henzinger and Peng~\cite{DBLP:journals/talg/HenzingerP22} obtained
randomized algorithms with \(O(1)\) expected amortized update time
against oblivious adversaries.
In 2025, Behnezhad, Rajaraman, and Wasim obtained an algorithm against
adaptive adversaries with
\(\widetilde O(n^{8/9})\) amortized update
time~\cite{DBLP:conf/soda/BehnezhadRW25}, which Flin and
Halld{\'o}rsson improved to \(\widetilde O(n^{2/3})\)
amortized update time~\cite{DBLP:conf/icalp/FlinH25}.
Very recently, two independent concurrent works achieved \(\widetilde O(n^{1/2})\) update time against adaptive
adversaries~\cite{GilHalldorsson27Coloring,AssadiYazdanyar27Coloring}.

For arboricity-dependent coloring, Ghaffari and Grunau gave a randomized
dynamic algorithm that maintains an implicit \(O(\alpha)\)-coloring,
where \(\alpha\) is the graph's arboricity, with polylogarithmic
worst-case update and query times against oblivious
adversaries~\cite{DBLP:conf/stoc/0001G24}.

\paragraph{Edge coloring.}
Bhattacharya, Chakrabarty, Henzinger, and
Nanongkai~\cite{BhattacharyaCHN18Coloring} gave a deterministic algorithm
maintaining a \((2\Delta-1)\)-edge coloring in \(O(\log\Delta)\) worst-case
update time.
Duan, He, and Zhang~\cite{DuanHeZhang19EdgeColoring} reduced the number of
colors to \((1+\epsilon)\Delta\) with
\(\operatorname{poly}(\log n,\epsilon^{-1})\) amortized update time when
\(\Delta=\Omega(\epsilon^{-2}\log^2 n)\).
Christiansen~\cite{Christiansen23VizingChains} removed this degree
restriction and obtained \(\operatorname{poly}(\log n,\epsilon^{-1})\)
worst-case update time against adaptive adversaries.
For fixed \(\epsilon>0\) and \(\Delta\) above an
\(\epsilon\)-dependent polylogarithmic threshold, Bhattacharya, Costa,
Panski, and Solomon~\cite{BhattacharyaCPS24Nibbling} achieved \(O(1)\)
expected update time against oblivious adversaries.
On the deterministic side, Christiansen~\cite{Christiansen26EdgeColoring}
obtained \((1+\epsilon)\Delta\)-edge coloring with \(n^{o(1)}\)
amortized update time for fixed \(\epsilon>0\).
For graphs of arboricity \(\alpha\), independent works of Bhattacharya,
Costa, Panski, and Solomon~\cite{BhattacharyaCPS24Arboricity} and
Christiansen, Rotenberg, and Vlieghe~\cite{ChristiansenRV24EdgeColoring}
give deterministic \((\Delta+O(\alpha))\)-edge coloring with
polylogarithmic amortized update time.
More recently, Haeupler et al.~\cite{HaeuplerMRST26DynamicLLL} developed
an approach based on the dynamic Lov{\'a}sz local lemma, obtaining
\((1+\epsilon)\Delta\)-edge coloring with polylogarithmic amortized
update time against adaptive adversaries.

\paragraph{The distinction for MIS.}
MIS differs from these problems in that a polynomial conditional lower bound is known against adaptive adversaries. Under the combinatorial Boolean matrix multiplication
hypothesis, Bernstein, Bhattacharya, Fischer, Kiss, and Saranurak proved
that every combinatorial dynamic MIS algorithm requires
\(n^{1-o(1)}\) amortized update time, even for edge insertions
only~\cite{DBLP:conf/soda/BernsteinBFKS26}.
This gives a conditional separation from the polylogarithmic bounds
against oblivious adversaries. For combinatorial dynamic MIS algorithms, the lower bound
rules out matching the oblivious-adversary bounds in general, while
leaving room to improve the polynomial dependence on \(m\).
Our result pursues this goal within the broader effort to develop faster
dynamic symmetry-breaking algorithms against adaptive adversaries.

\section{Technical Overview}\label{sec:overview}

We consider an initially empty simple undirected graph $G=(V,E)$ on a
fixed set of $n$ vertices, undergoing edge insertions and deletions.
Let $m$ be a fixed upper bound on the number of edges throughout the
update sequence. Initially, the maintained MIS is $\mis=V$.
Write $N(v)$ for the current neighbor set of $v$, $\deg(v)=|N(v)|$,
$N(S)=\bigcup_{v\in S}N(v)$, and $G[S]$ for the subgraph induced by
$S\subseteq V$.  An independent set $\mis$ is maximal precisely when
every vertex outside $\mis$ has a neighbor in $\mis$; we call such a
vertex \emph{dominated} and such an MIS neighbor a \emph{domination witness}.
The edge joining the vertex to its witness is a \emph{domination edge}.

The adversary is adaptive: it may choose each update using the entire
sequence of previously reported MISs.  It does not observe the
algorithm's private random bits, auxiliary data structures, memory
accesses, or internal running times. An event holds \emph{with high
probability} if its failure probability is at most $n^{-c}$ for an
arbitrarily large fixed constant $c>0$, with constants in the algorithm
chosen accordingly.  Our probabilistic guarantees apply over any fixed
polynomial-length update sequence, including one chosen adaptively.

\subsection{MIS Candidates and Domination Witnesses}
\label{sec:overview-candidates-witnesses}

\paragraph{The $O(m^{2/3})$ amortized algorithm.}
We begin with Gupta and Khan's deterministic
algorithm~\cite{DBLP:conf/sosa/GuptaK21}, which provides the
starting point for our improvement.  Fix a degree threshold $D$. Partition the vertices into
$V_\lo$, with degree at most $D$, and $V_\hi$, with degree greater than $D$.
Maintain an MIS $\mis_\lo$ of $G[V_\lo]$, giving low vertices priority over
high vertices, and extend it by an MIS $\mis_\hi$ of
$G[V_\hi\setminus N(\mis_\lo)]$.
We call $\mis_\lo$ and $\mis_\hi$ the low and high \emph{MIS layers}:
each consists of the vertices of the maintained MIS in its degree class.

The low layer can be repaired in $O(D)$ amortized time per update.
Maintain each vertex's number of neighbors in $\mis_\lo$, and greedily
include a low vertex in $\mis_\lo$ when its count becomes zero. For each such vertex, scan at most
$D$ edges to update these counts. Although many vertices can join $\mis_\lo$ during one update,
the potential decreases enough to pay for all their scans: assign potential $\Theta(D)$ to
each low vertex outside $\mis_\lo$, so that including it in $\mis_\lo$ releases enough
potential to cover its scan. An edge insertion forces at most one
low-MIS removal, and a deletion forces none. The potential increase
from these removals is therefore only $O(D)$ per update, giving the
claimed amortized bound.

\begin{figure}[!htbp]
\centering
\definecolor{overviewHigh}{RGB}{91,75,145}
\definecolor{overviewLow}{RGB}{70,133,96}
\definecolor{overviewRebuild}{RGB}{174,65,72}
\begin{tikzpicture}[
  x=1cm,y=1cm,
  font=\small,
  high vertex/.style={circle,draw=overviewHigh,fill=overviewHigh,
    minimum size=2.4mm,inner sep=0pt},
  unfilled high vertex/.style={circle,draw=overviewHigh,fill=white,
    minimum size=2.4mm,inner sep=0pt},
  excluded vertex/.style={circle,draw=overviewRebuild,fill=white,
    minimum size=2.4mm,inner sep=0pt},
  low vertex/.style={circle,draw=overviewLow,fill=white,
    text=overviewLow,minimum size=3.6mm,inner sep=0pt,font=\scriptsize},
  selected low vertex/.style={low vertex,fill=overviewLow,text=white},
]
  % Repeat the same irregular dense graph in both panels.  The fixed
  % layout and edge set keep the illustration reproducible.  Initially
  % {h2,h6,h11} is an MIS; after its exclusion, {h4,h10} is an MIS of
  % the remaining high graph.
  \foreach \dx/\panel in {0/before,7.2/after} {
    \begin{scope}[shift={(\dx,0)},name prefix=\panel-]
      \fill[overviewHigh!8,rounded corners=4pt]
        (0,1.48) rectangle (5.4,3.27);
      \fill[overviewLow!8,rounded corners=4pt]
        (0,0.08) rectangle (5.4,1.25);
      \node[anchor=west,text=overviewHigh,font=\bfseries]
        at (.18,3.02) {$V_\hi$};
      \node[anchor=west,text=overviewLow,font=\bfseries]
        at (.18,1.02) {$V_\lo$};
      \coordinate (h1) at (.80,2.48);
      \coordinate (h2) at (1.12,1.82);
      \coordinate (h3) at (1.46,2.98);
      \coordinate (h4) at (1.91,2.18);
      \coordinate (h5) at (2.30,2.77);
      \coordinate (h6) at (2.79,1.71);
      \coordinate (h7) at (2.88,2.34);
      \coordinate (h8) at (3.45,3.00);
      \coordinate (h9) at (3.84,2.04);
      \coordinate (h10) at (4.17,2.65);
      \coordinate (h11) at (4.77,1.83);
      \coordinate (h12) at (4.95,2.93);
      \foreach \a/\b in {
        h1/h2,h1/h3,h1/h4,h1/h5,h1/h7,h1/h10,
        h2/h3,h2/h4,h2/h5,h2/h9,
        h3/h4,h3/h5,h3/h6,
        h4/h5,h4/h6,h4/h7,h4/h9,
        h5/h6,h5/h8,h5/h12,
        h6/h7,h6/h8,h6/h9,
        h7/h8,h7/h9,h7/h10,
        h8/h9,h8/h10,h8/h11,h8/h12,
        h9/h10,h9/h11,h9/h12,
        h10/h11,h10/h12,h11/h12} {
        \draw[black!32,line width=.35pt] (\a) -- (\b);
      }
      \foreach \v in {h2,h6,h11} {
        \draw[overviewRebuild!75,line width=.65pt]
          (2.70,.63) -- (\v);
      }
      \foreach \i in {1,...,12} {
        \node[unfilled high vertex] at (h\i) {};
      }
      \node[low vertex] at (2.70,.63) {$v$};
    \end{scope}
  }

  \node[font=\bfseries] at (2.7,3.90) {Before low-layer repair};
  \foreach \v in {h2,h6,h11} {
    \node[high vertex] at (before-\v) {};
  }

  % Including v in the low MIS excludes its high-MIS neighbors; the classical
  % algorithm then recomputes the MIS on V_hi minus the neighbors of M_lo.
  \draw[->,overviewLow,line width=1pt] (5.65,.63) -- (6.98,.63);
  \node[above,text=overviewLow,font=\scriptsize,align=center] at (6.32,.72) {include $v$\\in $\mis_\lo$};
  \draw[->,overviewRebuild,line width=1pt] (5.65,2.28) -- (6.98,2.28);
  \node[above,text=overviewRebuild,font=\scriptsize] at (6.32,2.38) {rebuild};

  \node[font=\bfseries] at (9.90,3.90) {After $v\in\mis_\lo$};
  \node[text=overviewRebuild,font=\scriptsize\bfseries] at (9.90,3.58)
    {Full high-layer rebuild: $O(|V_\hi|^2)$};
  \draw[overviewRebuild,line width=1.05pt,rounded corners=4pt]
    (7.20,1.48) rectangle (12.60,3.27);
  \foreach \v in {h4,h10} {
    \node[high vertex] at (after-\v) {};
  }
  \foreach \v in {h2,h6,h11} {
    \node[excluded vertex] at (after-\v) {};
    \begin{scope}[shift={(after-\v)}]
      \draw[overviewRebuild,line width=.6pt]
        (-.065,-.065) -- (.065,.065)
        (-.065,.065) -- (.065,-.065);
    \end{scope}
  }
  \node[selected low vertex] at (9.90,.63) {$v$};

  \node[font=\scriptsize,anchor=west] at (.45,-.35)
    {\textcolor{overviewHigh}{$\bullet$} in $\mis_\hi$\quad
     \textcolor{overviewLow}{$\bullet$} in $\mis_\lo$\quad
     \textcolor{overviewRebuild}{$\times$} leaves $\mis_\hi$};
\end{tikzpicture}
\caption{The bottleneck in the $O(m^{2/3})$ amortized algorithm.  Before a low degree vertex $v$ joins $\mis_\lo$, its high-MIS neighbors (crosses) leave $\mis_\hi$.  The
algorithm rebuilds $\mis_\hi$ on $V_\hi\setminus N(\mis_\lo)$,
costing $O(|V_\hi|^2)=O((m/D)^2)$ per update.}
\label{fig:classical-high-rebuild}
\end{figure}
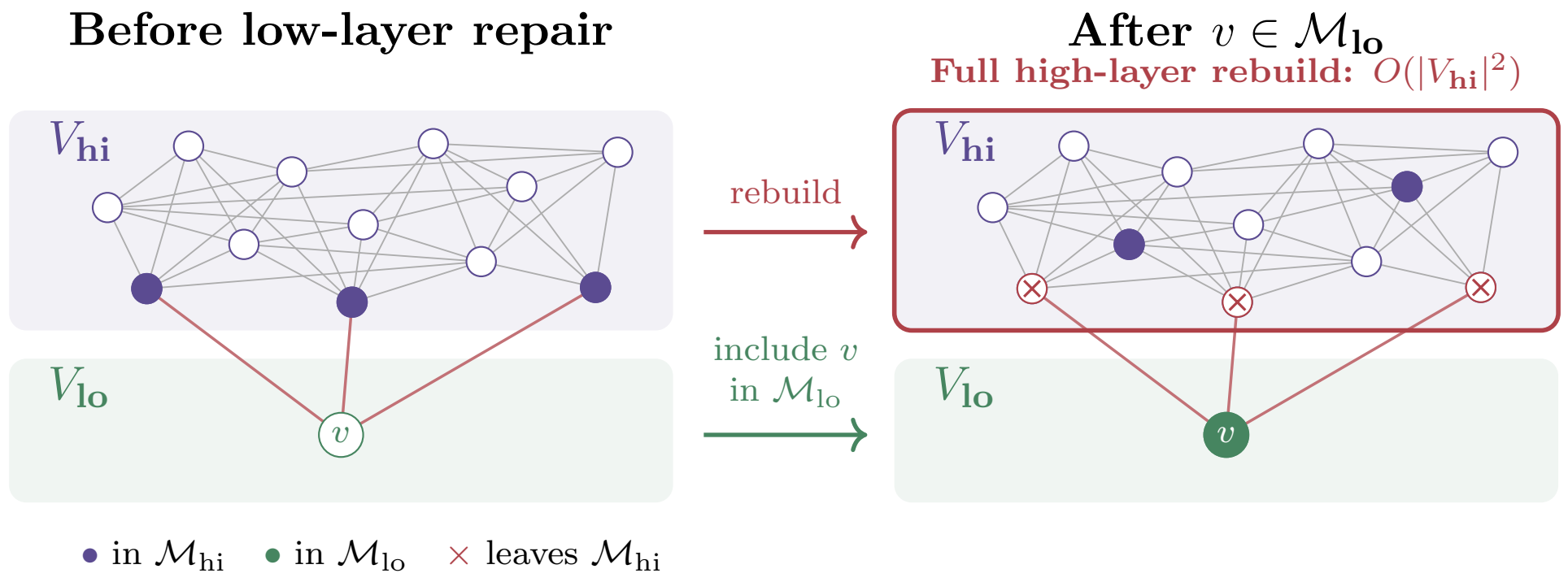

As illustrated in \Cref{fig:classical-high-rebuild}, including even one vertex in
$\mis_\lo$ can invalidate the current high-layer MIS.  After restoring
maximality in the low layer, rebuild $\mis_\hi$ from scratch.
The counters identify the high vertices with no neighbor in $\mis_\lo$, and
$|V_\hi|\leq 2m/D$, so this costs $O((m/D)^2)$ time.

Consequently, the amortized update time is
\[
 O\!\left(D+\frac{m^2}{D^2}\right),
\]
which is minimized at $D=\Theta(m^{2/3})$.

\paragraph{Lowering the degree threshold.}
To bypass the $m^{2/3}$ barrier, a natural starting point is to lower the
degree threshold from $m^{2/3}$ to $m^{2/3-\epsilon}$, for some constant
$\epsilon>0$. Maintaining the MIS on the low degree part would then cost
only $O(m^{2/3-\epsilon})$ amortized time.  The bottleneck is updating the
high degree part: when a vertex $v\in V_\lo$ joins $\mis_\lo$, it may be
adjacent to many vertices of $\mis_\hi$, all of which must leave the MIS.
These removals can affect most vertices in $V_\hi$. With the lower threshold, the bound becomes $|V_\hi|=O(m^{1/3+\epsilon})$, so rebuilding the high-layer MIS from scratch costs $O(m^{2/3+2\epsilon})$ time.

\paragraph{Leaving dominated low vertices outside the MIS.}
If a low vertex $v$ has many neighbors in $\mis_\hi$, we can leave it outside $\mis_\lo$.
Even if $v$ has no neighbor in $\mis_\lo$, its high-MIS neighbors already
certify that it is dominated.  We can therefore leave $v$ outside the MIS
and avoid removing all of these neighbors from $\mis_\hi$.  This relaxes the requirement
that $\mis_\lo$ be maximal in $G[V_\lo]$: a low vertex may instead rely on
the high layer for domination.  We only include such a vertex in the MIS when it has
few high-MIS neighbors, limiting the number of high-MIS vertices that must leave the MIS before it joins.

Sampling lets us distinguish vertices that should remain dominated from
those that can enter the MIS with few conflicts. To illustrate, suppose
every low vertex has either $\Theta(m^{1/3+\epsilon})$ high-MIS neighbors
or only $O(1)$. In the first case, a constant fraction of $V_\hi$
provides witnesses, so $O(\log n)$ uniform samples find one with high
probability for a fixed graph and MIS. A vertex for which no witness is
found therefore has only $O(1)$ high-MIS neighbors, with high probability.

These few conflicts allow local repair of the high layer in
$O(|V_\hi|)$ amortized time: maintain exact counters within that layer
and use the same potential argument as for the low layer to pay for
subsequent vertices that join $\mis_\hi$ greedily. Including the scan that finds the conflicts
and the low vertex's neighborhood scan, including a vertex in $\mis_\lo$ costs
$O(m^{1/3+\epsilon}+m^{2/3-\epsilon})$ amortized time. Both terms are
below $m^{2/3}$ for sufficiently small $\epsilon>0$.

\paragraph{Maintaining domination of candidate vertices.}
This idea leaves a \emph{primary candidate set}
$C_1\subseteq V_\lo\setminus\mis_\lo$
whose vertices have no neighbor in $\mis_\lo$ and are dominated solely by
$\mis_\hi$. The challenge is to preserve this domination guarantee as $\mis_\hi$ changes. When a high vertex $x$ leaves the MIS, a candidate
adjacent to $x$ may lose its last MIS neighbor and must be reconsidered
to restore maximality.  Maintaining exact high-MIS-neighbor counters for
all candidates would require updating every candidate adjacent to $x$.
Even a single high-MIS removal can therefore require scanning a large
fraction of $C_1$, with no bound below $m^{2/3}$; including a vertex in $\mis_\hi$
causes the same counter-maintenance problem.

Storing just one high-MIS neighbor as a domination witness for each
candidate does not immediately resolve this difficulty.  A single high
vertex may serve as the witness for many candidates.  When it leaves,
it no longer serves as an MIS neighbor for any of them, even if the affected candidates
still have many other neighbors in $\mis_\hi$.  Searching for a replacement
separately for each candidate can again be expensive.  Thus, limiting the
number of vertices removed from $\mis_\hi$ when a low vertex joins the MIS is only part of
the problem: we must also limit the work needed to discover which
candidates require attention and to maintain their domination witnesses.

\paragraph{Checkpoint: domination edge lists.}
We maintain a sample of each candidate's high-MIS neighbors so that
losing one witness need not trigger a new search. For each $v\in C_1$,
sample its edges to $\mis_\hi$ independently with probability $p$ and
store them in its \emph{high domination edge list} $A_v^\hi$. As long
as this list is nonempty, $v$ remains dominated and stays outside the MIS.

To find the candidates affected by a high-MIS removal, also collect the
sampled edges incident to each $x\in\mis_\hi$ in a \emph{reverse list}
$R_x^\hi$. To update the candidates' sampled lists after $x$ leaves
$\mis_\hi$, traverse $R_x^\hi$, remove the entries for its sampled edges
from the candidates' lists, and reconsider candidates whose lists become
empty. Such a candidate enters $\mis_\lo$ after its conflicting
high-MIS neighbors are removed from $\mis_\hi$. Thus the cost of finding affected
candidates depends on the reverse-list size, rather than on all of
$x$'s candidate neighbors.

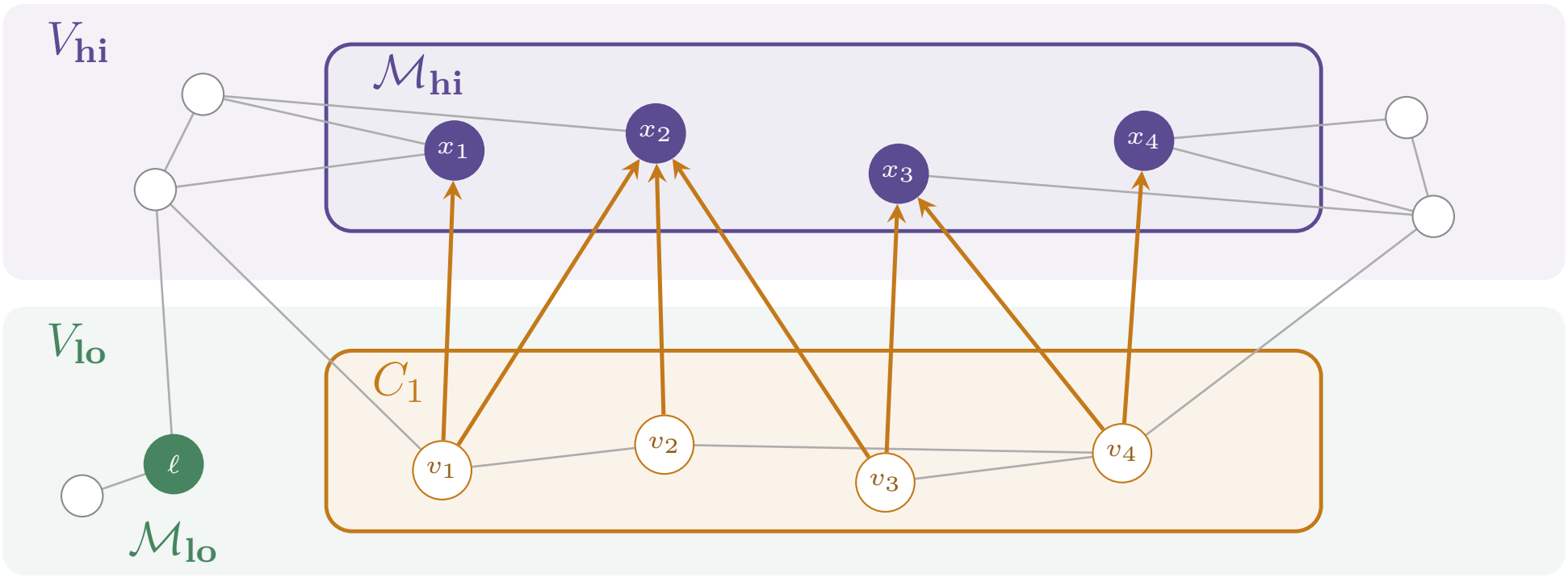
\begin{figure}[!htbp]
\centering
\definecolor{domHigh}{RGB}{91,75,145}
\definecolor{domLow}{RGB}{70,133,96}
\definecolor{domCandidate}{RGB}{196,121,25}
\begin{tikzpicture}[
  x=1cm,y=1cm,>=stealth,font=\small,
  high vertex/.style={circle,draw=domHigh,fill=domHigh,text=white,
    minimum size=4.8mm,inner sep=0pt,font=\tiny},
  candidate/.style={circle,draw=domCandidate,fill=white,
    text=domCandidate!75!black,minimum size=4.8mm,inner sep=0pt,
    font=\tiny},
  low vertex/.style={circle,draw=domLow,fill=domLow,text=white,
    minimum size=4.8mm,inner sep=0pt,font=\tiny},
  other vertex/.style={circle,draw=black!45,fill=white,
    minimum size=3.4mm,inner sep=0pt},
  edge/.style={draw=black!32,line width=.5pt},
  domination edge/.style={->,draw=domCandidate,line width=.95pt},
]
  % The degree layers contain the MIS and candidate subsets.
  \fill[domHigh!7,rounded corners=5pt]
    (0,2.42) rectangle (12.80,4.68);
  \fill[domLow!7,rounded corners=5pt]
    (0,0) rectangle (12.80,2.20);
  \node[anchor=west,font=\bfseries,text=domHigh]
    at (.22,4.37) {$V_\hi$};
  \node[anchor=west,font=\bfseries,text=domLow]
    at (.22,1.90) {$V_\lo$};

  \draw[rounded corners=6pt,draw=domHigh,line width=.9pt,
    fill=domHigh!10] (2.65,2.82) rectangle (10.80,4.35);
  \node[anchor=west,font=\bfseries,text=domHigh]
    at (2.89,4.10) {$\mis_\hi$};

  \draw[rounded corners=6pt,draw=domCandidate,line width=.9pt,
    fill=domCandidate!9] (2.65,.36) rectangle (10.80,1.84);
  \node[anchor=west,font=\bfseries,text=domCandidate]
    at (2.89,1.58) {$C_1$};
  \node[anchor=north,font=\bfseries,text=domLow]
    at (1.40,.57) {$\mis_\lo$};

  % Faint, irregular edges provide context; none join two MIS vertices
  % or a candidate to the low MIS.
  \coordinate (pXone) at (3.70,3.48);
  \coordinate (pXtwo) at (5.35,3.62);
  \coordinate (pXthree) at (7.34,3.29);
  \coordinate (pXfour) at (9.35,3.56);
  \coordinate (pVone) at (3.60,.86);
  \coordinate (pVtwo) at (5.42,1.07);
  \coordinate (pVthree) at (7.23,.76);
  \coordinate (pVfour) at (9.17,1.00);
  \coordinate (pHone) at (1.25,3.16);
  \coordinate (pHtwo) at (1.64,3.94);
  \coordinate (pHthree) at (11.50,3.75);
  \coordinate (pHfour) at (11.72,2.94);
  \foreach \a/\b in {
    pHone/pHtwo,pHone/pXone,pHtwo/pXone,pHtwo/pXtwo,
    pHthree/pHfour,pHthree/pXfour,pHfour/pXthree,pHfour/pXfour,
    pVone/pVtwo,pVtwo/pVfour,pVthree/pVfour,
    pVone/pHone,pVfour/pHfour} {
    \draw[edge] (\a) -- (\b);
  }

  \draw[edge] (.65,.65) -- (1.40,.91);
  \draw[edge] (1.40,.91) -- (pHone);
  \node[other vertex] at (pHone) {};
  \node[other vertex] at (pHtwo) {};
  \node[other vertex] at (pHthree) {};
  \node[other vertex] at (pHfour) {};
  \node[low vertex] at (1.40,.91) {$\ell$};
  \node[other vertex] at (.65,.65) {};
  \node[high vertex] (x1) at (pXone) {$x_1$};
  \node[high vertex] (x2) at (pXtwo) {$x_2$};
  \node[high vertex] (x3) at (pXthree) {$x_3$};
  \node[high vertex] (x4) at (pXfour) {$x_4$};
  \node[candidate] (v1) at (pVone) {$v_1$};
  \node[candidate] (v2) at (pVtwo) {$v_2$};
  \node[candidate] (v3) at (pVthree) {$v_3$};
  \node[candidate] (v4) at (pVfour) {$v_4$};

  % Each arrow is stored in A_v^hi and R_x^hi at its head.
  \draw[domination edge] (v1) -- (x1);
  \draw[domination edge] (v1) -- (x2);
  \draw[domination edge] (v2) -- (x2);
  \draw[domination edge] (v3) -- (x2);
  \draw[domination edge] (v3) -- (x3);
  \draw[domination edge] (v4) -- (x3);
  \draw[domination edge] (v4) -- (x4);
\end{tikzpicture}
\caption{Domination edges in the two-layer construction.  Gray lines show
other edges.  Arrows orient the sampled edges from $C_1$ toward
$\mis_\hi$: each edge
$v\to x$ is stored in the high domination edge list $A_v^\hi$ and in the reverse
list $R_x^\hi$.  For example, three high domination edge lists contribute edges to
$R_{x_2}^\hi$.}
\label{fig:checkpoint-domination-lists}
\end{figure}

When a candidate enters $C_1$, initialize its list by sampling edges to
the current high MIS. When a high vertex enters the MIS, sample its
edges to the current candidates with the same probability $p$.
A newly inserted graph edge is also sampled if it joins a high-MIS
vertex to a candidate. Each such sampling decision uses a fresh
independent bit. The algorithm does not resample this bit while the edge
exists, its high endpoint remains in the MIS, and the candidate's list
remains stored.
If any of these conditions stops holding, the bit is discarded.
When all three hold again, the edge is sampled with a fresh independent bit.
Each selected edge is stored in both lists, as illustrated in
\Cref{fig:checkpoint-domination-lists}.

An empty list suggests that including the vertex in the MIS requires few
high-MIS removals. To quantify this, choose $p=m^{\delta-1/3}$ with
$\epsilon<\delta<1/3$, and write $k_v=|N(v)\cap\mis_\hi|$.
For a fixed neighborhood and independent
sampling,
\[
 \Pr[A_v^\hi=\emptyset]=(1-p)^{k_v}\leq e^{-pk_v}.
\]
Thus, for any constant $c>0$, a vertex with
$k_v\geq c m^{1/3-\delta}\log n$ has an empty list with probability at
most $n^{-c}$.  This suggests that whenever we include a candidate with
an empty list in the MIS, it has only
$\widetilde O(m^{1/3-\delta})$ high-MIS neighbors.  Scanning its
neighborhood costs $O(m^{2/3-\epsilon})$, and removing these neighbors from $\mis_\hi$ and restoring maximality in the high layer costs
\[
 \widetilde O\!\left(
 m^{1/3-\delta}\cdot |V_\hi|
 \right)
 =\widetilde O(m^{2/3+\epsilon-\delta})
\]
amortized time, since each vertex that subsequently joins $\mis_\hi$ decreases the potential
as above. Both exponents are below $2/3$ when $\delta>\epsilon$;
choosing $\delta\geq2\epsilon$ makes this high-layer update cost at most
$\widetilde O(m^{2/3-\epsilon})$.

This preliminary algorithm leaves two challenges: bounding the maintenance cost and justifying the sampling guarantees under adaptive updates. We first address the maintenance cost and then turn to adaptivity.

\smallskip
\emph{Controlling the cost of maintaining candidate domination.}
The preceding time bound accounts only for work within the high layer.
For a fixed candidate neighborhood independent of the sample, a high vertex
has at most $p|C_1|$ sampled candidate edges in expectation. Since
$|C_1|$ can be $O(m)$, this calculation gives only an $O(pm)=O(m^{2/3+\delta})$
upper bound on the expected number of sampled candidate edges traversed
during a high-MIS removal. This bound is insufficient for the target
update time.

\smallskip
\emph{Making the sampling argument valid against adaptive updates.}
The empty-list calculation assumes a fixed neighborhood independent of
the sample.  Here the sampled lists influence which low vertices enter
the MIS, and an adversary can use the reported MIS to choose future
updates. Consequently, the current high-MIS neighborhood and the
edges stored in the sampled list need not be independent. We must show that
an empty list still implies few high-MIS neighbors with sufficiently
high probability under this dependence.

\subsection{Middle Degree Class}
\label{sec:overview-middle-degree}

\paragraph{Periodic candidate rebuild.}
Reverse-list maintenance becomes expensive when too many low vertices
depend on the high MIS. We therefore try to keep $C_1$ small by
periodically eliminating its dependence on that layer. Remove every
vertex from the high MIS and process the candidates in $C_1$, including each
in $\mis_\lo$ if it has no low-MIS neighbor. Whenever a candidate joins $\mis_\lo$, update
the low counters before processing the next candidate. Every former
candidate then belongs to $\mis_\lo$ or has a neighbor in $\mis_\lo$,
so $C_1$ is empty.
We can now rebuild $\mis_\hi$ on
$G[V_\hi\setminus N(\mis_\lo)]$.

Whenever a vertex joins $\mis_\lo$, the potential decreases enough to pay for its neighborhood
scan, as before. The high-layer reconstruction, however,
costs $O(|V_\hi|^2)=O(m^{2/3+2\epsilon})$, which exceeds our target
update time.  Thus it is essential to show that these rebuilds are not
too frequent.  For example, to obtain an amortized contribution of
$\widetilde O(m^{2/3-\epsilon})$, the number of reconstructions over
$T$ updates must be at most $\widetilde O(T/m^{3\epsilon})$.
Since each reconstruction empties $C_1$, a bound on how quickly new
candidates enter $C_1$ can give the required bound on the number of
reconstructions.

However, $C_1$ can grow substantially within a single update. Consider inserting an edge
$(u,v)$ with $u,v\in\mis_\lo$. To restore independence, suppose that
$u$ leaves the MIS. The vertex $u$ can have
$\Theta(m^{2/3-\epsilon})$ low degree neighbors for which it was the
only low-MIS neighbor.  If these vertices remain dominated by the high
layer, they can all become candidates in $C_1$ rather than enter
$\mis_\lo$.  Thus a single edge update can create
$\Theta(m^{2/3-\epsilon})$ new candidates, even if $C_1$ was empty
immediately beforehand. A small candidate threshold can therefore be
reached within a single update. To keep reverse lists short while
bounding the total rebuilding cost, we need a tighter bound on how
quickly candidates are created.

\paragraph{Middle degree class.}
To limit candidate growth, we lower the low degree threshold to $m^{3/8}$.
A low-MIS removal can then create only $O(m^{3/8})$ new low candidates.
We set the high degree threshold to $m^{5/8}$ and place the vertices
between these thresholds in a new middle degree class:
\[
\begin{aligned}
 V_\lo&=\{v:\deg(v)<m^{3/8}\},\\
 V_\mi&=\{v:m^{3/8}\leq\deg(v)<m^{5/8}\},\\
 V_\hi&=\{v:\deg(v)\geq m^{5/8}\}.
\end{aligned}
\]
As before, write $\mis_i=\mis\cap V_i$ for each MIS layer; their union
is independent, although a layer need not be maximal within its own
degree class. The degree bounds give
\[
 |V_\mi|=O(m^{5/8}),\qquad |V_\hi|=O(m^{3/8}).
\]
We still need to avoid expensive middle-MIS removals
when a low vertex joins the MIS. Suppose that, whenever a low
vertex $v$ loses its last low-MIS neighbor, we include it in $\mis_\lo$
and remove all of its neighbors from $\mis_\mi$. Even though $v$ has degree
$O(m^{3/8})$, each removed middle vertex can have degree $\Theta(m^{5/8})$.
Scanning their entire neighborhoods can therefore cost $O(m)$ for one
low vertex to join the MIS. We must therefore limit conflicts with the middle MIS or avoid creating them.

\subsection{Secondary Candidates and Exact Middle Domination}
\label{sec:overview-secondary-candidates}

The same principle that motivates $C_1$ applies here: if including a low vertex $v$ in the MIS would require removing its neighbors from $\mis_\mi$, those neighbors already dominate $v$, so we leave it outside the MIS. We introduce a \emph{secondary candidate set}
$C_2\subseteq V_\lo\setminus\mis$ for low vertices with no low-MIS
neighbor that remain outside the MIS because they are dominated by the
middle layer. Thus we allow a low vertex to be dominated by $\mis_\mi$
without being dominated by $\mis_\lo$.

\paragraph{Maintaining exact middle domination.}
The maximum degree of a middle vertex is $O(m^{5/8})$, which fits our
target update bound. We can therefore maintain the exact counter
\[
 d_\mi(v)=|N(v)\cap\mis_\mi|
\]
for every vertex, including low vertices. When a middle vertex $v$ joins $\mis_\mi$, increment $d_\mi(z)$ for every neighbor $z\in N(v)$; when $v$ leaves $\mis_\mi$, decrement $d_\mi(z)$ for every such neighbor.

The exact counters separate candidates according to which higher layer
must dominate them. Writing $d_\lo(v)=|N(v)\cap\mis_\lo|$, we maintain
\begin{align*}
 C_1&=\{v\in V_\lo\setminus\mis:
           d_\lo(v)=0\text{ and }d_\mi(v)=0\},\\
 C_2&=\{v\in V_\lo\setminus\mis:
           d_\lo(v)=0\text{ and }d_\mi(v)>0\}.
\end{align*}
The middle counter already certifies domination for $C_2$. Vertices in
$C_1$ must instead have high-MIS witnesses at the end of each update,
which we maintain through sampled high domination edges. Together,
$C_1\cup C_2$ is the \emph{candidate union}: every other low vertex
outside the MIS has a low-MIS neighbor.

A vertex in $C_2$ can remain there indefinitely. Its middle counter already
certifies domination, so it needs no high domination edge list. We do not
impose a size bound on $C_2$ and do not rebuild merely because $C_2$ is
large. When its last middle-MIS neighbor disappears, it moves to $C_1$.
This transition changes which layer must dominate the vertex, but it does
not create a new member of $C_1\cup C_2$.

To bound growth of the candidate union, consider both changes in degree
class and the loss of a low vertex's last low-MIS neighbor. We distinguish
new entries into the union from work on vertices that are already candidates:
\begin{itemize}
 \item \emph{Degree-class changes create few new candidates.}
 Only the updated edge's endpoints can change class. An endpoint entering
 the low degree class contributes at most one candidate. When a low-MIS
 vertex moves to the middle degree class, only its $O(m^{3/8})$ neighbors
 can lose a low-MIS neighbor. Other degree-class changes do not enlarge
 the candidate union.

 \item \emph{A single low-MIS removal creates few new candidates.}
 Inserting $(s,t)$ with $s,t\in\mis_\lo$ removes one of $s,t$ from
 the MIS. The removal affects only the removed vertex and its
 $O(m^{3/8})$ neighbors. Thus at most
 $O(m^{3/8})$ low vertices enter the candidate union. Deleting
 an edge incident to a low-MIS vertex can directly create only one new
 candidate.

 \item \emph{Middle- and high-MIS removals do not directly enlarge the
 candidate union.}
 These removals do not change any low-MIS-neighbor counter. A middle-MIS
 removal can move many vertices from $C_2$ to $C_1$, but all of them were
 already candidates. A high-MIS removal can invalidate witnesses of
 existing vertices in $C_1$, but again it does not create new candidates.

 \item \emph{No vertex leaves $\mis_\lo\cup\mis_\mi$ during maximality restoration.}
 A low vertex $v$ enters the MIS only when $d_\lo(v)=d_\mi(v)=0$,
 so no middle-MIS vertex needs to leave when $v$ joins the MIS. Similarly,
 a middle vertex joins the MIS only when it has no low- or middle-MIS neighbor.
 In either case, only high-MIS neighbors may need to leave the MIS.
 Thus restoration creates no further candidates through low-MIS removals.
\end{itemize}
The number of new entries into $C_1\cup C_2$ per edge update is therefore
$O(m^{3/8})$. Assigning potential to each member of this union therefore
gives a bounded potential increase per update. Transitions from $C_2$
to $C_1$ leave the union unchanged and do not increase this potential.

\paragraph{Checkpoint: the algorithm when $C_1$ is small.}
Keeping $C_1$ small makes sampled witness maintenance affordable: a
high-MIS vertex needs to maintain sampled edges only to the middle
vertices and to $C_1$. For now, assume $|C_1|<B=m^{1/2}$ throughout an
update. Since $|V_\mi|=O(m^{5/8})$, there are only $O(m^{5/8})$
vertices in $V_\mi\cup C_1$ whose lists may store sampled edges to
high-MIS vertices. We sample each edge joining
$\mis_\hi$ to $V_\mi\cup C_1$ with probability
\[
 p_\hi=\Theta(m^{-1/4}),
\]
so the expected reverse-list size for a fixed sampling universe is
$O(m^{3/8})$, matching the number of high degree neighbors. An empty list
should then imply only $\widetilde O(m^{1/4})$ high-MIS neighbors.
These two bounds let us remove all conflicting high-MIS neighbors of a low or
middle vertex from $\mis_\hi$ in
$\widetilde O(m^{1/4}m^{3/8})=\widetilde O(m^{5/8})$ time; the
vertex can then join the MIS. The probabilistic analysis below justifies
these bounds under adaptive updates.

\paragraph{Requests and the repair queue.}
Repairs may propagate: including a low or middle vertex in the MIS can require removing its high-MIS neighbors from $\mis_\hi$
and leave other vertices needing attention. We keep these pending checks,
called \emph{requests}, in a queue $Q$, with at most one request per
vertex. To check domination, we track low- and middle-MIS neighbors by
exact counters at every vertex; high vertices also track their high-MIS
neighbors exactly. A vertex outside the MIS is requested when these
counters and its sampled witnesses no longer certify domination. Before
acting on a request, we check again, since the vertex may have acquired
an MIS neighbor while waiting.

Every current member of $\mis_\lo\cup\mis_\mi$ remains in the MIS during maximality restoration. A low or middle
vertex enters the MIS only when it has no low- or middle-MIS neighbor
and its sampled high domination edge list is empty. Its remaining
conflicts are therefore all in the high MIS, and we remove them from $\mis_\hi$ before
including the vertex in the MIS. A high vertex joins the MIS only when its exact counters
show that it has no MIS neighbor at all. Whenever a vertex joins the MIS, update the
corresponding counters and witnesses, and the candidate sets follow
their definitions above.

To begin repairing an edge update, update the affected counters and stored domination edges to reflect the edge change. If an insertion joins two MIS vertices, keep the endpoint
with higher priority, giving priority to low vertices, then middle
vertices, then high vertices, with a fixed vertex order within each
class. Process the resulting requests in this same order until $Q$ is
empty.

A high-MIS removal finds the vertices needing attention by visiting its
high degree neighbors and traversing its reverse list. These two traversals
identify every counter and sampled witness that must change, allowing us
to issue the resulting requests without scanning all low degree and middle degree
neighbors. If a high vertex $x$ is removed from $\mis_\hi$ so that a low degree or middle degree neighbor $v$ can join the MIS, then $x$ needs no request: $v$ joins the MIS before the next request is extracted and then dominates $x$.

\paragraph{Amortized update time.}
The queue may have many extractions in one update. To bound their total
amortized cost, assign potential to each vertex outside the MIS, with weights
\[
 W_\lo=W_\mi=\widetilde\Theta(m^{5/8}),\qquad
 W_\hi=\widetilde\Theta(m^{3/8}).
\]
Including a vertex in the MIS decreases the potential by its weight; removing it from the MIS increases the potential by the same weight.
Assuming the sampling bounds justified below, the following calculations
include both the work of removing conflicting vertices from $\mis_\hi$ and the potential
needed for these vertices to rejoin the MIS later.
\begin{itemize}
 \item A change in $\mis_\hi$ visits $O(m^{3/8})$ high degree neighbors. Its reverse
 list has size
 \[
 \widetilde O\bigl(p_\hi(|V_\mi|+|C_1|)\bigr)
 =\widetilde O(m^{3/8}),
 \]
 so a removal, including its potential increase, costs $O(W_\hi)$.
 When a high vertex $x$ joins the MIS, independently sample $V_\mi$ and $C_1$ with probability $p_\hi$ using \Cref{lem:output-sensitive-bernoulli}. Test each sampled vertex $z$ for adjacency to $x$ and store each selected edge in both $R_x^\hi$ and $A_z^\hi$. This constructs $R_x^\hi$ within the same time bound, and the decrease of $W_\hi$ pays for the work.

 \item A vertex with an empty high domination edge list has only
 $\widetilde O(p_\hi^{-1})
 =\widetilde O(m^{1/4})$ actual high-MIS neighbors. Thus removing these
 neighbors from $\mis_\hi$ before a low or middle vertex joins the MIS costs
 \[
 \widetilde O(m^{1/4}W_\hi)=\widetilde O(m^{5/8}).
 \]
 Its full neighborhood scan is also within $\widetilde O(m^{5/8})$.
 The potential decrease of $W_\lo$ or $W_\mi$ offsets both terms.

 \item A stored high domination edge list has size
 \[
 \widetilde O(p_\hi|V_\hi|)=\widetilde O(m^{1/8}).
 \]
 Add $\beta=\widetilde\Theta(m^{1/8})$ to the potential for each
 vertex in $C_1$. When a vertex $z$ leaves $C_1$ during queue processing,
 the potential decreases by $\beta$, paying for deletion of its high domination edge list $A_z^\hi$ and the entry for every stored edge $(z,x)$ in $R_x^\hi$. A vertex in $C_2$ has no such list
 and contributes nothing to this potential term.
\end{itemize}

\subsection{Periodic Candidate Rebuild}
\label{sec:overview-candidate-rebuild}

We now use the slow growth of $C_1\cup C_2$ to pay for a periodic
rebuild, called a \emph{flush}, when $C_1$ becomes large. The flush
must remove candidates from the union altogether: merely moving them
from $C_1$ to $C_2$ would allow them to return without any new candidate
being created. We therefore ensure that every current member of $C_1$
joins $\mis_\lo$ or has a neighbor in $\mis_\lo$.

After applying the edge update and restoring independence, test whether
$|C_1|\geq B=m^{1/2}$, before processing any queued request. If so,
perform a flush; otherwise, construct the missing high domination edge
lists and process the queue. These subsequent repairs can only shrink
$C_1$, so no further threshold test is needed. New low lists are
constructed only when $|C_1|<B$, after this test, ensuring that fewer
than $B$ low lists are stored throughout the execution even when $C_1$
temporarily becomes larger. A flush handles the new candidates directly
without first building their lists.

\paragraph{Flushing while preserving middle domination.}
Let $C_1^0$ be the set $C_1$ at the start of a flush. Remove every current member of $\mis_\hi$ from the MIS and clear the sampled high domination edge lists. Retain every current member of $\mis_\lo\cup\mis_\mi$ in the MIS throughout the flush. Process the vertices of $C_1^0$ in a fixed order. If a vertex has
no low-MIS neighbor, include it in $\mis_\lo$; otherwise it has a neighbor
in $\mis_\lo$ and is removed from $C_1$. Whenever a vertex joins $\mis_\lo$, update
low counters before processing the next vertex.

Every vertex of $C_1^0$ had zero middle count, and the middle MIS is unchanged
during this pass. Hence including these low vertices in $\mis_\lo$ creates no conflict with
$\mis_\mi$. At the end, every former member of $C_1^0$ is either in
$\mis_\lo$ or dominated by it, and therefore leaves $C_1\cup C_2$.
No new candidate is created, because no vertex is removed from $\mis_\lo$.

Next process the middle degree vertices $v$ in fixed order. If $v$ is not in the MIS
and both its low- and middle-MIS-neighbor counters are zero, include $v$ in $\mis_\mi$
and increment its neighbors' middle counters. Complete these changes before processing
the next vertex. After this pass, rebuild an MIS of
\[
 G[V_\hi\setminus N(\mis_\lo\cup\mis_\mi)].
\]
At this point $C_1$ is empty. No vertex is removed from $\mis_\mi$
during the flush, so every vertex left in $C_2$ has a middle-MIS neighbor
throughout the flush.

\paragraph{Amortizing a flush through the candidate union.}
Clearing and rebuilding the high layer costs $\widetilde O(m^{3/4})$:
there are $O(m^{3/8})$ high vertices, and each vertex leaving or joining $\mis_\hi$ costs
$\widetilde O(m^{3/8})$, including witness maintenance. The potential
increase from removals fits the same bound, so the total charge is
\[
 O(|V_\hi|W_\hi)
 =\widetilde O(m^{3/4}).
\]

Starting from an empty candidate union, the total number of times vertices
join $C_1\cup C_2$ over $T$ edge updates is $O(Tm^{3/8})$, counting a
vertex again whenever it rejoins the union. Every flush removes at least
$B=m^{1/2}$ vertices from the union and creates no new candidates.
Consequently, the total number of flushes is at most
\[
 O\!\left(\frac{Tm^{3/8}}{B}\right)=O(Tm^{-1/8}).
\]
Their high-layer reconstruction cost per update is therefore
\[
 \widetilde O\!\left(m^{-1/8}\cdot m^{3/4}\right)
 =\widetilde O(m^{5/8}).
\]
In other words, this bounds the total number of rebuilds; consecutive rebuilds need not
be separated by many updates. Large transfers from $C_2$ to $C_1$ do
not invalidate the bound because they create no new members of the union.

Equivalently, add $\alpha|C_1\cup C_2|$ to the potential, where
$\alpha=\widetilde\Theta(m^{1/4})$. New entries increase this term by
at most $O(m^{3/8})\alpha=\widetilde O(m^{5/8})$ per update, while
each flush decreases it by at least
$\alpha B=\widetilde\Theta(m^{3/4})$, offsetting the reconstruction
cost. Moving vertices between $C_1$ and $C_2$ leaves this potential unchanged.

These estimates explain the choice of $B$: initializing lists for fewer
than $B$ candidates costs $\widetilde O(Bm^{1/8})$, while charging
rebuilds to candidate growth gives $\widetilde O(m^{9/8}/B)$ per
update. Setting $B=m^{1/2}$ balances these contributions at
$\widetilde O(m^{5/8})$.

\paragraph{Checkpoint: the full algorithmic structure.}
To summarize, the algorithm maintains an independent set across the low, middle, and
high degree classes, with boundaries $m^{3/8}$ and $m^{5/8}$. Exact
counters record low- and middle-MIS neighbors at every vertex, and
high-MIS neighbors at high vertices. For middle vertices and primary
candidates in $C_1$, sampled high domination edge lists $A_z^\hi$, together with the entry for each sampled edge $(z,x)$ in $R_x^\hi$, maintain high-MIS witnesses using sampling
probability $p_\hi=\Theta(m^{-1/4})$. The secondary candidates in $C_2$
rely on their exact middle counters instead. After each completed update,
a low vertex outside the MIS has a neighbor in $\mis_\lo$, belongs to
$C_2$ and has a neighbor in $\mis_\mi$, or belongs to $C_1$ and has a
neighbor in $\mis_\hi$.

Each edge update has a preparation phase followed by one of two ways
to restore maximality:
\begin{enumerate}[(1),leftmargin=*]
 \item \emph{Prepare the update.}
 Apply the edge insertion or deletion, and update the affected counters,
 membership in $C_1$ and $C_2$, and stored witnesses. If an insertion joins two MIS
 vertices, remove the lower-priority endpoint from the MIS. Add requests to $Q$ for
 vertices that may no longer be dominated, but initialize new high
 domination edge lists at low vertices only after the flush test of whether $|C_1| < B$.

 \item[(2a)] \emph{Regular mode when $|C_1|<B=m^{1/2}$.}
 Initialize the high domination edge lists of vertices which have newly joined $C_1$, then process requests in $Q$
 in the fixed low--middle--high priority order. Discard requests for
 vertices already in the MIS or whose domination is confirmed by the
 exact counters or a stored domination edge. Include a low vertex in $\mis_\lo$ or a middle vertex in $\mis_\mi$
 only when its exact low and middle counters equal zero and
 its high domination edge list is empty, first removing all of its actual
 high-MIS neighbors from $\mis_\hi$. Include a high vertex in $\mis_\hi$ only when all its exact counters
 equal zero. Update counters and witnesses after each change and process
 the resulting requests until $Q$ is empty.

 \item[(2b)] \emph{Flush mode when $|C_1|\geq B$.}
 Let $C_1^0$ be the set $C_1$ at the start of the flush. Before
 initializing any missing high domination edge lists at low vertices
 or extracting requests, remove every vertex of $\mis_\hi$ from the MIS. Every current member of $\mis_\lo\cup\mis_\mi$ remains in the MIS throughout the flush. Greedily process $C_1^0$ so that each vertex
 joins the low MIS or has a low-MIS neighbor. Next process the middle
 degree vertices $v$ in fixed order: if $v$ is not in the MIS and both
 its low and middle counters are zero, include $v$ in $\mis_\mi$, then increment its neighbors'
 middle counters before processing the next vertex. Finally, rebuild
 the high MIS on the high vertices with no low- or middle-MIS neighbor.
 This branch ends with $C_1=Q=\emptyset$. No vertex is removed from
 $\mis_\mi$ during the flush, so every vertex remaining in $C_2$ has
 a middle-MIS neighbor throughout the flush.
\end{enumerate}
The threshold is tested only once, after preparation. Regular mode
can only remove vertices from $C_1$, while flush mode makes $C_1$ empty. Both modes finish
with an MIS, whose membership changes are then reported.

\paragraph{The potential function.}
We analyze the running time of this algorithm using the potential function
\[
\begin{aligned}
 \Phi={}&W_\lo|V_\lo\setminus\mis|
        +W_\mi|V_\mi\setminus\mis|
        +W_\hi|V_\hi\setminus\mis|\\
       &+\alpha|C_1\cup C_2|+\beta|C_1|+\Theta(\log n)|Q|,
\end{aligned}
\]
with
\[
\begin{gathered}
 W_\lo=W_\mi=\widetilde\Theta(m^{5/8}),\qquad
 W_\hi=\widetilde\Theta(m^{3/8}),\\
 \alpha=\widetilde\Theta(m^{1/4}),\qquad
 \beta=\widetilde\Theta(m^{1/8}).
\end{gathered}
\]
Including a vertex in the MIS while processing the queue decreases one of
the first three terms enough
to pay for its neighborhood scan and any conflicting high-MIS removals.
If including the vertex in the MIS also removes low vertices from $C_1$, the resulting
decrease in $\beta|C_1|$ pays for deleting their high domination edge
lists. A flush uses the same vertex weights to pay for including low and middle
vertices in the MIS, while the decrease in $\alpha|C_1\cup C_2|$ covers clearing
and rebuilding the high layer.

The queue term covers requests discarded without a change in MIS
membership. Increasing the potential by $\Theta(\log n)$ when a request is issued
pays for its later extraction and domination tests, even when no vertex
potential decreases.

The flush test also bounds the cost of vertices entering
$C_1$. If $|C_1|<B$, the time to initialize their missing lists plus the
increase in $\beta|C_1|$ is at most
$\widetilde O(Bm^{1/8})=\widetilde O(m^{5/8})$; the potential then pays
for the queue repairs. Otherwise, the flush empties $C_1$ and cancels
any increase in this term earlier in the update, with its other costs
covered by the vertex weights and the candidate-union decrease.
Thus even when many vertices move from $C_2$ to $C_1$, their total
amortized cost is $\widetilde O(m^{5/8})$.
Together with the $\widetilde O(m^{5/8})$ running time and
vertex-potential increase for the edge change and any forced MIS removal, these bounds
give $\widetilde O(m^{5/8})$ amortized time for the entire update.
Initially $\mis=V$ and $C_1=C_2=Q=\emptyset$, so $\Phi=0$.
Since $\Phi$ is nonnegative, summing the actual running time plus the potential
change over all updates bounds the total running time.

This analysis requires small sampled lists and the guarantee that an empty
high domination edge list implies few actual high-MIS neighbors. The reported MISs reveal information
about these samples, so ordinary independent-sampling estimates do not
yet justify the bounds. We next show why both guarantees hold against
an adaptive adversary.

\subsection{Membership Query Probabilistic Game}
\label{sec:overview-membership-game}

\paragraph{How an adversary could enlarge a reverse list.}
We first show that a high vertex's reverse list remains small throughout
its stay in the MIS. The difficulty is that observations of the MIS may
reveal which incident edges were sampled and let the adversary influence
which sampled edges remain in the reverse list. Fix a high vertex $z$,
and follow $R_z^\hi$ from
its entry into $\mis_\hi$ until its next departure. To see how too
much information could enlarge this list, model the edges from $z$ to
vertices whose high domination edge lists are stored
as a set $U$, with $|U|\leq K$,
and the sampled edges as a subset $S\subseteq U$.  Each new element
receives an independent Bernoulli bit with success probability $p$.
When the fixed universe has the maximum allowed size, the expected value of $|S|$ is $pK$.

If the adversary sees every sampling bit, it can delete each
element outside $S$ and replace it with a fresh element, while keeping
every element in $S$.  After $\Theta(K/p)$ insertion attempts in
expectation, it can fill all $K$ positions with sampled elements.
Thus $|S|$ can reach $K$, a factor $1/p$ larger than its fixed-set
expectation.  Independence of the bits when they are first generated
does not prevent this attack: the set $U$ has been chosen
using those bits.

The attack above relies on a crucial ability: the adversary must
identify nonsampled edges while keeping $z$ in the MIS, so that it
can repeatedly replace them and accumulate sampled edges in
$R_z^\hi$.  This is where observing the MIS differs from seeing
the sampling bits directly.  Our key observation is that, when the
MIS reveals that an edge $(z,v)$ was not sampled, the same event can
cause $v$ to enter the MIS and $z$ to leave it.  The adversary then
loses $R_z^\hi$, rather than gaining
another opportunity to alter its contents.  To turn this observation
into a size bound, we must characterize exactly which membership
answers the MIS can reveal while $z$ remains in $\mis_\hi$.

\paragraph{The membership-query game.}
We formalize the information available to an adaptive player as the
following game, played in discrete steps.
\begin{description}[
  leftmargin=3.4cm,
  labelwidth=3.0cm,
  style=multiline,
  font=\normalfont\itshape]
 \item[Universe and sample.]
 Initially, the universe is empty; thereafter it is a set $U$ with $|U|\leq K$.
 Each newly inserted element $e$ receives a hidden, independent bit
 $X_e\sim\operatorname{Bernoulli}(p)$, where $p<1/2$, and
 $S=\{e\in U\mid X_e=1\}$ is the sampled subset of $U$.  Both $U$ and
 $S$ refer to the current step of the game.

 \item[Updates.]
 The player may insert a fresh element into $U$ or delete an element
 $e\in U$ from $U$ and $S$. A deletion reveals $X_e$ and discards it;
 every later insertion, including one with the same name, receives
 a fresh independent bit.

 \item[Membership query.]
 The player may choose an element $e\in U$ and ask whether
 $e\in S$.  The answer reveals $X_e$.

 \item[Termination.]
 If a membership query returns yes, the game continues.  The first no
 answer terminates the game immediately.  Before termination, the
 player may choose each operation adaptively from the revealed history.
\end{description}

This restriction prevents the attack above.  The player cannot test
an element $e\in U$, discover that $e\notin S$, and replace it
while continuing the game. For any $q$, the probability that membership
queries on $q$ distinct elements all return yes before termination is at
most $p^q$.  Consequently, only $O(\log n)$ distinct elements have
been identified as sampled by membership queries, with high probability.
Deletion operations may reveal additional bits equal to one, but their
elements immediately leave the sample and cannot return with those bits.
Thus, before termination, only $O(\log n)$ elements $e\in U$ have
revealed bits $X_e=1$. Conditional on the history, the unrevealed bits
of elements in $U$ are mutually independent Bernoulli-$p$ random variables.
Concentration therefore gives
\[
 |S|=O(pK+\log n)=\widetilde O(1+pK)
\]
simultaneously over a polynomial number of operations. The terminating
query does not change $S$, so the bound also holds after that query.
\Cref{lem:adaptive-stopped-membership} formalizes this argument.

\paragraph{Why the algorithm removes the high vertex after a no answer.}
Before including a low or middle vertex $v$ in the MIS, the algorithm removes every neighbor of $v$ in $\mis_\hi$, including $z$, from the MIS. Thus, when the decision to include $v$ depends only on whether $(z,v)$ was sampled, a no answer establishes that $v$ has no sampled witness. The algorithm then removes $z$ from $\mis_\hi$ before including $v$ in the MIS, ending the game for $R_z^\hi$. A yes answer establishes that $v$ is dominated; the algorithm keeps $z$ in $\mis_\hi$ and leaves $v$ outside the MIS. This is precisely the
stopping rule of the membership game.

To apply the game, let $U$ contain the graph edges from $z$ to vertices
whose high domination edge lists are stored, and let $S$ be its reverse list:
\[
 U=\{(z,v)\in E(G):A_v^\hi\text{ is stored}\},\qquad S=R_z^\hi.
\]
Fix the adversary's internal randomness and all sampling coins except
those defining $R_z^\hi$. Simulate the algorithm and adversary while
keeping these excluded coins hidden. The same coin determines whether
$(z,v)$ appears in $R_z^\hi$ and $A_v^\hi$, so the simulator cannot
inspect the copy in $A_v^\hi$ separately.
The simulator can decide whether
to include $v$ in the MIS without inspecting $(z,v)$ whenever MIS membership, exact
counters, or other sampled witnesses already settle that decision.
It needs a membership query only when this one bit determines whether
$v$ enters the MIS. Inserting a graph edge $(z,v)$ while $A_v^\hi$
is stored, or initializing $A_v^\hi$ at a neighbor of $z$, uses the
game's insertion operation. Deleting that edge or discarding $A_v^\hi$
uses its reveal-and-delete operation. For an edge $(z,v)$ still in $U$,
the simulator must defer the membership query until $v$ is extracted
from the queue.

\begin{figure}[!htbp]
\centering
\definecolor{simPlayer}{RGB}{52,103,135}
\definecolor{simList}{RGB}{176,107,40}
\begin{tikzpicture}[x=1cm,y=1cm,>=stealth,font=\small]
  % The simulator contains the adversary, the simulated data structure,
  % and all sampling coins except those defining R_z^hi.
  \draw[rounded corners=5pt,draw=simPlayer,line width=.9pt,
    fill=simPlayer!6] (0,.15) rectangle (5.4,4.35);
  \node[font=\bfseries,text=simPlayer] at (2.7,4.02) {Simulator (game player)};
  \draw[rounded corners=4pt,draw=simPlayer!70,fill=white]
    (.4,2.68) rectangle (2.4,3.56);
  \draw[rounded corners=4pt,draw=simPlayer!70,fill=white]
    (3.0,2.68) rectangle (5.0,3.56);
  \node[font=\footnotesize,align=center] at (1.40,3.12)
    {Adaptive\\adversary};
  \node[font=\footnotesize,align=center] at (4.00,3.12)
    {MIS data\\structure};
  \draw[<->,draw=simPlayer,line width=.8pt]
    (2.42,3.12) -- (2.98,3.12);
  \draw[rounded corners=4pt,draw=simPlayer!70,fill=white]
    (.5,.72) rectangle (4.9,1.65);
  \node[font=\footnotesize,align=center] at (2.7,1.19)
    {Sampling coins except those\\defining $R_z^\hi$};
  \draw[->,draw=simPlayer,line width=.8pt]
    (4.0,1.67) -- (4.0,2.66);

  % U contains the edges from z to vertices with stored high domination edge lists;
  % its sampled subset S is the entire reverse list.
  \draw[rounded corners=5pt,draw=simList,line width=.9pt,
    fill=simList!6] (9.8,.15) rectangle (15.2,4.35);
  \node[font=\bfseries,text=simList] at (12.5,4.02)
    {Reverse list $R_z^\hi$};
  \draw[rounded corners=4pt,draw=simList!75,fill=white]
    (10.15,.67) rectangle (14.85,3.59);
  \node[font=\footnotesize,text=simList,align=center] at (12.50,3.20)
    {$U$: edges $(v,z)$\\with $A_v^\hi$ stored};
  \draw[rounded corners=4pt,draw=simList,line width=.8pt,
    fill=simList!12] (10.90,1.35) rectangle (14.10,2.62);
  \node[font=\small] at (12.50,1.98)
    {$S=R_z^\hi$};

  % Requests go to the reverse list. Only the game's permitted
  % answers return to the player.
  \draw[->,draw=black!70,line width=.9pt] (5.60,3.32) -- (9.60,3.32);
  \node[font=\scriptsize,align=center] at (7.60,3.70)
    {update $U$: insert / delete $e$};
  \draw[->,draw=black!70,line width=.9pt] (5.60,2.34) -- (9.60,2.34);
  \node[font=\scriptsize] at (7.60,2.60)
    {query: $e\in S$?};
  \draw[<-,draw=simList,line width=.9pt] (5.60,1.40) -- (9.60,1.40);
  \node[font=\scriptsize] at (7.60,1.67)
    {reveal-and-delete: reveal $X_e$};
  \draw[<-,draw=simList,line width=.9pt] (5.60,.49) -- (9.60,.49);
  \node[font=\scriptsize] at (7.60,.76)
    {yes: continue; no: stop};
\end{tikzpicture}
\caption{Simulation interface for the reverse list $R_z^\hi$. The simulator runs
the adaptive adversary and the MIS data structure with the adversary's
internal randomness and all sampling coins except those defining $R_z^\hi$
fixed. It changes $U$ and learns about $S$ only through
the membership game's permitted operations and answers.}
\label{fig:membership-game-simulation}
\end{figure}
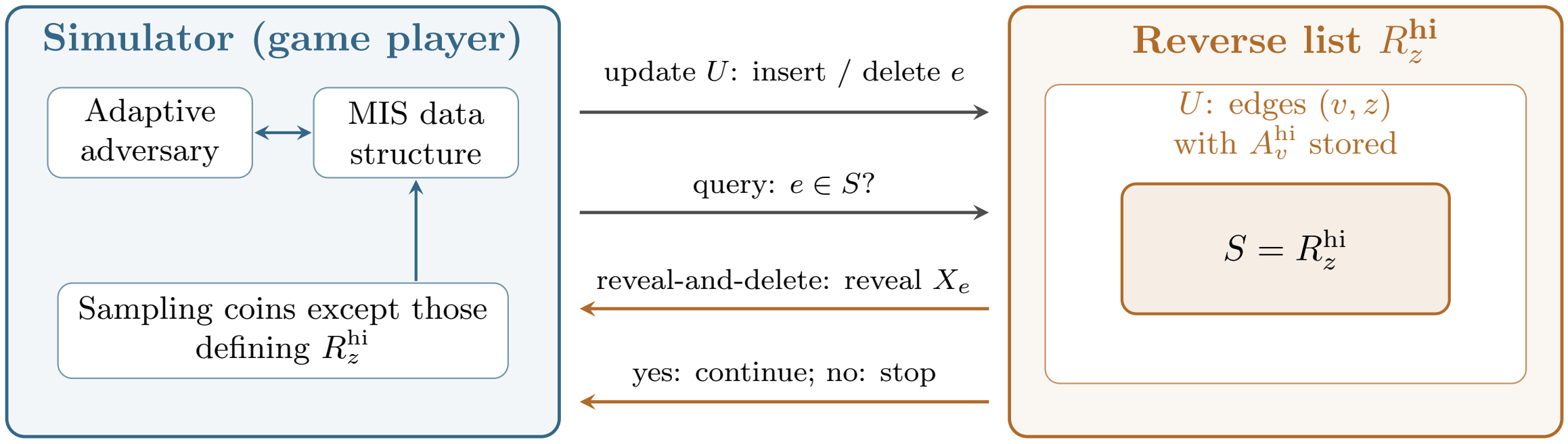

Querying while issuing a request would be too early: after a no answer, the algorithm would only queue $v$, while $z$ could remain in the MIS. The simulator
therefore queues $v$ without inspecting the hidden bit and postpones
the query until extraction.
Every vertex outside the MIS with no positive maintained MIS-neighbor
counter and no nonempty stored high domination edge list is already in
the real queue before extraction. A request present only in the simulation
must therefore find its vertex already in the MIS or certified as
dominated by a counter or stored witness, so processing it changes
only the simulated queue. Using the same fixed priority order therefore
preserves all MIS changes and reported outputs. This
lets the simulator defer each membership query until a no answer is followed by the algorithm removing $z$ from $\mis_\hi$. The full simulation is proved in
\Cref{lem:adaptive-reverse-list-size};
\Cref{fig:membership-game-simulation} shows its interface.

\paragraph{The resulting reverse-list bound.}
The membership game gives the $\widetilde O(m^{3/8})$ reverse-list
bound needed for efficient high-MIS removals, even under adaptive updates.
Indeed, a reverse list samples edges to $O(m^{5/8})$ middle vertices
and fewer than $B=m^{1/2}$ low vertices with stored lists. Its universe
therefore has capacity $K=O(m^{5/8})$, and the game gives
\[
 |R_z^\hi|=\widetilde O(1+p_\hi K)
          =\widetilde O(m^{-1/4}m^{5/8})
          =\widetilde O(m^{3/8})
\]
with high probability throughout the execution. Together with the
$O(m^{3/8})$ bound on the number of high degree neighbors, this ensures that
removing a vertex from $\mis_\hi$ and finding the resulting requests takes
only $\widetilde O(m^{3/8})$ time.

\subsection{Emptiness Query Probabilistic Game}
\label{sec:overview-emptiness-game}

\paragraph{Modeling a high domination edge list.}
To make including a low vertex in the MIS affordable, an empty sampled list must imply
that the vertex has few actual high-MIS neighbors. Fix a low candidate
$v\in C_1$ whose high domination edge list is stored, and follow that
list until it is discarded. Let
\[
 U=\{(x,v):x\in N(v)\cap\mis_\hi\},
 \qquad S=A_v^\hi\subseteq U.
\]
Then $|U|$ counts exactly the high-MIS neighbors that must be removed
before including $v$ in the MIS. With $p=p_\hi=\Theta(m^{-1/4})$, our goal is
\[
 S=\emptyset\quad\Longrightarrow\quad
 |U|=\widetilde O(p^{-1})=\widetilde O(m^{1/4}).
\]
For a fixed universe, the probability of missing every edge is
$(1-p)^{|U|}\leq e^{-p|U|}$, which gives this implication with high
probability. We must prove the same implication when observations of the
MIS reveal whether the list is empty and guide the adversary's later updates.

\paragraph{Why unrestricted emptiness queries are too powerful.}
Suppose the player may continue after either answer to an emptiness
query.  Starting from $U=S=\emptyset$, it inserts one fresh element
$e$ and asks whether $S$ is empty.  If the answer is nonempty, then
$e$ must be sampled, so the player deletes $e$.  If the answer is
empty, then $e$ is not sampled, so the player keeps it in $U$.
In either case $S$ is empty again at the end of the round, and every
element left in $U$ is known to be unsampled.  Repeating the procedure
therefore reveals the sampling bit of each new element using only
emptiness queries.  The player can accumulate arbitrarily many
unsampled elements while keeping $S$
empty after every round. Consequently, unrestricted access to emptiness answers cannot justify the claimed bound on the universe size when the sample is empty.

\paragraph{The emptiness-query game.}
We will simulate the algorithm so that it queries a list's emptiness
only when its vertex is being processed, is outside the MIS, and has
no domination certified by the exact counters. At that point, an empty answer establishes that the vertex has no sampled high-MIS witness. The algorithm then removes all its neighbors in $\mis_\hi$ from the MIS before including the vertex in the MIS. The universe of possible high witnesses becomes empty,
ending the current game. This motivates the following stopping rule,
which prevents the repeated empty queries used in the attack above.

\begin{description}[
  leftmargin=3.4cm,labelwidth=3.0cm,style=multiline,
  font=\normalfont\itshape]
 \item[Universe and sample.]
 The game begins with $U=S=\emptyset$. Each newly inserted element
 receives a hidden, independent Bernoulli-$p$ bit, and $S$ consists of
 the elements in $U$ whose bits equal one.
 \item[Updates.]
 The player may insert a fresh element into $U$ or delete an element
 $e\in U$ from $U$ and $S$. Deletion reveals no bit, and every later
 insertion, including one with the same name, receives a fresh
 independent bit.
 \item[Emptiness query.]
 The player may ask whether $S$ is empty. The answer reveals only whether
 $S=\emptyset$ and is the only information about the hidden bits.
 \item[Termination.]
 A nonempty answer allows the game to continue. The first empty answer
 ends the game immediately. Operations may depend on the preceding
 answers.
\end{description}

In the attack above, the first query that identifies an unsampled element
now ends the game, preventing its repeated use to build a large unsampled
universe.

\paragraph{Why the stopping rule gives the required bound.}
Before the game stops, every query has the same answer: nonempty. Thus,
once the player's randomness is fixed, there is only one continuing
sequence of operations, and the set $U$ at each step of that
sequence is fixed. Reaching a step requires certain earlier sets to
contain sampled elements. These requirements favor sampling bits equal
to one, whereas an empty current sample requires $X_e=0$ for every $e\in U$. For a
step reached with positive probability, the event $H$ that it is
reached is increasing in the sampling bits, while the event $E$ that
its sample is empty is decreasing. Harris's inequality therefore gives
\[
 \Pr[E\mid H]\leq\Pr[E]
 =(1-p)^{|U|}\leq e^{-p|U|},
\]
whether or not the player queries $S$ at this step. A union bound over
polynomially many steps shows that, with high probability, after every
update or query, $S=\emptyset$ implies $|U|=\widetilde O(p^{-1})$.

The sample-size bound uses the same fixed sequence. For each fixed universe
on that sequence, the sampling bits are independent before conditioning
on the event that its step is reached. Ordinary Chernoff concentration
therefore gives $|S|=\widetilde O(1+p|U|)$ before this conditioning.
Reaching the step may bias the bits, but a reached step with an oversized
sample is still contained in the corresponding fixed-set failure event.
A union bound therefore also bounds the sample size throughout the game.
\Cref{lem:adaptive-hidden-sample} gives the formal argument.

\begin{figure}[!htbp]
	\centering
	\definecolor{emptyPlayer}{RGB}{52,103,135}
	\definecolor{emptyList}{RGB}{176,107,40}
	\begin{tikzpicture}[x=1cm,y=1cm,>=stealth,font=\small]
		\draw[rounded corners=5pt,draw=emptyPlayer,line width=.9pt,
		fill=emptyPlayer!6] (0,.15) rectangle (5.4,4.35);
		\node[font=\bfseries,text=emptyPlayer] at (2.7,4.02)
		{Simulator (game player)};
		\draw[rounded corners=4pt,draw=emptyPlayer!70,fill=white]
		(.4,2.68) rectangle (2.4,3.56);
		\draw[rounded corners=4pt,draw=emptyPlayer!70,fill=white]
		(3.0,2.68) rectangle (5.0,3.56);
		\node[font=\footnotesize,align=center] at (1.40,3.12)
		{Adaptive\\adversary};
		\node[font=\footnotesize,align=center] at (4.00,3.12)
		{MIS data\\structure};
		\draw[<->,draw=emptyPlayer,line width=.8pt]
		(2.42,3.12) -- (2.98,3.12);
		\draw[rounded corners=4pt,draw=emptyPlayer!70,fill=white]
		(.5,.72) rectangle (4.9,1.65);
		\node[font=\footnotesize,align=center] at (2.7,1.19)
		{Sampling coins except those\\defining $A_v^\hi$};
		\draw[->,draw=emptyPlayer,line width=.8pt]
		(4.0,1.67) -- (4.0,2.66);
		
		\draw[rounded corners=5pt,draw=emptyList,line width=.9pt,
		fill=emptyList!6] (9.8,.15) rectangle (15.2,4.35);
		\node[font=\bfseries,text=emptyList] at (12.5,4.02)
		{List $A_v^\hi$};
		\draw[rounded corners=4pt,draw=emptyList!75,fill=white]
		(10.15,.67) rectangle (14.85,3.59);
		\node[font=\footnotesize,text=emptyList,align=center] at (12.50,3.20)
		{$U$: edges from $v$\\to $\mis_\hi$};
		\draw[rounded corners=4pt,draw=emptyList,line width=.8pt,
		fill=emptyList!12] (10.90,1.35) rectangle (14.10,2.62);
		\node[font=\small] at (12.50,1.98)
		{$S=A_v^\hi$};
		\node[font=\scriptsize,text=emptyList] at (12.50,.97)
		{updates reveal no sampling bits};
		
		\draw[->,draw=black!70,line width=.9pt] (5.60,3.32) -- (9.60,3.32);
		\node[font=\scriptsize,align=center] at (7.60,3.70)
		{update $U$: insert / delete $e$};
		\draw[->,draw=black!70,line width=.9pt] (5.60,2.34) -- (9.60,2.34);
		\node[font=\scriptsize] at (7.60,2.60)
		{query: $S=\emptyset$?};
		\draw[<-,draw=emptyList,line width=.9pt] (5.60,1.40) -- (9.60,1.40);
		\node[font=\scriptsize] at (7.60,1.67)
		{nonempty: continue};
		\draw[<-,draw=emptyList,line width=.9pt] (5.60,.49) -- (9.60,.49);
		\node[font=\scriptsize] at (7.60,.76)
		{empty: stop};
	\end{tikzpicture}
	\caption{Simulation interface for the high domination edge list $A_v^\hi$.
		The simulator runs the adaptive adversary and the MIS data structure with
		the adversary's internal randomness and all sampling coins except those
		defining $A_v^\hi$ fixed. Updates reveal no sampling bits. An
		emptiness query reveals only whether $S$ is empty, and the first empty
		answer ends the game.}
	\label{fig:emptiness-game-simulation}
\end{figure}
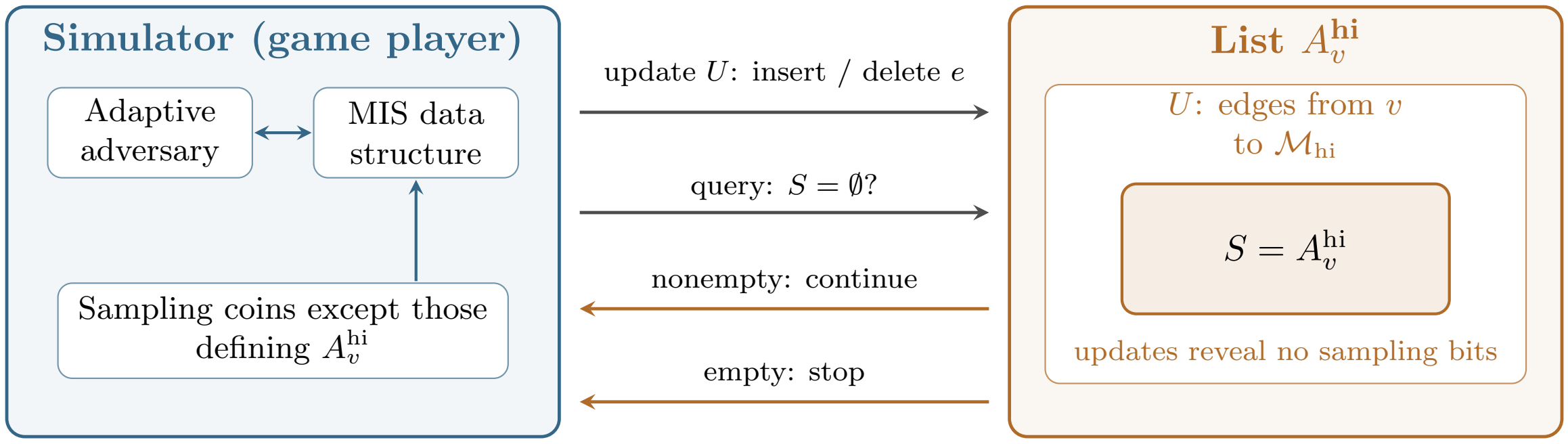

To apply this game, fix the adversary's internal randomness and all
sampling coins except those defining $A_v^\hi$.
The simulator must reproduce the MIS outputs using
only emptiness answers about $A_v^\hi$. In particular, deleting a
potential witness must not reveal whether its edge was sampled. We
keep the entries of $A_v^\hi$ and the corresponding entries in the lists $R_x^\hi$ at its high-MIS neighbors hidden and allow extra requests
for $v$, so that the simulator can arrange to reconsider $v$ without
inspecting individual sampling bits.
\Cref{fig:emptiness-game-simulation} depicts the permitted information.

\paragraph{Deferring a query until the request is processed.}
As in the membership simulation, whenever the real algorithm would test
$A_v^\hi$ before queuing $v$, the simulator checks $v\notin\mis$ and
$d_\lo(v)=d_\mi(v)=0$. If both conditions hold, it queues $v$ without
querying the list and postpones the emptiness query until extraction.
The same queue argument applies: any request absent from the real queue
is rejected by the repair checks, and the fixed priority order preserves
the sequence of vertices joining the MIS and hence all reported outputs.

At extraction, the simulator discards the request if $v\in\mis$ or
$d_\lo(v)+d_\mi(v)>0$. Otherwise, it asks whether $A_v^\hi$ is empty.
A nonempty answer establishes that $v$ is dominated. On an empty answer, the algorithm removes all of $v$'s neighbors in $\mis_\hi$ from the MIS and then includes $v$ in the MIS. Thus every queried empty
answer ends the game, as required. For a low vertex, discarding its
list also ends the game, and a later list starts with fresh samples.
\Cref{lem:adaptive-high-list-size} gives the full simulation argument.

\paragraph{The high domination edge list of a middle vertex.}
The same argument applies to middle vertices: including a middle vertex
in the MIS removes all its high-MIS neighbors, so both $U$ and $S$ are
empty. The algorithm does not delete the list object $A_v^\hi$ or
change its pointer, but the list contains no edges and all its sampling
bits have been discarded. A new game therefore starts with an empty
universe and an empty sample; every insertion uses a fresh independent
bit. A flush creates the same reset by clearing the high MIS. Applying the game bound between
successive resets, starting at list initialization, covers the entire
period during which the middle vertex's list is stored.

The emptiness game supplies both remaining estimates needed to include low and middle
vertices in the MIS at low cost. Since each list samples edges to at most
$|V_\hi|=O(m^{3/8})$ high vertices, the following bounds hold while
the list is stored:
\[
 |A_v^\hi|=\widetilde O(1+p_\hi|V_\hi|)
           =\widetilde O(m^{1/8}),
 \qquad
 A_v^\hi=\emptyset\ \Longrightarrow\
 |N(v)\cap\mis_\hi|=\widetilde O(m^{1/4}).
\]
The first estimate bounds the cost of deleting a candidate's stored
list. The second ensures that at most $\widetilde O(m^{1/4})$ high-MIS neighbors
need to leave before $v$ joins the MIS. Each such removal costs $\widetilde O(m^{3/8})$ by
the reverse-list bound, so all conflicts together cost
$\widetilde O(m^{5/8})$. These are precisely the sampling guarantees
used in the amortized analysis above.

\clearpage

\section{Preliminaries}\label{sec:prelim}

\paragraph{Output-sensitive independent sampling.}
We repeatedly sample elements of an indexed set independently with probability \(p\). Generating one sampling bit per element would make the running time depend on the size of the input set. The following standard geometric-skipping procedure makes the running time depend only on the number of selected elements, up to logarithmic factors.

An \emph{indexed set} is a set \(S\) of size at most $n$ equipped with an arbitrary
order and supporting \(\operatorname{size}(S)\), insertion, deletion, and
\(\operatorname{select}(S,i)\), which returns the \(i\)-th element in the
current order. A balanced binary search tree augmented with subtree sizes supports all these operations in \(O(\log n)\) time.

For \(p\in(0,1]\), let \(\operatorname{Geo}(p)\) denote a geometric random variable counting the number of failures before the next success:
\[
\Pr[\operatorname{Geo}(p)=k]=(1-p)^kp,
\qquad k=0,1,2,\ldots.
\]
It can be generated in \(\widetilde O(1)\) time from independent random
bits, for example by inverse transform, to the \(O(\log n)\) bits of
precision needed to distinguish positions in a set of polynomial size.

\begin{algorithm}[H]
	\DontPrintSemicolon
	\KwIn{An indexed set \(S\) and a probability \(p\in(0,1]\)}
	\KwOut{A subset \(T\subseteq S\)}
	\(T\gets\emptyset\), \(i\gets0\)\;
	\While{\(i<\operatorname{size}(S)\)}{
		\(g\gets\operatorname{Geo}(p)\), \(i\gets i+g+1\)\;
		\If{\(i\leq\operatorname{size}(S)\)}{
			\(T\gets T\cup\{\operatorname{select}(S,i)\}\)\;
		}
	}
	\Return \(T\)\;
	\caption{\(\operatorname{BernoulliSample}(S,p)\)}
	\label{alg:bernoulli-sample}
\end{algorithm}

\begin{lemma}[Bernoulli sampling]
	\label{lem:output-sensitive-bernoulli}
	For any fixed indexed set \(S\) and \(p\in(0,1]\),
	\(\operatorname{BernoulliSample}(S,p)\) includes each element of \(S\)
	independently with probability \(p\).  If the returned set has size \(s\),
	the running time is
	\(
	\widetilde O(1+s).
	\)
\end{lemma}

\begin{proof}
	Consider a sequence of independent zero--one random variables, each equal to one with probability \(p\). The number of zeros before its first one has
	the distribution \(\operatorname{Geo}(p)\).  After that one, the number of
	zeros before the next one is a fresh independent copy of
	\(\operatorname{Geo}(p)\), and so on.  Consequently, the successive values
	\[
	i_1=g_1+1,\qquad
	i_2=i_1+g_2+1,\qquad\ldots
	\]
	generated by \Cref{alg:bernoulli-sample} are exactly the positions of the
	ones in the independent coin sequence.  Hence each element is selected
	independently with probability \(p\).

	The procedure generates one geometric value and makes at most one call to $\operatorname{select}(S,i)$ for every returned element, with at most one additional geometric value that jumps beyond the end of \(S\). Its time is therefore
	\(\widetilde O(1+s)\).
\end{proof}

\paragraph{Probabilistic games.}
We use two probabilistic game lemmas to capture robustness against adaptive
adversaries.  The first bounds the size of a hidden sample under membership queries.  The second uses emptiness queries to bound both
the sample size throughout the game and the universe size whenever the sample
is empty.
In either game, one step consists of one permitted update or query.
In our applications, a single edge update of the graph may generate
several game steps.

\begin{samepage}
\begin{lemma}
	\label{lem:adaptive-stopped-membership}
	Fix \(p\in(0,1/2)\), integers \(k\geq2\) and \(M\geq1\), and a failure
	probability \(\eta\in(0,1)\).  A player takes at most \(M\) steps in the
	following game.
	\begin{description}[
	  leftmargin=3.4cm,
	  labelwidth=3.0cm,
	  style=multiline,
	  font=\normalfont\itshape]
		\item[Universe and sample.]
		The set \(U\) starts empty and always satisfies
		\(|U|\leq k\).  Each new element \(e\) receives a hidden, independent
		bit \(X_e\sim\operatorname{Bernoulli}(p)\), and the sampled subset of \(U\) is
		\(S=\{e\in U:X_e=1\}\).

		\item[Updates.]
		The player may insert a fresh element into \(U\) or \emph{reveal and delete}
		an element \(e\in U\): this reveals \(X_e\), removes \(e\) from
		\(U\) and \(S\), and discards \(X_e\).
		Every later insertion, including one with the same name, creates a new element with
		a fresh independent bit.

		\item[Membership query.]
		The player may choose an element \(e\in U\) and ask whether it belongs to
		\(S\). The answer reveals \(X_e\); repeated queries of the same
		element are allowed while it belongs to \(U\).

		\item[Termination.]
		A yes answer lets the game continue; the first no answer ends it
		immediately.  The player may choose each operation adaptively from
		all preceding query answers and deletion revelations.
	\end{description}
	Then, with probability at least \(1-\eta\), simultaneously after every
	update or query, including the query that terminates the game,
	\[
		|S|=O\left(pk+\log\frac{M+1}{\eta}\right).
	\]
\end{lemma}
\end{samepage}

\begin{proof}
	Fix the player's internal randomness and assign the independent bit \(X_e\)
	on insertion of each element \(e\).
	Let \(R\) be the set of elements for which a membership query has returned
	yes. Querying an element already in \(R\) reveals no new bit. A reveal-and-delete operation may reveal a
	bit, but it immediately deletes the corresponding element from \(U\)
	and \(S\). The deleted element cannot be queried again, and its bit
	is never used in a later insertion.

	Consider the first membership query to an element outside \(R\).  Conditional
	on the entire preceding history, including the outcomes of all
	reveal-and-delete operations, the queried element has an unrevealed
	Bernoulli\((p)\) bit. Indeed, an element whose bit was revealed by a reveal-and-delete operation is no
	longer available.  It follows inductively that
	\[
		\Pr[\textnormal{membership queries on at least \(q\) distinct elements
		return yes}]\leq p^q.
	\]
	Since \(p<1/2\), this bound implies that,
	except with probability at most \(\eta/2\), only
	\(O(\log(1/\eta))\) distinct elements ever receive a yes answer to
	their first membership query. In particular, \(|S\cap R|\leq |R|=O(\log(1/\eta))\)
	after every update or query.

	Fix a step reached before termination and condition on all preceding
	operations and revealed answers. This history fixes \(U\) and \(R\).
	Every revealed bit belongs to an element in \(R\) or to a deleted element.
	Consequently, conditional on this history, the bits of the elements in
	\(U\setminus R\) remain mutually independent Bernoulli variables with
	parameter \(p\).
	Hence
	\[
		Y:=|S\cap(U\setminus R)|
		\sim\operatorname{Bin}(|U\setminus R|,p),
		\qquad
		\mathbb E[Y]\leq pk.
	\]
	For a binomial variable \(Z\) with mean \(\mu\), we use the Chernoff bound
	\[
		\Pr[Z>2\mu+t]\leq e^{-t}\qquad(t\geq0).
	\]
	Put \(\lambda=\log(2(M+1)/\eta)\).  Applying this bound conditional
	on the preceding history gives
	\[
		\Pr[Y>2pk+\lambda\mid\textnormal{the preceding history}]
		\leq e^{-\lambda}.
	\]
	Averaging over the history and taking a union bound over all steps shows
	that this bound fails before termination with probability at most
	\((M+1)e^{-\lambda}=\eta/2\).

	When both bounds hold, at every step before termination we have
	\[
		|S|
		=|S\cap R|+|S\cap(U\setminus R)|
		\leq |R|+Y
		=O\left(pk+\log\frac{M+1}{\eta}\right).
	\]
	A terminating no answer does not change \(S\), so the same estimate holds
	after that query.
	The union of the two failure events has probability at most \(\eta\). Averaging over
	the fixed internal randomness proves the result for a randomized player as
	well.
\end{proof}

\begin{definition}[Increasing and decreasing events]
\label{def:monotone-events}
For two binary vectors \(x,y\in\{0,1\}^m\), write \(x\leq y\) if
\(x_i\leq y_i\) for every \(i\).  Thus, \(y\) can be obtained from \(x\) by
changing some zeroes into ones.  An event \(\mathcal E\subseteq\{0,1\}^m\) is
\emph{increasing} if
\[
	x\in\mathcal E\text{ and }x\leq y
	\quad\Longrightarrow\quad y\in\mathcal E;
\]
once an increasing event holds, changing zeroes into ones cannot make it fail.
An event \(\mathcal D\) is \emph{decreasing} if
\[
	x\in\mathcal D\text{ and }y\leq x
	\quad\Longrightarrow\quad y\in\mathcal D;
\]
once a decreasing event holds, changing ones into zeroes cannot make it fail.
\end{definition}

\begin{lemma}[Harris's inequality]
\label{lem:harris-inequality}
	Let \(X_1,\ldots,X_m\) be independent Bernoulli random variables.  If
	\(\mathcal E_1\) and \(\mathcal E_2\) are increasing events, then
	\[
		\Pr[\mathcal E_1\cap\mathcal E_2]
		\geq \Pr[\mathcal E_1]\Pr[\mathcal E_2].
	\]
	Equivalently, if \(\mathcal E\) is increasing and \(\mathcal D\) is
	decreasing, then
	\[
		\Pr[\mathcal E\cap\mathcal D]
		\leq \Pr[\mathcal E]\Pr[\mathcal D].
	\]
\end{lemma}

The second game provides two guarantees using only emptiness queries.

\begin{lemma}
\label{lem:adaptive-hidden-sample}
Fix $p\in(0,1]$, integers $k\geq1$ and $M\geq1$, and a failure
probability $\eta\in(0,1)$. A player takes at most $M$ steps in the
following game.
\begin{description}[
 leftmargin=3.4cm,labelwidth=3.0cm,style=multiline,
 font=\normalfont\itshape]
 \item[Universe and sample.]
 The set $U$ starts empty and always satisfies $|U|\leq k$.
 Each newly inserted element $e$ receives a hidden, independent bit
 $X_e\sim\operatorname{Bernoulli}(p)$, and
 $S=\{e\in U:X_e=1\}$ is the sampled subset of $U$.
 \item[Updates.]
 The player may insert a fresh element into $U$ or delete an element
 $e\in U$ from $U$ and $S$. Deletion discards $X_e$ without revealing it.
 Every later insertion, including one with the same name, creates a new
 element with a fresh independent bit.
 \item[Emptiness query.]
 The player may ask whether $S$ is empty. This answer is the only
 information it receives about the hidden bits.
 \item[Termination.]
 A nonempty answer lets the game continue; the first empty answer ends it
 immediately. The player may choose each operation adaptively from all
 preceding answers, and may also stop voluntarily.
\end{description}
Then, with probability at least $1-\eta$, simultaneously after every
update or query, including the query that terminates the game,
\begin{align*}
 |S|&=O\left(pk+\log\frac{M+1}{\eta}\right),\\
 S=\emptyset&\quad\Longrightarrow\quad
 |U|=O\left(p^{-1}\log\frac{M+1}{\eta}\right).
\end{align*}
The second implication holds whether or not the player asks an emptiness
query at that state.
\end{lemma}

\begin{proof}
Fixing the player's internal randomness reduces the analysis to a deterministic strategy. Give each freshly inserted element a distinct
index $i$, and let $X_i$ be the independent Bernoulli coin assigned to
that element at insertion. An insertion with the same name as a deleted
element receives a new index and an independent Bernoulli random variable.

Write $U_t$ and $S_t$ for the sets $U$ and $S$ after step $t$,
with $U_0=S_0=\emptyset$, and put $\lambda=\log(2(M+1)/\eta)$.
Fix $t\in\{1,\ldots,M\}$ for which the player performs at least $t$
steps with positive probability. Every earlier query must have answered
nonempty, since an empty answer would have terminated the game. These
answers and the fixed strategy determine all operations through step $t$,
and hence the sets of element indices $U_s$ for $s\leq t$. The event that
the player performs at least $t$ steps is
\[
 H_t:=\bigcap_{\substack{s<t\\s\text{ is a query step}}}
 \left\{\sum_{i\in U_s}X_i\geq1\right\}.
\]
This event is increasing: changing any coin from zero to one can only
increase each sum in its definition. The event that the sampled set after
step $t$ is empty is
\[
 E_t:=\{X_i=0\text{ for every }i\in U_t\},
\]
which is decreasing, whether or not step $t$ is a query. Applying
\Cref{lem:harris-inequality} to the increasing event $H_t$ and the
decreasing event $E_t$, we obtain
\[
 \Pr[E_t\mid H_t]
 \leq\Pr[E_t]
 =(1-p)^{|U_t|}
 \leq e^{-p|U_t|}.
\]
Thus, if $|U_t|\geq\lambda/p$, the probability that the player performs
step $t$ and $S_t=\emptyset$ is at most $e^{-\lambda}$. A union bound
over the at most $M$ steps gives failure probability at most
$Me^{-\lambda}\leq\eta/2$. Before the first step, $|U_0|=0$.
The union bound includes steps at which the player does not query $S$,
as well as the query that terminates the game.

For the sample-size bound, let $\mathcal B_t$ be the event that the player
performs at least $t$ steps and $|S_t|>2pk+\lambda$. If no sampling
outcome allows the player to perform $t$ steps, then $\mathcal B_t$ is
empty. Otherwise, as above, the fixed strategy and the common preceding
nonempty answers determine the same set $U_t$ on all outcomes
that reach this step. Put
\[
 Z_t:=\sum_{i\in U_t}X_i.
\]
For this fixed $U_t$, the sampling bits are independent before conditioning
on the event that step $t$ is reached. Thus, before this conditioning,
$Z_t\sim\operatorname{Bin}(|U_t|,p)$, with $|U_t|\leq k$.
Conditioning on the player performing $t$ steps may
destroy this independence, but no conditional bound is needed: on
$\mathcal B_t$, we have $|S_t|=Z_t>2pk+\lambda$. Thus, using the
Chernoff bound $\Pr[Z>2\mathbb E[Z]+x]\leq e^{-x}$,
\[
 \Pr[\mathcal B_t]
 \leq\Pr[Z_t>2pk+\lambda]
 \leq e^{-\lambda}.
\]
A union bound over $t=0,\ldots,M$ gives failure probability at most
$(M+1)e^{-\lambda}=\eta/2$. The two bounds therefore hold simultaneously
with probability at least $1-\eta$. Averaging over the player's initially
fixed internal randomness proves the result for a randomized player as well.
\end{proof}

\section{Dynamic Maximal Independent Set}
\label{sec:algorithm}

\subsection{Data Structures}
\label{sec:mis-data-structures}

We describe the maintained state after a completed update. The next
subsection specifies how each update changes these structures.
For the current graph $G=(V,E)$, write $N(v)$ for the neighbors of $v$
and $\deg(v)=|N(v)|$ for its degree. Since the graph is undirected,
$(v,x)$ and $(x,v)$ denote the same edge.

\paragraph{Degree partition.}
We fix the two degree boundaries
\begin{equation}\label{eq:degree-boundaries}
 b_0=m^{3/8},\qquad b_1=m^{5/8},
\end{equation}
and partition the vertex set $V$ into the low degree class $V_\lo$, the middle degree class $V_\mi$, and the high degree class $V_\hi$. These maintained classes are disjoint, although their permitted degree ranges overlap. A vertex remains in its current degree class until a migration condition in \Cref{sec:degree-migration} is met. The ranges satisfy:
\begin{itemize}
 \item For every $v\in V_\hi$, we have $\deg(v)\geq b_1$.
 \item For every $v\in V_\mi$, we have
 $b_0\leq\deg(v)\leq2b_1$.
 \item For every $v\in V_\lo$, we have $\deg(v)\leq2b_0$.
\end{itemize}
The degree sum gives
\begin{equation}\label{eq:class-sizes}
 |V_\mi|=O(m^{5/8}),\qquad |V_\hi|=O(m^{3/8}).
\end{equation}

\paragraph{Maximal independent set.}
We maintain a maximal independent set $\mis\subseteq V$ as the output, with $\mis_i=\mis\cap V_i$ for $i\in\{\lo,\mi,\hi\}$.
We call $\mis_i$ an \emph{MIS layer}; the shorthand low, middle, and high
MIS refers to these intersections. An MIS layer need not be maximal in
$G[V_i]$.
For every $v\in V$, maintain a bit indicating whether $v\in\mis$.

\paragraph{Exact MIS-neighbor counters.}
We maintain the following exact counters:
\begin{align*}
 d_\lo(v)&:=|N(v)\cap\mis_\lo| &&\text{for every }v\in V,\\
 d_\mi(v)&:=|N(v)\cap\mis_\mi| &&\text{for every }v\in V,\\
 d_\hi(v)&:=|N(v)\cap\mis_\hi| &&\text{for every }v\in V_\hi.
\end{align*}
Thus $d_\lo$ and $d_\mi$ count low- and middle-MIS neighbors at every
vertex, while $d_\hi$ counts high-MIS neighbors only at high vertices.

\begin{definition}[Candidate sets]\label{def:candidate-sets}
The \emph{primary candidate set} $C_1$ consists of low vertices outside $\mis$ with
no low- or middle-MIS neighbor. The secondary candidate set $C_2$
consists of low vertices outside $\mis$ with no low-MIS neighbor but
at least one middle-MIS neighbor. Equivalently,
\begin{align}
 C_1&=\{v\in V_\lo\setminus\mis:
 d_\lo(v)=d_\mi(v)=0\},\label{eq:c1-definition}\\
 C_2&=\{v\in V_\lo\setminus\mis:
 d_\lo(v)=0, d_\mi(v)>0\}.\label{eq:c2-definition}
\end{align}
Members of $C_1$ are \emph{primary candidates}, and members of $C_2$ are
\emph{secondary candidates}. The sets are disjoint; we use \emph{candidates}
for members of either set. Their \emph{candidate union} $C_1\cup C_2$
consists exactly of the low vertices outside $\mis$ with no low-MIS neighbor.
The \emph{candidate label} of a low vertex specifies whether it belongs
to $C_1$, to $C_2$, or to neither set.
\end{definition}

After each update, the following properties hold:
\begin{itemize}
 \item $\mis_\lo$ is a maximal independent set of
 $G[V_\lo\setminus(C_1\cup C_2)]$.
 \item Each vertex of $C_1$ has no low- or middle-MIS neighbor and has
 a high-MIS neighbor.
 \item Each vertex of $C_2$ has no low-MIS neighbor and has a middle-MIS
 neighbor.
 \item Every middle vertex outside $\mis$ is dominated by one of the
 three MIS layers; high vertices have the same guarantee.
\end{itemize}
The sets $C_1,C_2,V_\mi,V_\hi$, and $\mis_\hi$ are stored as dynamic
sets supporting membership, insertion, deletion, cardinality, and traversal.
The sets used for sampling also support the indexed selection operation defined in \Cref{sec:prelim}.

\paragraph{Adjacency lists.}
For each vertex $v$, maintain three separate adjacency lists containing
$N(v)\cap V_\lo$, $N(v)\cap V_\mi$, and $N(v)\cap V_\hi$,
respectively. The lists group the neighbors of $v$ by their degree classes,
regardless of the class of $v$ itself. When a neighbor $z$ changes class,
move its entry to the corresponding list at $v$. An adjacency dictionary
supports edge tests. Each edge has one adjacency-list entry at each
endpoint, and each entry stores a pointer to the other entry.
Write
\[
 \mathcal E_v^\hi:=\{(v,z)\in E:z\in V_\hi\}
\]
for the list of edges from $v$ to its high degree neighbors. Traversing this list
examines exactly the edges from $v$ to vertices in $V_\hi$.

\paragraph{Domination edges and sampled lists.}
A \emph{domination edge} for a vertex $z\notin\mis$ is an edge
$(z,x)\in E(G)$ with $x\in\mis$; the MIS neighbor $x$ is its
\emph{domination witness}. In figures, we draw the edge with an arrow
from $z$ to $x$ solely to indicate which vertex dominates the other.
Such an edge certifies that $z$ is dominated by the MIS layer containing
$x$. The algorithm stores a sample of the high domination edges.
A nonempty stored list certifies domination only because every entry
represents a current graph edge to a current high-MIS vertex.

Set the sampling probability to
\begin{equation}\label{eq:sampling-probability}
 p_\hi=\tfrac12m^{-1/4}.
\end{equation}
For every $z\in V_\mi\cup C_1$, maintain the high domination edge list
$A_z^\hi$ consisting of sampled edges from $z$ to $\mis_\hi$.
For each $x\in\mis_\hi$, define its \emph{reverse high domination edge list},
or \emph{reverse list}, by
\[
 R_x^\hi:=\{(z,x):z\in V_\mi\cup C_1,\ (z,x)\in A_z^\hi\}.
\]
Each sampled edge $(z,x)$ has a \emph{forward copy} in $A_z^\hi$ and a
\emph{reverse copy} in $R_x^\hi$. Each copy stores a pointer to the other copy.
The two copies represent one edge selected by one sampling coin.
The adjacency-list entries for an edge
point to its sampled copies, if present, so a graph-edge deletion can
locate them without searching either list.

A \emph{domination certificate} is either a positive exact MIS-neighbor
counter or a nonempty high domination edge list. Either certifies that
the vertex has a current MIS neighbor. A zero counter certifies the absence
of neighbors in the layer it counts, whereas an empty sampled list does
not rule out unsampled high-MIS neighbors.

The high domination edge lists at low vertices are stored in an indexed
collection, including empty lists; each entry identifies the corresponding low vertex.
There are no high domination edge lists at vertices of $C_2$. Every middle
vertex has a high domination edge list, including middle MIS vertices,
whose lists are empty by independence.

% Original Figure 5, retained for reference.
% \begin{figure}[!htbp]
% \centering
% \input{figures/data-structures-overview}
% \caption{The maintained domination data structures, showing selected graph
% edges. Every displayed non-MIS vertex has a displayed domination witness.
% Filled vertices belong to the MIS. Solid arrows show edges counted by the
% exact MIS-neighbor counters; dashed orange arrows show sampled high
% domination edges. Each arrow points toward
% the MIS witness. In particular, the dashed edge from $v$ to $x$ has a
% forward copy in $A_v^\hi$ and a reverse copy in $R_x^\hi$; the two entries
% use the same sampling bit. Low vertices in $C_2$ need no high domination edge list.}
% \label{fig:data-structures-overview}
% \end{figure}

\begin{figure}[!htbp]
\centering
% Three-layer domination data structures (Figure 5).
% Self-contained palette; uses the manuscript's \lo, \mi, \hi, and \mis.
% Arrows orient undirected graph edges toward an MIS witness.
\definecolor{domHigh}{RGB}{91,75,145}
\definecolor{domMid}{RGB}{43,120,144}
\definecolor{domLow}{RGB}{70,133,96}
\definecolor{domCandidate}{RGB}{196,121,25}
\definecolor{domSecondary}{RGB}{187,69,96}
\begin{tikzpicture}[
  x=1cm,y=1cm,>=stealth,font=\small,line cap=round,
  set/.style={rounded corners=6pt,line width=.9pt},
  vertex/.style={circle,draw=black!43,fill=white,
    minimum size=3.6mm,inner sep=0pt,line width=.4pt},
  selected/.style={vertex,minimum size=4.9mm,
    text=white,font=\scriptsize},
  high/.style={selected,draw=domHigh,fill=domHigh},
  middle/.style={selected,draw=domMid,fill=domMid},
  low/.style={selected,draw=domLow,fill=domLow},
  primary/.style={vertex,minimum size=4.9mm,draw=domCandidate,
    text=domCandidate!75!black,font=\scriptsize},
  secondary/.style={vertex,minimum size=4.9mm,draw=domSecondary,
    text=domSecondary!75!black,font=\scriptsize},
  edge/.style={draw=black!25,line width=.5pt,
    shorten <=1.9mm,shorten >=1.9mm},
  sample/.style={->,draw=domCandidate,line width=.95pt},
  middle witness/.style={->,draw=domMid,line width=.95pt},
  note/.style={font=\scriptsize,inner sep=1pt},
  list arc/.style={draw=domCandidate!85!black,line width=.6pt},
]
  % Three degree classes; contextual vertices share the central space.
  \fill[domHigh!7,rounded corners=5pt] (0,5.75) rectangle (15.7,8.08);
  \fill[domMid!6,rounded corners=5pt]  (0,2.90) rectangle (15.7,5.50);
  \fill[domLow!7,rounded corners=5pt]  (0,0) rectangle (15.7,2.65);
  \node[anchor=west,font=\bfseries,text=domHigh] at (.24,7.77) {$V_\hi$};
  \node[anchor=west,font=\bfseries,text=domMid] at (.24,5.16) {$V_\mi$};
  \node[anchor=west,font=\bfseries,text=domLow] at (.24,2.32) {$V_\lo$};

  \draw[set,draw=domHigh,fill=domHigh!10]
    (1.15,6.10) rectangle (6.35,7.75);
  \draw[set,draw=domCandidate,fill=domCandidate!9]
    (1.15,.35) rectangle (6.35,1.99);
  \draw[set,draw=domMid,fill=domMid!10]
    (9.35,3.38) rectangle (14.55,5.03);
  \draw[set,draw=domSecondary,fill=domSecondary!8]
    (9.35,.35) rectangle (14.55,1.99);
  \draw[set,draw=domLow,fill=domLow!10]
    (6.70,.35) rectangle (9.00,1.99);
  \node[anchor=west,font=\bfseries,text=domHigh]
    at (1.40,7.46) {$\mis_\hi$};
  \node[anchor=west,font=\bfseries,text=domCandidate]
    at (1.40,1.70) {$C_1$};
  \node[anchor=west,font=\bfseries,text=domMid]
    at (9.60,4.74) {$\mis_\mi$};
  \node[anchor=west,font=\bfseries,text=domSecondary]
    at (9.60,1.70) {$C_2$};
  \node[font=\bfseries,text=domLow]
    at (7.85,1.70) {$\mis_\lo$};

  % Coordinates allow the faint contextual graph to be drawn behind nodes.
  \coordinate (px1) at (1.90,6.80);
  \coordinate (px2) at (3.75,6.99);
  \coordinate (px3) at (5.60,6.75);
  \coordinate (pv1) at (2.30,.88);
  \coordinate (pv2) at (3.75,1.08);
  \coordinate (pv3) at (5.60,.88);
  \coordinate (pm1) at (10.10,3.92);
  \coordinate (pm2) at (11.95,4.12);
  \coordinate (pm3) at (13.80,3.92);
  \coordinate (pz1) at (10.50,.88);
  \coordinate (pz2) at (11.95,1.08);
  \coordinate (pz3) at (13.80,.88);
  % One irregular pool replaces the two separate outer vertex groups.
  \coordinate (ph1) at (7.20,6.55);
  \coordinate (ph2) at (7.65,7.55);
  \coordinate (ph3) at (9.15,6.30);
  \coordinate (ph4) at (9.80,7.35);
  \coordinate (pw) at (7.15,4.35);
  \coordinate (pt1) at (6.95,3.30);
  \coordinate (pt2) at (8.60,3.85);
  \coordinate (pt3) at (8.25,4.95);
  \coordinate (pll) at (7.25,.95);
  \coordinate (plr) at (8.45,.95);
  \coordinate (pl1) at (6.60,2.35);
  \coordinate (pl2) at (9.10,2.35);

  % Ordinary graph edges: no MIS--MIS edge and no forbidden candidate edge.
  % Internal edges are allowed inside C1 and C2.
  \foreach \a/\b in {
    ph1/ph2,ph1/ph3,ph2/ph4,ph3/ph4,
    ph1/px3,ph2/px3,ph1/pm1,ph3/pm2,ph4/pm3,
    pv1/pv2,pv2/pv3,pz1/pz2,pz2/pz3,
    pv2/px3,pv3/px3,
    pt1/pw,pt2/pw,pt3/pw,pt2/pt3,
    pt1/pll,pt2/pm1,pt3/pm1,pt3/px3,
    pl1/pll,pl2/plr,pl1/pl2,
    pv3/ph1,pz2/pt2%
  } {
    \draw[edge] (\a) -- (\b);
  }
  \foreach \p in {ph1,ph2,ph3,ph4,pt1,pt2,pt3,pl1,pl2}
    \node[vertex] at (\p) {};
  \node[vertex,font=\scriptsize] (w) at (pw) {$w$};
  \node[high] (x1) at (px1) {$x_1$};
  \node[high] (x2) at (px2) {$x_2$};
  \node[high] (x3) at (px3) {$x_3$};
  \node[primary] (v1) at (pv1) {$v_1$};
  \node[primary] (v2) at (pv2) {$v_2$};
  \node[primary] (v3) at (pv3) {$v_3$};
  \node[middle] (m1) at (pm1) {$y_1$};
  \node[middle] (m2) at (pm2) {$y_2$};
  \node[middle] (m3) at (pm3) {$y_3$};
  \node[secondary] (z1) at (pz1) {$z_1$};
  \node[secondary] (z2) at (pz2) {$z_2$};
  \node[secondary] (z3) at (pz3) {$z_3$};
  \node[low] at (pll) {$\ell_1$};
  \node[low] at (plr) {$\ell_2$};

  % Three well-separated vertices in each highlighted set.
  \draw[sample] (v1) -- (x1);
  \draw[sample] (v1) -- (x2);
  \draw[sample] (v2) -- (x2);
  \draw[sample] (v3) -- (x2);

  % Middle rows also own sampled high lists; the reverse copy shares its coin.
  \draw[sample] (w) -- (x2);
  \draw[sample] (w) -- (x3);
  % The thin arcs group sampled edges near their common endpoint.
  % These are set annotations, not additional graph edges.
  \draw[list arc] ($(px2)+(-116:.95)$)
    arc[start angle=-116,end angle=-26,radius=.95];
  \node[note,text=domCandidate!85!black,anchor=south west]
    at (4.62,6.52) {$R_{x_2}^\hi$};
  \draw[list arc] ($(pv1)+(69:1.50)$)
    arc[start angle=69,end angle=101,radius=1.50];
  \node[note,text=domCandidate!85!black,anchor=west]
    at (2.98,2.34) {$A_{v_1}^\hi$};
  \draw[list arc] ($(pw)+(117:.90)$)
    arc[start angle=117,end angle=148,radius=.90];
  \node[note,text=domCandidate!85!black,anchor=south west]
    at (6.90,5.18) {$A_w^\hi$};

  % Ordinary graph edges highlighted only to illustrate C2 membership.
  \draw[middle witness] (z1) -- (m1);
  \draw[middle witness] (z1) -- (m2);
  \draw[middle witness] (z2) -- (m2);
  \draw[middle witness] (z3) -- (m2);
  \draw[middle witness] (z3) -- (m3);

  \node[font=\scriptsize,text=domCandidate!85!black]
    at (3.75,.14) {$d_\lo=d_\mi=0$};
  \node[font=\scriptsize,text=domSecondary!85!black]
    at (11.95,.14) {$d_\lo=0,\quad d_\mi>0$};

  % Compact key, with ordinary graph edges kept visually subordinate.
  \node[high,minimum size=3.6mm] at (.58,-.48) {};
  \node[anchor=west,font=\scriptsize] at (.89,-.48) {in MIS};
  \draw[black!32,line width=.5pt] (3.05,-.48) -- (3.61,-.48);
  \node[anchor=west,font=\scriptsize] at (3.80,-.48) {other graph edges};
  \draw[sample] (7.20,-.48) -- (7.76,-.48);
  \node[anchor=west,font=\scriptsize] at (7.95,-.48) {sampled high edges};
  \draw[middle witness] (11.82,-.48) -- (12.38,-.48);
  \node[anchor=west,font=\scriptsize] at (12.57,-.48) {middle domination};
\end{tikzpicture}
\caption[The three-layer domination data structures]{%
The maintained domination data structures.
Only selected vertices and edges are shown. Filled vertices belong to MIS $\mis$.
Arrows point toward MIS witnesses.
\textup{(i)} Candidates in $C_1$ have $d_\lo=d_\mi=0$;
those in $C_2$ have $d_\lo=0$ and $d_\mi>0$.
Blue arrows highlight middle-MIS domination edges counted by $d_\mi$.
\textup{(ii)} Orange arrows show sampled high domination edges from
$V_\mi\cup C_1$ to $\mis_\hi$.
Each sampled edge $(u,x)$ has a forward copy in $A_u^\hi$ and a reverse
copy in $R_x^\hi$, determined by the same sampling bit.
\textup{(iii)} Gray lines show other graph edges, including the unsampled
high domination edges $(v_2,x_3)$ and $(v_3,x_3)$.
No high domination edge list is stored at a vertex of $C_2$.
}
\label{fig:data-structures-overview}
\end{figure}
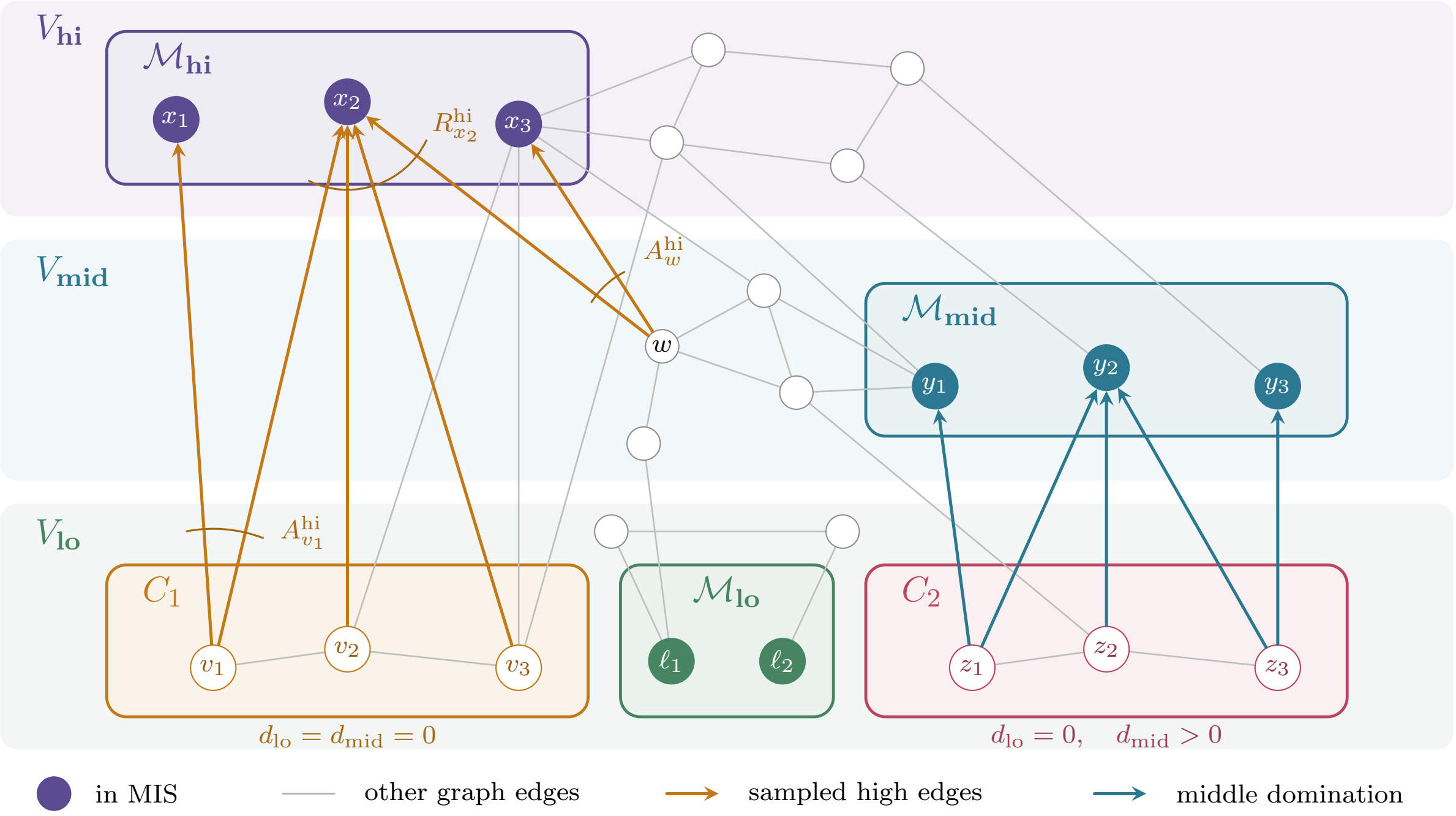

\subsection{Algorithm Description}

Throughout the update algorithm, maximality repairs are organized by a
priority queue $Q$. At the beginning of each edge update, $Q=\emptyset$.
Order vertices first by degree class $V_\lo,V_\mi,V_\hi$, then by a
fixed vertex order. Write $v\prec w$ when $v$ comes before $w$. During preparation, update the position of any vertex in $Q$ that changes degree class. Classes and priorities remain fixed during maximality restoration.
A \emph{request for $v$} is a pending instruction, represented by $v\in Q$,
to check from the counters and domination edge lists whether $v$ should
be included in its current MIS layer. Whenever we say to \emph{request} a
vertex $v$, first check that $v\notin\mis$ and that the following
condition holds:
\begin{itemize}
 \item if $v\in V_\lo\cup V_\mi$, then
 $d_\lo(v)+d_\mi(v)=0$ and its high domination edge list $A_v^\hi$
 is stored and empty;
 \item if $v\in V_\hi$, then
 $d_\lo(v)+d_\mi(v)+d_\hi(v)=0$.
\end{itemize}
Place $v$ in $Q$ only if these tests pass and $v$ is not already
present. If a vertex of $C_1$ has no stored high domination edge list
during preparation, defer this test until its list is initialized after
the flush test. If a flush starts, its low pass processes $C_1$ directly.

Issuing a request changes no other data structure. In particular, a queued
low vertex stays in whichever candidate set its exact counts specify.
A low or middle vertex with an empty high domination edge list may still
have unsampled high-MIS neighbors; these are removed from $\mis_\hi$ before it enters
the MIS. A queued vertex may also acquire domination before extraction,
so the algorithm repeats the tests when it processes the request.

Each update has two phases. Preparation handles degree migrations of the two endpoints of the updated edge, applies the edge update, and restores independence.
To restore maximality, we test once whether $|C_1|\geq B$, with $B$
defined in \Cref{sec:restore-maximality}. If so, a flush operation rebuilds
the high MIS after processing the low candidates and middle vertices.
Otherwise, we initialize the missing high domination edge lists at vertices
of $C_1$ and process requests in queue order, including vertices in the
appropriate MIS layer and making the resulting repairs. These repairs
may issue further requests. When the queue becomes empty, the update ends.

\subsubsection{Preparation}\label{sec:degree-migration}

The input is an insertion or deletion $e=(s,t)$. This edge update changes
vertex degrees, so we may have to update the degree partition
$V_\lo,V_\mi,V_\hi$; we call this process \emph{degree migration}.
We first deal with degree migration, then process the edge insertion or
deletion. Before changing the graph, compute, for $w\in\{s,t\}$,
\[
 \widehat{\deg}(w)=
 \begin{cases}
  \deg(w)+1,&e\text{ is an insertion},\\
  \deg(w)-1,&e\text{ is a deletion}.
 \end{cases}
\]
The migration routine uses $\widehat{\deg}$ only to choose the degree
class for $w$ after the edge update. All neighbor scans, counter
changes, and sampling during migration use the current graph. Thus an edge
being deleted is still present during migration, while an edge being
inserted is not yet present. All preparation operations finish before the
flush test or any queue extraction.

\paragraph{Overview of degree migration.}
Some maintained data structures are defined only for vertices in a particular degree class. When $w\in\{s,t\}$ changes degree class, we delete the data structures required only by $w$'s old class and initialize those required by its new class. If $w\in\mis$, we also remove $w$ from its old MIS layer and include it in its new MIS layer, then update
the exact counters, candidate membership, and domination edge lists at its neighbors. The case analysis below spells out
these changes for every data structure.

Using the boundaries in \eqref{eq:degree-boundaries}, the transitions are
\[
 \begin{aligned}
 V_\lo&\xrightarrow{\,\widehat{\deg}>2b_0\,}V_\mi,
 &V_\mi&\xrightarrow{\,\widehat{\deg}<b_0\,}V_\lo,\\
 V_\mi&\xrightarrow{\,\widehat{\deg}>2b_1\,}V_\hi,
 &V_\hi&\xrightarrow{\,\widehat{\deg}<b_1\,}V_\mi.
 \end{aligned}
\]
Process $s$ and $t$ in fixed order, completing one migration before
starting the next. The separation of the boundaries implies at most one
migration for each of $s,t$ per update. Since $\deg(w)$ and
$\widehat{\deg}(w)$ differ by one, all degree bounds used during
preparation allow this additive difference.

Every case below follows the same three steps.
\begin{enumerate}[(1),leftmargin=*]
 \item \emph{Delete objects belonging to the old class.}
 Delete any high domination edge list $A_z^\hi$ or reverse list $R_x^\hi$ that is no longer
 maintained in the new class. When deleting $A_z^\hi$, remove the entry
 for every edge $(z,x)$ from $R_x^\hi$ as well. When deleting $R_x^\hi$,
 remove the entry for every edge $(z,x)$ from $A_z^\hi$ as well.
 \item \emph{Change the degree class and, if necessary, the MIS layer.}
 Move the vertex between degree classes and update its entries in its
 neighbors' adjacency lists. A migrating MIS vertex remains in $\mis$,
 but moves between $\mis_\lo,\mis_\mi$, and $\mis_\hi$.
 If $w\in\mis$, then for each neighbor $z$, decrease the counter for $w$'s old MIS layer
 by one and increase the counter for its new MIS layer by one, whenever
 those counters are maintained at $z$. Apply both changes before
 updating a low degree neighbor's membership in $C_1$ and $C_2$.
 \item \emph{Initialize new objects and issue requests.}
 Initialize any newly required counter, high domination edge list at a
 middle vertex, or reverse high domination edge list. Update the migrated
 vertex's priority in $Q$, if it is already queued, and request that vertex
 and the affected neighbors. Construction of high domination edge lists
 at low vertices waits until after the flush test.
\end{enumerate}

Throughout preparation, whenever the MIS membership of a low vertex $z$
or either counter $d_\lo(z),d_\mi(z)$ changes, use its MIS-membership bit
and updated counters to determine whether $z$ belongs to $C_1$, $C_2$,
or neither set, according to \eqref{eq:c1-definition}--\eqref{eq:c2-definition}.
If $z$ leaves $C_1$, delete $A_z^\hi$, if stored, and remove the entry for every stored edge $(z,x)$ from $R_x^\hi$.
Leaving the low degree class also removes the vertex from the candidate sets.
For each vertex $z$ entering or reentering $C_1$, construct its high domination edge list $A_z^\hi$ using fresh independent samples after the flush test, provided that the update enters regular mode. Thus membership in the
candidate sets satisfies \Cref{def:candidate-sets} during preparation, but a new
member of $C_1$ need not yet have a list or a high-MIS neighbor.

To distinguish a missing list from a stored empty list, each vertex $z$
has a pointer to $A_z^\hi$, which is null exactly when no list is stored.
Whenever $A_z^\hi$ is deleted, remove the entry for each edge $(z,x)$
from $R_x^\hi$ as well and set this
pointer to null; if $z$ is low, also remove the entry for $A_z^\hi$ from the
indexed collection. Preparation only deletes lists from this collection.

To initialize the high domination edge list $A_z^\hi$ of a middle
vertex $z$, sample the indexed set $\mis_\hi$ with probability $p_\hi$ using
\Cref{lem:output-sensitive-bernoulli}. Test adjacency for each returned
vertex $x$, store each edge that passes the adjacency test in both $A_z^\hi$ and $R_x^\hi$,
and set $z$'s list pointer. To initialize a reverse high domination edge list
$R_x^\hi$, independently sample $V_\mi$ and the indexed collection of
high domination edge lists stored at low vertices with probability $p_\hi$.
For each sampled middle vertex $z$ or sampled list $A_z^\hi$ at a low vertex $z$, test adjacency to $x$ and store both copies of each edge that passes the test. This does not scan all of $N(x)$ or enumerate low vertices whose
lists have not yet been initialized. The cost includes every sampled middle
vertex, every sampled list at a low vertex, and the corresponding adjacency
tests, including tests that fail. We use the same reverse-list initialization whenever a
vertex enters $\mis_\hi$ during maximality restoration.

For all sampled lists, use the following sampling rule. Each edge
$(z,x)$ with $x\in\mis_\hi$ and a stored list $A_z^\hi$ receives a
fresh independent Bernoulli-$p_\hi$ bit. Use the same bit without
resampling as long as the graph edge is present, $x$ remains continuously in $\mis_\hi$, and the
same list $A_z^\hi$ remains stored. If any condition ceases to hold,
discard the bit; when all conditions hold again, use a fresh independent
bit, even within the same update. Changes to exact counters alone do not
resample edges. This rule applies to zero bits as well, although only
edges whose sampling bits equal one are stored. Because degree migrations finish one at
a time and delete old lists before initializing new ones, an edge
whose two lists are newly created is sampled when the later list is
initialized.

\paragraph{Specific case analysis.}
The following four cases specify the counter and list updates required by the three steps above. Let $w\in\{s,t\}$ be the vertex being migrated.
\begin{enumerate}[(a),leftmargin=*]
 \item\label{case:migration-low-middle} \emph{$V_\lo\to V_\mi$.}
 Move $w$ when $\widehat{\deg}(w)>2b_0$. Delete any stored $A_w^\hi$, removing the entry for every stored edge $(w,x)$ from $R_x^\hi$, and remove $w$ from $C_1\cup C_2$. For every neighbor $z$
 of $w$, move the entry for $w$ from $z$'s list of low degree neighbors
 to $z$'s list of middle degree neighbors. The counters $d_\lo(w)$ and $d_\mi(w)$ remain
 defined and require no reinitialization merely because its class changed.

 \emph{If $w\notin\mis$.}
 Changing $w$'s degree class does not change any neighbor's MIS-neighbor
 counter, because $w$ is outside the MIS. Initialize the high domination edge
 list $A_w^\hi$ by sampling the current $\mis_\hi$, and request $w$.

 \emph{If $w\in\mis$.}
 Remove $w$ from $\mis_\lo$ and include it in $\mis_\mi$.
 For every $z\in N(w)$, decrease $d_\lo(z)$ by one and increase
 $d_\mi(z)$ by one. After both changes, update the candidate membership
 of $z$ if it is low, and request $z$. Initialize the high domination edge
 list $A_w^\hi$ and request $w$. This scan visits all
 $\deg(w)=O(b_0)$ neighbors of $w$.

 \item\label{case:migration-middle-low} \emph{$V_\mi\to V_\lo$.}
 Move $w$ when $\widehat{\deg}(w)<b_0$. Delete $A_w^\hi$ and remove the entry for every stored edge $(w,x)$ from $R_x^\hi$. For every neighbor $z$ of $w$, move the entry for $w$ from
 $z$'s list of middle degree neighbors to $z$'s list of low degree neighbors.

 \emph{If $w\notin\mis$.}
 If $d_\lo(w)>0$, leave it outside the candidate union. If
 $d_\lo(w)=0$ and $d_\mi(w)>0$, put it in $C_2$. Otherwise put it
 in $C_1$, with its high domination edge list still uninitialized. Request $w$.

 \emph{If $w\in\mis$.}
 Remove $w$ from $\mis_\mi$ and include it in $\mis_\lo$.
 For every neighbor $z$, increase $d_\lo(z)$ by one and decrease
 $d_\mi(z)$ by one. Apply both changes before updating the candidate
 membership of $z$ if it is low. Every low candidate adjacent to $w$ leaves
 the union and deletes its high domination edge list if present. Request
 every neighbor $z$ and $w$. This scan visits all $\deg(w)=O(b_0)$ neighbors of $w$.

 \item\label{case:migration-middle-high} \emph{$V_\mi\to V_\hi$.}
 Move $w$ when $\widehat{\deg}(w)>2b_1$. Delete its high domination edge
 list $A_w^\hi$ and remove the entry for every stored edge $(w,x)$ from $R_x^\hi$. For every neighbor $z$ of $w$, move
 the entry for $w$ from $z$'s list of middle degree neighbors to $z$'s
 list of high degree neighbors. Initialize
 $d_\hi(w)=|N(w)\cap\mis_\hi|$ by scanning its high degree neighbors.

 \emph{If $w\notin\mis$.}
 Changing $w$'s degree class does not change any neighbor's MIS-neighbor
 counter, because $w$ is outside the MIS. Request $w$ with the priority assigned to
 its new degree class $V_\hi$.

 \emph{If $w\in\mis$.}
 Remove $w$ from $\mis_\mi$ and include it in $\mis_\hi$.
 For every neighbor $z$, decrease $d_\mi(z)$ by one; if $z\in V_\hi$,
 also increase $d_\hi(z)$ by one. Then update the candidate membership
 of $z$ if it is low, and request $z$. Initialize the new $R_w^\hi$ by sampling $V_\mi$ and the
 indexed collection of high domination edge lists stored at low vertices.
 Vertices moved from $C_2$ to $C_1$ by this migration have no stored high
 domination edge list, so they are not in the collection sampled to initialize $R_w^\hi$ at this step. Request $w$.
 This scan visits all $\deg(w)=O(b_1)$ neighbors of $w$.

 \item\label{case:migration-high-middle} \emph{$V_\hi\to V_\mi$.}
 Move $w$ when $\widehat{\deg}(w)<b_1$. For every neighbor $z$ of $w$,
 move the entry for $w$ from $z$'s list of high degree neighbors to $z$'s
 list of middle degree neighbors. Discard $d_\hi(w)$ after completing
 the applicable case below.

 \emph{If $w\notin\mis$.}
 Initialize its high domination edge list $A_w^\hi$ by sampling
 $\mis_\hi$ and request $w$.

 \emph{If $w\in\mis$.}
 For every edge $(z,w)$ stored in $R_w^\hi$, remove its entries from
 $R_w^\hi$ and $A_z^\hi$, and request $z$.
 For every $z\in N(w)\cap V_\hi$, decrease $d_\hi(z)$ by one.
 Remove $w$ from $\mis_\hi$ and include it in $\mis_\mi$.
 For every neighbor $z$, increase $d_\mi(z)$ by one, update the candidate
 membership of $z$ if it is low, and request $z$. Candidates
 thereby moving from $C_1$ to $C_2$ delete their high domination edge lists.
 Initialize the high domination edge list $A_w^\hi$ of $w$ and request it.
 This scan visits all $\deg(w)=O(b_1)$ neighbors of $w$, even though a general
 high vertex can have much larger degree.
\end{enumerate}
All list deletions and initializations use the sampling rule
above. Class migration itself changes no graph edge or MIS membership,
so it preserves independence.

\paragraph{Edge insertion.}\label{rule:edge-insertion-domination-list}
After migrating $s$ and $t$ as needed, insert $e=(s,t)$ into $G$ and perform the
following steps, using the updated degree classes.
\begin{enumerate}[(1),leftmargin=*]
 \item \emph{Class-indexed adjacency lists.}
 Store an entry for $t$ in the appropriate adjacency list at $s$, and
 an entry for $s$ in the appropriate adjacency list at $t$. Update the
 adjacency dictionaries and store in each entry a pointer to the other entry.

 \item \emph{Exact MIS-neighbor counters.}
 If $s\in\mis$, update the counter at $t$ corresponding to the layer
 of $s$: increment $d_\lo(t)$ if $s\in\mis_\lo$; increment
 $d_\mi(t)$ if $s\in\mis_\mi$; and increment $d_\hi(t)$ if
 $s\in\mis_\hi$ and $t\in V_\hi$. Apply the symmetric rules at
 $s$ if $t\in\mis$. These are precisely the counters whose definitions
 include the new edge.

 \item \emph{Restoring independence.}
 If both $s$ and $t$ belong to the MIS, remove $t$ from $\mis$ when $s\prec t$,
 and remove $s$ from $\mis$ when $t\prec s$. Thus the order $\prec$ gives priority
 to $V_\lo$ before $V_\mi$ before $V_\hi$, using the fixed vertex order
 when both vertices belong to the same degree class. The complete removal
 rules are given immediately below. If at most one of $s,t$ belongs to the MIS, no vertex needs to be removed from it.

 \item \emph{Candidate membership and sampling the inserted edge.}
 For each low degree endpoint, recompute membership in $C_1$ and $C_2$ from its MIS-membership bit and updated counters, according to \Cref{def:candidate-sets}. If the edge now
 joins $x\in\mis_\hi$ to $z$ with a stored list $A_z^\hi$, sample it once
 with probability $p_\hi$ and, if the sampling bit equals one, store
 the edge in both $A_z^\hi$ and $R_x^\hi$.
 The eligibility of the edge for sampling is determined using the endpoints' MIS memberships after Step~(3). A low vertex
 whose list has not yet been initialized waits for the flush test.

\end{enumerate}
If neither $s$ nor $t$ belongs to the MIS, the insertion changes only
adjacency lists, adjacency dictionaries, and degrees. When Step~(3) removes $t$ from $\mis$, the vertex $s$
stays in the MIS and dominates $t$, so $t$ is not placed in $Q$; the
case of removing $s$ from $\mis$ is symmetric. All steps of the edge insertion finish
before any request is extracted or any MIS is reported.

\medskip
\noindent\emph{Removing the conflicting MIS vertex.}
Let $w\in\{s,t\}$ be the vertex removed from $\mis$ in Step~(3). Clear its MIS-membership bit and perform the following updates for its degree class.
\begin{enumerate}[(a),leftmargin=*]
 \item \emph{$w\in V_\lo$.}
 Scan every $z\in N(w)$ and decrement $d_\lo(z)$. Recompute each low degree neighbor's membership in $C_1$ and $C_2$ using the updated counters and its MIS-membership bit, and request each affected neighbor. Then recompute $w$'s membership in $C_1$ and $C_2$ in the same way. For every vertex
 newly entering $C_1$, initialize its high domination edge list after the
 flush test if the update enters regular mode.

 \item \emph{$w\in V_\mi$.}
 Scan every $z\in N(w)$, including low degree neighbors, and decrement $d_\mi(z)$. Update each low degree neighbor's membership in $C_1,C_2$
 according to \Cref{def:candidate-sets}, and request a neighbor only if
 the counter and list tests above pass. For each vertex
 $z$ moving from $C_2$ to $C_1$, initialize its high domination edge list
 $A_z^\hi$ using new independent samples after the flush test if the
 update enters regular mode (see \Cref{sec:restore-maximality}). Keep the high domination edge list
 $A_w^\hi$ of $w$ stored.

 \item \emph{$w\in V_\hi$.}
 Scan $\mathcal E_w^\hi$ and decrement $d_\hi(z)$ at each high degree
 neighbor, requesting it. Traverse $R_w^\hi$ and, for each stored edge
 $(z,w)$, remove its entries from $R_w^\hi$ and $A_z^\hi$, and request $z$. Remove $w$ from
 the indexed high MIS. Its low degree and middle degree neighbors are not scanned in full.
\end{enumerate}

\paragraph{Edge deletion.}
After migrating $s$ and $t$ as needed, process the deletion of $e=(s,t)$ using
the updated degree classes. The edge-deletion step preserves independence
and leaves MIS membership unchanged, but it may remove a domination witness.
The subsequent maximality-restoration procedure may change MIS membership.
Perform the following steps.
\begin{enumerate}[(1),leftmargin=*]
 \item If the edge has stored sampled copies, let $z\in\{s,t\}$
 be the vertex whose high domination edge list $A_z^\hi$ contains the edge.
 Remove the entry for the edge from $A_z^\hi$ and from $R_x^\hi$, where $x$ is the other endpoint, and request $z$.
 \item Remove the edge from both endpoints' adjacency lists and adjacency dictionaries. If $t\in\mis_\lo$, decrement $d_\lo(s)$; if $t\in\mis_\mi$, decrement $d_\mi(s)$; and if $t\in\mis_\hi$ and $s\in V_\hi$, decrement $d_\hi(s)$. Apply the symmetric rules at $t$ using the MIS layer of $s$.
 \item Recompute membership in $C_1$ and $C_2$ for each $w\in\{s,t\}\cap V_\lo$ using its MIS-membership bit and updated counters,
 and request both $s$ and $t$, whether or not the deleted edge was sampled.
 If a low vertex $w\in\{s,t\}$ enters
 $C_1$, initialize its high domination edge list after the flush test if
 the update enters regular mode.
\end{enumerate}

\subsubsection{Restoring Maximality}\label{sec:restore-maximality}

To restore maximality of $\mis$, the algorithm uses either \emph{regular mode} or \emph{flush mode}. Define the flush threshold
\[
 B:=m^{1/2}.
\]
After preparation, check whether $|C_1|\geq B$. If this condition holds,
enter flush mode. Otherwise, initialize the missing high domination edge
lists of vertices in $C_1$ as described below, and enter regular mode.
The flush test is performed before these initializations, since flush
mode removes every vertex of $\mis_\hi$ from the MIS and handles the vertices of $C_1$ directly.

For each $z\in C_1$ whose list pointer is null, call
$\mathsf{InitializeHigh}(z)$. This procedure independently samples
$\mis_\hi$ with probability $p_\hi$ using
\Cref{lem:output-sensitive-bernoulli}. For each returned vertex $x$, test
whether $(z,x)\in E(G)$ and, if so, store an entry for the edge in both $A_z^\hi$
and $R_x^\hi$. Place $A_z^\hi$ in the indexed collection of high
domination edge lists at low vertices and set $z$'s list pointer. Place $z$
in $Q$ only if $A_z^\hi$ is empty; $d_\lo(z)=d_\mi(z)=0$
because $z\in C_1$. Every initialization uses fresh
independent coins, and its
cost includes adjacency tests for returned vertices that are not
neighbors of $z$. Thus every vertex of $C_1$ has its high domination edge
list before regular mode begins.

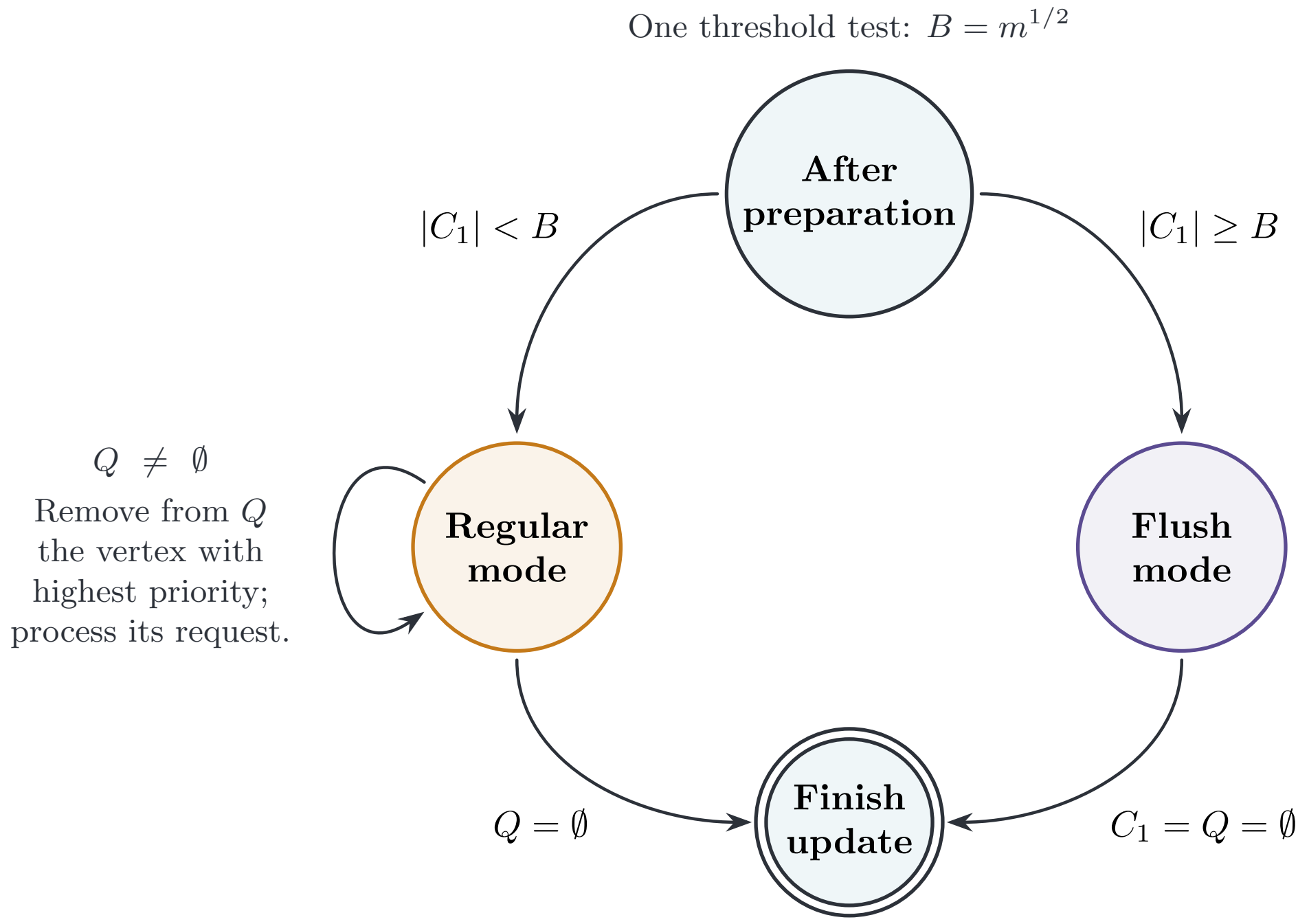
\begin{figure}[!htbp]
  \centering
  % Mode overview for the m^{5/8} algorithm. Uses the paper's ds* palette.
\begin{tikzpicture}[
  x=1cm,y=1cm,font=\small,line cap=round,line join=round,
  mrm mode/.style={circle,minimum size=20mm,inner sep=2pt,
    line width=1pt,align=center,font=\small\bfseries},
  mrm terminal/.style={double=white,double distance=1.8pt},
  mrm arrow/.style={-{Stealth[length=2.5mm,width=1.8mm]},
    line width=.85pt,draw=dsInk,shorten <=2pt,shorten >=2pt},
  mrm condition/.style={align=center,fill=white,inner sep=3pt},
  mrm note/.style={align=center,font=\footnotesize,text=dsInk}
]
  \node[mrm mode,draw=dsInk,fill=dsLayerMid] (mrm-entry) at (6.3,3.4)
    {After\\preparation};
  \node[mrm note,above=5pt] at (mrm-entry.north)
    {One threshold test: $B=m^{1/2}$};
  \node[mrm mode,draw=dsCOne,fill=dsCOne!9] (mrm-regular) at (3.1,0)
    {Regular\\mode};
  \node[mrm mode,draw=dsHighSet,fill=dsHighSet!8] (mrm-flush) at (9.5,0)
    {Flush\\mode};
  \node[mrm mode,mrm terminal,minimum size=17mm,
    draw=dsInk,fill=dsLayerMid] (mrm-finish) at (6.3,-2.65)
    {Finish\\update};

  % The choice is made once. Initialization details are in the caption/text.
  \draw[mrm arrow] (mrm-entry.west) to[out=180,in=90]
    node[mrm condition,pos=.48,above left=2pt] {$|C_1|<B$}
    (mrm-regular.north);
  \draw[mrm arrow] (mrm-entry.east) to[out=0,in=90]
    node[mrm condition,pos=.48,above right=2pt] {$|C_1|\geq B$}
    (mrm-flush.north);

  % Smooth self-loop with its label outside the curve.
  \draw[mrm arrow] (mrm-regular.145)
    to[out=145,in=215,looseness=3.2] (mrm-regular.215);
  \node[mrm note,anchor=east,inner sep=0pt,text width=30mm]
    at ([xshift=-10mm]mrm-regular.west)
    {$Q\neq\emptyset$\\[3pt]Remove from $Q$\\the vertex with\\highest priority;\\process its request.};
  \draw[mrm arrow] (mrm-regular.south) to[out=270,in=180]
    node[mrm condition,pos=.55,below left=3pt] {$Q=\emptyset$}
    (mrm-finish.west);

  % A flush terminates after completing the reconstruction described in text.
  \draw[mrm arrow] (mrm-flush.south) to[out=270,in=0]
    node[mrm condition,pos=.55,below right=3pt] {$C_1=Q=\emptyset$}
    (mrm-finish.east);
  % Add invisible space on the right, so the picture is centered on x=6.3.
  % The left annotation remains within the picture's bounding box.
  \path let \p1=(current bounding box.west), \p2=(mrm-entry.center)
    in ({2*\x2-\x1},0);
\end{tikzpicture}
  \caption{The two modes of maximality restoration, with $B=m^{1/2}$.
After preparation, the threshold is checked once; missing high domination
edge lists at low vertices are initialized only on the regular branch.
In regular mode, $C_1$ can only shrink, so there is no further flush test.
The flush processes $C_1^0$ (the set $C_1$ at its start), then $V_\mi$,
and finally $V_\hi$. The double circle marks termination of the update.}
  \label{fig:maximality-restoration-modes}
\end{figure}

Figure~\ref{fig:maximality-restoration-modes} summarizes the two modes and their termination conditions.

\paragraph{Regular mode.}
Regular mode repeatedly removes the highest-priority vertex $v$ from $Q$ and processes its request. We call this removal from $Q$ an \emph{extraction}.
If $v\in\mis$, one of its maintained exact MIS-neighbor counters is
positive, or its stored high domination edge list is nonempty, stop processing this request without changing MIS membership or any maintained
list or resampling any edge, and continue with the next request in $Q$, if any. Otherwise, include $v$ in the MIS
according to its degree class as follows. The steps below spell out the
domination tests and the subsequent change in MIS membership. Complete all steps for $v$
before extracting the next vertex from $Q$.

\begin{enumerate}[(1),leftmargin=*]
 \item\label{case:regular-low} $v\in V_\lo$.\nopagebreak

 First test whether the maintained counters or sampled list certify that
 $v$ has an MIS neighbor.
 \begin{enumerate}[(a),leftmargin=*,beginpenalty=10000]
  \item If $d_\lo(v)+d_\mi(v)>0$, then $v$ is dominated by the low or
  middle MIS, so stop processing this request and continue with the next request in $Q$, if any. Its membership in
  $C_1,C_2$ and any required list deletion were already updated when
  these counters changed.

  \item We are left with $d_\lo(v)=d_\mi(v)=0$, so $v\in C_1$.
  Check the list $A_v^\hi$. If $A_v^\hi\neq\emptyset$, then it
  contains a domination edge from $v$ to a vertex of $\mis_\hi$,
  so stop processing this request and continue with the next request in $Q$, if any.
 \end{enumerate}

 If neither test ends processing of this request, then
 $d_\lo(v)+d_\mi(v)=0$ and $A_v^\hi=\emptyset$. Thus $v$ has no
 neighbor in $\mis_\lo\cup\mis_\mi$, but it may still have neighbors
 in $\mis_\hi$ whose edges were not sampled. We remove these neighbors from $\mis_\hi$
 before including $v$ in $\mis_\lo$, as follows.
 \begin{enumerate}[(a),leftmargin=*]
  \setcounter{enumii}{2}
  \item Scan $N(v)$ and compute $X_\hi=N(v)\cap\mis_\hi$.
  For every $x\in X_\hi$, in fixed vertex order, remove $x$ from
  $\mis_\hi$ and update its neighbors as follows. Scan every high degree
  neighbor $z\in N(x)\cap V_\hi$ and decrement $d_\hi(z)$.
  Place $z$ in $Q$ only if $z\notin\mis$ and
  $d_\lo(z)+d_\mi(z)+d_\hi(z)=0$. Then traverse $R_x^\hi$ and,
  for each stored edge $(z,x)$, remove its entries from $R_x^\hi$ and $A_z^\hi$.
  Place $z$ in $Q$ only if $z\notin\mis$,
  $d_\lo(z)+d_\mi(z)=0$, and $A_z^\hi=\emptyset$.
  Do not place $x$ in $Q$: including $v$ in $\mis_\lo$ in the next step dominates
  $x$, and no request is extracted before that step finishes.

  \item Remove $v$ from $C_1$, delete $A_v^\hi$, and include $v$ in
  $\mis_\lo$. For every neighbor $z\in N(v)$, increment
  $d_\lo(z)$. If $z\in C_1\cup C_2$, remove it
  from these sets and delete its high domination edge list $A_z^\hi$, if present, removing the entry for every stored edge $(z,x)$ from $R_x^\hi$.
 \end{enumerate}

 \item\label{case:regular-middle} $v\in V_\mi$.

 We process $v$ in the following steps.
 \begin{enumerate}[(a),leftmargin=*,beginpenalty=10000]
  \item First check whether $v$ has a neighbor in
  $\mis_\lo\cup\mis_\mi$. If $d_\lo(v)+d_\mi(v)>0$, then $v$
  is dominated by the low or middle MIS, so stop processing this request and continue with the next request in $Q$, if any.

  \item We are left with $d_\lo(v)+d_\mi(v)=0$. Check the list
  $A_v^\hi$. If $A_v^\hi\neq\emptyset$, then it contains a
  domination edge from $v$ to a vertex of $\mis_\hi$, so stop processing this request and continue with the next request in $Q$, if any.
 \end{enumerate}

 If neither test ends processing of this request, then
 $d_\lo(v)+d_\mi(v)=0$ and $A_v^\hi=\emptyset$. We include $v$ in
 $\mis_\mi$ after removing any remaining high-MIS neighbors from $\mis_\hi$, in the
 following steps.
 \begin{enumerate}[(a),leftmargin=*]
  \setcounter{enumii}{2}
  \item Scan $N(v)$ and compute $X_\hi=N(v)\cap\mis_\hi$.
  Remove every vertex of $X_\hi$ from $\mis_\hi$ in fixed vertex order, updating
  counters, domination edge lists, and $Q$ exactly as in the low-vertex
  case above. Leave the pointer to $A_v^\hi$ unchanged; the list is empty because all of $v$'s high-MIS neighbors have been removed from $\mis_\hi$.

  \item Include $v$ in $\mis_\mi$. Scan every neighbor $z\in N(v)$,
  including low degree neighbors, and increment $d_\mi(z)$.
  If $z\in C_1$, move it to $C_2$ and delete $A_z^\hi$, removing the entry for every stored edge $(z,x)$ from $R_x^\hi$.
 \end{enumerate}

 \item\label{case:regular-high} $v\in V_\hi$.

 We process $v$ in the following steps.
 \begin{enumerate}[(a),leftmargin=*,beginpenalty=10000]
  \item First check whether $v$ is already dominated by $\mis$. If
  \[
   d_\lo(v)+d_\mi(v)>0\qquad\text{or}\qquad d_\hi(v)>0,
  \]
  stop processing this request and continue with the next request in $Q$, if any.

  \item Otherwise, include $v$ in $\mis_\hi$. For each high degree neighbor
  $z\in N(v)\cap V_\hi$, increment $d_\hi(z)$.

  \item Finally, initialize $R_v^\hi$. Independently compute
  \[
   S_v^\mi\gets\operatorname{BernoulliSample}(V_\mi,p_\hi),
   \qquad
   S_v^\lo\gets\operatorname{BernoulliSample}(C_1,p_\hi).
  \]
  For each returned $z$, test whether $(z,v)\in E(G)$ and, if so,
  store an entry for the edge in both $R_v^\hi$ and $A_z^\hi$.
 \end{enumerate}
\end{enumerate}
No vertex is removed from $\mis_\lo$ or $\mis_\mi$ during regular
mode. Consequently, $C_1$ can only shrink, every vertex remaining in
$C_1$ keeps its high domination edge list, and $|C_1|<B$ holds throughout.
When a vertex joins the MIS, it only increases domination at its neighbors and
therefore creates no queue requests. Requests arise only from the
high-MIS removals described above.
There is no further flush test. When $Q$ becomes empty, terminate the
update.

\paragraph{Flush mode.}\label{mode:flush}
The purpose of a flush is to empty the primary candidate set $C_1$. To do so,
the algorithm removes every vertex of $\mis_\hi$ from the MIS, extends the low and
middle MISs, and then reconstructs the high MIS. Every vertex currently in $\mis_\lo\cup\mis_\mi$ remains in the MIS throughout the flush. The algorithm performs the following steps.

\begin{enumerate}[(1),leftmargin=*]
 \item\label{step:flush-clear}
 Make a copy $C_1^0$ of $C_1$ at the start of the flush and set
 $Q\gets\emptyset$. Keep $C_1^0$ unchanged throughout the flush.
 The following steps examine the vertices of $C_1^0$, $V_\mi$, and
 $V_\hi$ directly; no vertex is placed in $Q$ during flush mode.
 Remove every vertex $x\in\mis_\hi$ from the MIS. For each such $x$, scan its
 high degree neighbors and decrement their $d_\hi$ counters, then traverse $R_x^\hi$
 and, for each stored edge $(z,x)$, remove its entries from $R_x^\hi$
 and $A_z^\hi$. Thus $\mis_\hi=\emptyset$ and all high domination edge
 lists are empty. Delete the high domination edge lists stored at low
 vertices, set their list pointers to null, and clear the indexed collection.
 Keep the empty lists $A_z^\hi$ for $z\in V_\mi$.

 \item\label{step:flush-low}
 Process all vertices $v\in C_1^0$ in fixed vertex order. If
 $d_\lo(v)>0$, then $v$ is already dominated by $\mis_\lo$ and
 nothing needs to be done. Otherwise, remove $v$ from $C_1$ and include
 it in $\mis_\lo$. Scan every neighbor $z\in N(v)$, increment
 $d_\lo(z)$, and remove $z$ from $C_1\cup C_2$ if it belongs there.
 Every vertex of $C_1^0$ has no neighbor in $\mis_\mi$, and this MIS
 does not change during the step, so including $v$ in $\mis_\lo$ preserves independence.
 At the end of this step, $C_1=\emptyset$.

 \item\label{step:flush-middle}
 Process all vertices $v\in V_\mi$ in fixed vertex order. If
 $v\in\mis$ or $d_\lo(v)+d_\mi(v)>0$, nothing needs to be done.
 Otherwise, include $v$ in $\mis_\mi$. Scan every neighbor $z\in N(v)$,
 including low degree neighbors, and increment $d_\mi(z)$. This extends
 $\mis_\mi$ to a maximal independent set of
 \[
  G[V_\mi\setminus N(\mis_\lo)].
 \]
 The high MIS is still empty. Since $C_1=\emptyset$ and no vertex leaves $\mis_\lo\cup\mis_\mi$, the candidate sets remain unchanged.

 \item\label{step:flush-high}
 Rebuild $\mis_\hi$ from scratch as a maximal independent set of
 \[
  G[V_\hi\setminus N(\mis_\lo\cup\mis_\mi)].
 \]
 To do so, process all vertices $v\in V_\hi$ in fixed vertex order. If
 $d_\lo(v)+d_\mi(v)+d_\hi(v)=0$, include $v$ in $\mis_\hi$ and
 increment $d_\hi(z)$ for every high degree neighbor $z$ of $v$.
 For every vertex $v$ included in $\mis_\hi$, initialize $R_v^\hi$ by independently sampling
 $V_\mi$ with probability $p_\hi$. For each returned vertex $z$, test
 whether $(z,v)\in E(G)$ and, if so, store an entry for the edge in both $R_v^\hi$
 and $A_z^\hi$.
 Since $C_1=\emptyset$, no high domination edge list at a low vertex
 needs initialization.
\end{enumerate}
No vertex is removed from $\mis_\mi$ during the flush, so every vertex
remaining in $C_2$ has a middle-MIS neighbor throughout the flush.
The flush therefore does not enumerate $C_2$.
The update terminates after this reconstruction,
with $C_1=Q=\emptyset$.

\paragraph{Initialization and reporting.}
The graph starts empty. Set $\mis=\mis_\lo=V_\lo=V$, all other
sets and lists to empty, all counters to zero, and all list pointers to null.
This initialization takes $O(n)$ time using the fixed order of vertex identifiers. Each update reports the final changes in MIS membership; returning a complete explicit listing incurs an additional cost proportional to its output size. The adversary observes reported
MISs, but not private lists, coin outcomes, traversal orders, or running times.

\subsection{Proof of Correctness}

Fix an arbitrary realization of the algorithm's random choices, and let
$e=(s,t)$ be the updated edge. We prove
that $\mis$ is independent and maximal when the update terminates.
We follow the algorithm's order: first complete any degree migrations of $s$ and $t$
one at a time on the graph before the edge update, then process the edge
insertion or deletion using the resulting degree classes and MIS-layer
labels, and finally restore maximality.
During queue processing, the exact counters rule out low- and middle-MIS
neighbors of a low or middle vertex. The algorithm then finds
and removes all of its neighbors from $\mis_\hi$ before including the vertex
in the MIS. A high vertex joins the MIS only when all three exact counters
equal zero. These checks and removals establish independence, regardless of
which domination edges were sampled.

For maximality, we show that every vertex outside $\mis$ is dominated when
the update terminates. The exact counters and domination edge lists provide domination certificates: for $z\in V_\mi\cup C_1$, every stored edge in
$A_z^\hi$ joins $z$ to a current vertex of $\mis_\hi$. A nonempty list
therefore certifies an MIS neighbor, whereas an empty list does not imply
that no such neighbor exists. We verify that these certificates remain valid
and that, in regular mode, every vertex outside $\mis$ without a maintained
domination certificate belongs to $Q$ immediately before each queue
extraction and when the update terminates. In flush mode, the low and middle passes establish domination
by $\mis_\lo\cup\mis_\mi$, and high vertices are handled by the final
reconstruction of $\mis_\hi$.

\paragraph{The restoration invariant.}
A vertex $z\notin\mis$ has a \emph{domination certificate} in regular mode
if one of the following conditions holds:
\begin{itemize}
 \item if $z\in V_\lo$, then
 \[
 d_\lo(z)>0\qquad\text{or}\qquad d_\mi(z)>0
 \qquad\text{or}\qquad
 \bigl(z\in C_1\ \text{and}\ A_z^\hi\neq\emptyset\bigr);
 \]
 \item if $z\in V_\mi$, then
 \[
 d_\lo(z)+d_\mi(z)>0\qquad\text{or}\qquad A_z^\hi\neq\emptyset;
 \]
 \item if $z\in V_\hi$, then
 \[
 d_\lo(z)+d_\mi(z)+d_\hi(z)>0.
 \]
\end{itemize}
Every vertex with a domination certificate has an actual neighbor in $\mis$: each exact counter counts current MIS neighbors, and each sampled list entry corresponds to a current graph edge to a vertex in $\mis_\hi$.

Throughout regular mode, we maintain the following invariants at the
beginning of every queue-extraction iteration and at the termination of
the update.
\begin{enumerate}[(I)]
 \item \emph{Independence.}
 The set $\mis=\mis_\lo\cup\mis_\mi\cup\mis_\hi$ is an independent
 set, not necessarily maximal.
 \item \emph{Correctness of the data structures.}
 The exact counters, adjacency lists, indexed sets, and domination edge
 lists satisfy their definitions in \Cref{sec:mis-data-structures}.
 In particular, every domination edge used to certify domination joins
 a vertex to a current MIS neighbor.
 \item \emph{Candidate membership.}
 The sets $C_1$ and $C_2$ satisfy \Cref{def:candidate-sets}. Every vertex
 of $C_1$ has a high domination edge list, and $|C_1|<B$.
 \item \emph{Queue coverage of vertices without domination certificates.}
 Every vertex outside $\mis$ without a domination certificate belongs to $Q$. A vertex in $Q$ might acquire a domination certificate before its request is processed.
\end{enumerate}
A vertex without a domination certificate may have actual high-MIS neighbors whose edges were
not sampled. Requiring it to belong to $Q$ ensures that losing such an
edge cannot leave an undominated vertex unnoticed.

In flush mode, the high MIS is empty during the passes over $C_1^0$ and
$V_\mi$. These passes handle low and middle vertices directly;
high vertices are deliberately postponed until $\mis_\hi$ is rebuilt.
The exact counters remain correct, but high domination edge lists at low
vertices are absent during these passes. The final reconstruction restores \textup{(I)}--\textup{(IV)} with $C_1=Q=\emptyset$.

During preparation, a new member of $C_1$ may have no stored high
domination edge list, and $|C_1|$ may exceed the flush threshold. Thus
\textup{(II)}--\textup{(IV)} are asserted at the start of regular mode,
after the flush test and the required list initializations. A flush
performs no queue-extraction iteration; for this mode, the lemmas assert
these properties at completion of the final reconstruction.

All four properties hold at initialization. The following four lemmas
verify them in order.

\begin{lemma}\label{lem:restore-independence}
Assume that \textup{(I)}--\textup{(IV)} hold before an update. The
preparation phase, every completed regular iteration, and the flush
passes preserve \textup{(I)}.
\end{lemma}
\begin{proof}
\noindent\textbf{Preparation.}
\begin{enumerate}[(a),leftmargin=*]
\item \emph{Degree migration.}
A degree migration only transfers an MIS vertex between layer sets, so it
changes neither the graph nor $\mis$ and preserves independence.

\item \emph{Edge insertion.}
After both migrations, inserting $(s,t)$ can create a conflict only
between $s$ and $t$. If both belong to $\mis$, the algorithm removes
$t$ from $\mis$ when $s\prec t$ and removes $s$ from $\mis$ otherwise, using the degree classes
after migration. The earlier vertex under $\prec$ stays in the MIS,
and independence is restored.

\item \emph{Edge deletion.}
After both migrations, deleting the edge changes no MIS membership and
cannot violate independence.
\end{enumerate}
Thus preparation preserves \textup{(I)}.

\medskip
\noindent\textbf{Regular mode.}
Let $v$ be the extracted vertex. Consider the degree class of $v$.
\begin{itemize}[leftmargin=*]
 \item If $v\in V_\lo$, \Cref{case:regular-low} permits including $v$ in $\mis_\lo$
 only when $d_\lo(v)=d_\mi(v)=0$ and removes every vertex of
 $N(v)\cap\mis_\hi$ from $\mis_\hi$ before $v$ joins $\mis_\lo$. Hence no conflict involving
 $v$ remains.
 \item If $v\in V_\mi$, \Cref{case:regular-middle} permits including
 $v$ in $\mis_\mi$ only when $d_\lo(v)+d_\mi(v)=0$ and removes every high-MIS
 neighbor of $v$ from $\mis_\hi$ before $v$ joins $\mis_\mi$.
 \item If $v\in V_\hi$, \Cref{case:regular-high} permits including $v$ in $\mis_\hi$
 only when $d_\lo(v)+d_\mi(v)+d_\hi(v)=0$.
\end{itemize}
If $v$ already belongs to $\mis$, has a positive maintained MIS-neighbor
counter, or has a nonempty stored high domination edge list, the iteration
ends without changing $\mis$. Thus every completed regular iteration
preserves \textup{(I)}.

\medskip
\noindent\textbf{Flush mode.}
Step~\ref{step:flush-clear} removes every vertex from $\mis_\hi$ and therefore preserves
independence of $\mis_\lo\cup\mis_\mi$. Every vertex of $C_1^0$ has no
middle-MIS neighbor, and $\mis_\mi$ is unchanged during
Step~\ref{step:flush-low}. A vertex joins $\mis_\lo$ in this step only when its
low counter is zero, so this change creates no conflict.
Step~\ref{step:flush-middle} next extends $\mis_\mi$ inside
$V_\mi\setminus N(\mis_\lo)$. Finally,
Step~\ref{step:flush-high} constructs an independent set inside
$V_\hi\setminus N(\mis_\lo\cup\mis_\mi)$. Each construction excludes
every conflict with $\mis_\lo$ in the middle pass and with
$\mis_\lo\cup\mis_\mi$ in the high pass.
Consequently their union satisfies \textup{(I)} after the flush.
\end{proof}

\begin{lemma}[Correctness of the data structures]\label{lem:data-correctness}
Assume that \textup{(I)}--\textup{(IV)} hold before an update. The
preparation phase together with the list initializations preceding regular
mode, every completed regular iteration, and the completed flush preserve
\textup{(II)}.
\end{lemma}
\begin{proof}
\noindent\textbf{Preparation.}
\begin{enumerate}[(a),leftmargin=*]
\item \emph{Degree migration.}
For each $w\in\{s,t\}$ that changes degree class, the algorithm first deletes lists that are no longer
maintained. For every edge $(z,x)$ in a deleted list $A_z^\hi$ or
$R_x^\hi$, it also removes the entry for that edge from the other list.
It then changes the degree
class, moves the entries for $w$ in its neighbors' adjacency lists,
and updates the indexed class sets. If $w\in\mis$, the algorithm
moves $w$ to the corresponding MIS layer, updates the indexed high MIS
when necessary. In that case, at each neighbor it decreases the counter for $w$'s
old MIS layer by one and increases the counter for its new MIS layer by
one, whenever those counters are maintained there.
The counters $d_\lo$ and $d_\mi$ remain defined at every vertex;
$d_\hi$ is initialized when a vertex enters $V_\hi$ and discarded
when it leaves. All scans use the graph before the edge update, so the
maintained counters equal the numbers of neighbors in the resulting
MIS layers of that graph.

Finally, migration initializes each newly required high domination edge
list at a middle vertex or reverse high domination edge list at a
high-MIS vertex, using \Cref{lem:output-sensitive-bernoulli} with
probability $p_\hi$. For each sampled vertex that passes the adjacency
test, the edge $(z,x)$ is stored in both $A_z^\hi$ and $R_x^\hi$.
The second degree migration uses the
classes and lists resulting from the first. Thus an edge whose two lists
are newly created is sampled when the later list is initialized.
The edge being inserted is still absent, and the edge being deleted is
still present. Lists at low vertices leaving $C_1$ are deleted together
with the entry for each stored edge $(z,x)$ in $R_x^\hi$ and the entry for each list in the indexed collection; lists at new members of
$C_1$ remain uninitialized until after the flush test.

\item \emph{Edge insertion.}
With the degree migrations complete, the algorithm inserts the edge into the
adjacency lists and adjacency dictionaries using the resulting degree classes and increments precisely the
counters at $s$ and $t$ whose definitions include the new edge. If both
$s,t\in\mis$, the algorithm then removes from $\mis$ whichever of $s,t$ comes
later under $\prec$. Removing a vertex from the low or middle MIS decrements the corresponding
counter at every neighbor. Removing a vertex from the high MIS decrements the high
counters at its high degree neighbors and deletes both copies of every edge in
its reverse list. The indexed high MIS is updated with its membership.
After recomputing membership in $C_1$ and $C_2$ from the updated counters and MIS-membership bits, the algorithm samples the new
edge only if it joins a current high-MIS vertex to a vertex with a stored
high domination edge list. Hence it stores no edge incident to the
vertex that has just left the high MIS.

\item \emph{Edge deletion.}
The algorithm first deletes both sampled copies of the edge, if present.
It then removes the edge from the adjacency lists and adjacency dictionaries, decrements the counters at $s$ and $t$
using the degree classes after migration, and recomputes the endpoints' membership in $C_1$ and $C_2$ from their updated counters and MIS-membership bits. Thus the counters and adjacency lists now describe
the graph after deletion, and no domination edge list contains the deleted
edge.
\end{enumerate}

\medskip
\noindent\textbf{Regular mode.}
When $|C_1|<B$ after preparation, the algorithm initializes the missing high
domination edge lists at vertices of $C_1$ by sampling their current
high-MIS neighbors. It sets the list pointer and places the list in the indexed collection even
if the list is empty. This establishes \textup{(II)} before the first
queue extraction.

Consider a completed regular iteration. If the extracted vertex already
belongs to the MIS, has a positive maintained MIS-neighbor counter, or
has a nonempty stored high domination edge list, the iteration ends
without changing any of these data structures. To include a low or middle
vertex in the MIS, the algorithm first
removes each conflicting vertex $x$ from $\mis_\hi$ and its indexed set. For each such $x$, it decrements $d_\hi(z)$ at every neighbor $z\in V_\hi$ and removes both stored entries of every edge $(z,x)$ in $R_x^\hi$, one from $R_x^\hi$ and one from $A_z^\hi$. It then includes
the extracted vertex in $\mis_\lo$ or $\mis_\mi$, according to its degree
class, and increments the corresponding
counter at every neighbor. For each low vertex $z$ leaving $C_1$, the algorithm deletes $A_z^\hi$ and removes the entry for every stored edge $(z,x)$ from $R_x^\hi$. At every middle vertex
$v$, the high domination edge list $A_v^\hi$ remains stored, including
when $v$ joins $\mis_\mi$. When a vertex joins the high MIS,
the algorithm first updates the indexed high MIS and the counters at high degree
neighbors, then initializes the new reverse high domination edge list
$R_x^\hi$ of the new high-MIS member $x$. For each sampled edge $(z,x)$,
it stores an entry in both $R_x^\hi$ and $A_z^\hi$.

In all these operations, a stored edge $(z,x)\in A_z^\hi$ joins $z$
to a current vertex $x\in\mis_\hi$ and is stored in $A_z^\hi$ if
and only if it is stored in $R_x^\hi$. Whenever a low vertex's high
domination edge list is deleted, its pointer is set to null and its
entry in the indexed collection is removed. Each edge $(z,x)$ receives one fresh
sampling bit when the conditions $(z,x)\in E(G)$, $x\in\mis_\hi$,
and $A_z^\hi$ being stored first hold simultaneously; the bit remains
unchanged until one of these conditions ceases
to hold. Thus the sampling rule and all maintained
data structures satisfy \textup{(II)} after every completed iteration.

\medskip
\noindent\textbf{Flush mode.}
Step~\ref{step:flush-clear} empties $\mis_\hi$, removes both stored entries
of every sampled edge by traversing the reverse lists, and discards the high domination edge lists
at low vertices. The low and middle passes use the same counter updates verified
above. Finally, Step~\ref{step:flush-high} rebuilds $\mis_\hi$, updates
its indexed set and the counters $d_\hi$, and initializes the reverse
high domination edge lists $R_x^\hi$, storing each sampled edge $(z,x)$ in both $R_x^\hi$ and the middle vertex's list $A_z^\hi$. At this
point $C_1=\emptyset$, so no high domination edge list at a low vertex is required.
The graph and degree classes do not change during a flush, so the adjacency
lists remain correct. Hence \textup{(II)} holds when the flush terminates.
\end{proof}

\begin{lemma}[Candidate membership]\label{lem:low-list-count}
Assume that \textup{(I)}--\textup{(IV)} hold before an update. The
preparation phase together with the list initializations preceding regular
mode, every completed regular iteration, and the completed flush preserve
\textup{(III)}. Moreover, fewer than $B$ high domination edge lists are
stored at low vertices at any time, and a flush finishes with
$C_1=\emptyset$ and no such stored lists.
\end{lemma}
\begin{proof}
We consider each phase separately.

\medskip
\noindent\textbf{Preparation.}
\begin{enumerate}[(a),leftmargin=*]
\item \emph{Degree migration.}
A vertex leaving the low degree class is removed from $C_1\cup C_2$. A vertex
$v$ entering the low degree class is assigned to $C_1$, $C_2$, or neither
according to whether $v\in\mis$ and the values of $d_\lo(v)$ and
$d_\mi(v)$. If an MIS vertex changes
class, the algorithm completes both counter changes at each neighbor
before updating that neighbor's candidate membership. Thus the candidate
sets satisfy \Cref{def:candidate-sets} after each degree migration
on the graph before the edge update.

\item \emph{Edge insertion.}
After updating the counters at $s$ and $t$ and removing $s$ or $t$
from the MIS if necessary, the algorithm recomputes membership in $C_1$ and $C_2$ for each low degree endpoint and every low degree neighbor $z$ whose $d_\lo(z)$ or $d_\mi(z)$ changes, using the updated counters and MIS-membership bits. Thus candidate membership uses the updated graph and MIS.

\item \emph{Edge deletion.}
After deleting the edge and decrementing the counters at $s$ and $t$,
the algorithm recomputes membership in $C_1$ and $C_2$ for each $w\in\{s,t\}\cap V_\lo$ using its updated counters and MIS-membership bit.
For every $z\notin\{s,t\}$, MIS membership, degree class, $d_\lo(z)$,
and $d_\mi(z)$ remain unchanged at this step. Hence the candidate sets again
satisfy their definitions.
\end{enumerate}

Preparation deletes a low vertex's high domination edge list when the vertex
leaves $C_1$ and never initializes a new one. It therefore preserves the
list-count bound from the previous update.

\medskip
\noindent\textbf{Regular mode.}
When $|C_1|<B$ after preparation, the algorithm initializes exactly the
missing high domination edge lists at vertices in $C_1$. Thus
\textup{(III)} holds when regular mode begins.

When a low vertex $v$ joins $\mis_\lo$, the algorithm removes $v$ and its low degree neighbors from the candidate
union. When a middle vertex joins $\mis_\mi$, the algorithm moves each of its neighbors in $C_1$ to $C_2$.
A high-MIS change affects neither $d_\lo(z)$ nor $d_\mi(z)$ at any
vertex $z$. These are all
possible candidate-set changes during regular mode, and the algorithm
performs each one explicitly. No vertex leaves $\mis_\lo\cup\mis_\mi$,
so neither $d_\lo(z)$ nor $d_\mi(z)$ decreases at any vertex $z$.
Thus $C_1$ can only shrink,
and the lists of vertices leaving it are deleted. This proves
\textup{(III)} throughout regular mode.

\medskip
\noindent\textbf{Flush mode.}
Every vertex of $C_1^0$, the set $C_1$ at the start of the flush,
enters $\mis_\lo$ or acquires a neighbor in $\mis_\lo$.
Every low vertex outside $C_1^0$ already belongs to the MIS or has
a neighbor in $\mis_\lo\cup\mis_\mi$ that remains in the MIS throughout the flush.
Thus $C_1$ becomes empty
after the low pass and cannot grow during the later passes. Membership in $C_1$ and $C_2$ is recomputed whenever $d_\lo$ or $d_\mi$ changes, all high domination edge
lists at low vertices are deleted, and no new one is initialized. Hence
\textup{(III)} holds after the flush.
\end{proof}

\begin{lemma}[Queue coverage of vertices without domination certificates]\label{lem:request-coverage}
Assume that \textup{(I)}--\textup{(IV)} hold before an update. The
preparation phase together with the list initializations preceding regular
mode, every completed regular iteration, and the completed flush preserve
\textup{(IV)}.
\end{lemma}
\begin{proof}
At the end of the previous update, every vertex outside the MIS has a domination certificate.
We check every operation that can change this fact.

\medskip
\noindent\textbf{Preparation.}
\begin{enumerate}[(a),leftmargin=*]
\item \emph{Degree migration.}
A migration changes neither the graph nor MIS membership, but it can
change which counters or sampled edges certify domination. When an MIS
vertex changes layer, the algorithm scans the neighbors whose exact
counters change and, if it leaves $\mis_\hi$, traverses its reverse
high domination edge list. It tests the affected vertices and places a
vertex in $Q$ only when its counters and stored list no longer certify
domination. After initializing the structures for the migrated vertex's
new class, it applies the same tests to that vertex. If the vertex is
already in $Q$, its position is updated according to its new degree
class. A new member of $C_1$ without a stored list is tested after
list initialization, as described below. These operations are completed
for both $s$ and $t$ before the graph edge is changed.

\item \emph{Edge insertion.}
Write the inserted edge as $e=(s,t)$, with $s\prec t$ in the order
after degree migration. Inserting $e$ updates the counters at $s$ and
$t$. If at most one of $s,t$ belongs to $\mis$, no vertex is removed from the MIS and no vertex loses a domination certificate.

If both $s$ and $t$ belong to $\mis$, the algorithm keeps $s$ in
$\mis$ and removes $t$ from $\mis$. Since $s\prec t$, a low or middle vertex
$t$ has $s\in\mis_\lo\cup\mis_\mi$ as a neighbor, so
$d_\lo(t)+d_\mi(t)>0$. If $t$ is high, the neighbor $s$ contributes
to one of the three exact counters at $t$, so
$d_\lo(t)+d_\mi(t)+d_\hi(t)>0$. Thus $t$ has a domination certificate and need
not be placed in $Q$.

Removing $t$ from $\mis$ may leave a neighbor $z$ without a domination certificate. If $t$ is low or
middle, the algorithm scans $N(t)$ and decrements the corresponding
counter at each neighbor $z$. If $t$ is high, the algorithm decrements
$d_\hi(z)$ at every high degree neighbor $z$ and, for each edge $(z,t)$ stored
in $R_t^\hi$, removes its entries from $R_t^\hi$ and $A_z^\hi$. After each change,
the algorithm requests the affected vertex $z$ only if $z\notin\mis$
and the counter and list tests for a request pass. Low vertices without
a stored high domination edge list are checked after list initialization.
Finally, sampling the inserted edge $(s,t)$ can only add a domination
certificate.

\item \emph{Edge deletion.}
The algorithm first deletes the sampled copies, removes the edge from the
adjacency lists and adjacency dictionaries, and decrements $d_\lo(s)$ if $t\in\mis_\lo$, $d_\mi(s)$ if $t\in\mis_\mi$, or $d_\hi(s)$ if $t\in\mis_\hi$ and $s\in V_\hi$, with the symmetric rules at $t$. It tests $s$ and $t$ after these
changes. Only $s$ and $t$ can lose a domination certificate because
of the deleted edge. For an endpoint with a stored high domination edge
list or a maintained high-MIS-neighbor counter, these tests determine
whether it has lost all maintained domination certificates. A low endpoint
newly entering $C_1$ without a stored list is checked after list
initialization if regular mode is used.
\end{enumerate}

\medskip
\noindent\textbf{Regular mode.}
When $|C_1|<B$ after preparation, a low vertex whose high domination edge list
is missing is checked after that list is initialized: if the list is
empty, the vertex is put in $Q$
before regular mode begins. These checks establish queue coverage at
the start of regular mode, proving \textup{(IV)} there.

During regular mode, only high-MIS vertices are removed from the MIS.
Suppose that $x$ is removed from $\mis_\hi$. The algorithm checks every
high degree neighbor of $x$ and every vertex $z$ for which $(z,x)$ is
stored in $R_x^\hi$: it scans $N(x)\cap V_\hi$ and traverses $R_x^\hi$.
Every affected vertex is tested, and it is put in $Q$ if its remaining
counters and high domination edge list no longer certify domination.
These operations are performed for every neighbor removed from $\mis_\hi$ before a
low or middle vertex joins the MIS. The removed vertex $x$ itself is dominated by
that new MIS member before the next extraction and needs no
request. When a high vertex joins the MIS, no vertex is removed from the MIS, and all
other branches either add domination or leave the MIS unchanged.

For a low degree or middle degree neighbor $z$ of the removed high-MIS vertex $x$,
an edge $(z,x)$ that was not stored requires no separate check. Either
$z$ already had no domination certificate and belonged to $Q$, or it had a domination certificate from $d_\lo(z)+d_\mi(z)>0$ or a different sampled edge in $A_z^\hi$.
Removing $x$ from $\mis_\hi$ changes neither condition.
This observation applies to high-MIS
removals in preparation as well.

A request is discarded without changing MIS membership only when its vertex
already belongs to the MIS, has a positive maintained MIS-neighbor
counter, or has a nonempty stored high domination edge list. Otherwise,
the iteration includes the vertex in the MIS and makes all resulting requests
before the next extraction. No other pending request is removed, and
duplicate requests leave the original one present.
Therefore every completed regular iteration preserves \textup{(IV)}.

\medskip
\noindent\textbf{Flush mode.}
Step~\ref{step:flush-clear} empties $Q$ and removes every vertex of $\mis_\hi$ from the MIS.
The subsequent passes process the affected vertices directly, so
\textup{(IV)} is not asserted until the high layer has been reconstructed.
Every low vertex outside $\mis$ with no low- or middle-MIS neighbor at
the start of the flush belongs to $C_1^0$.
Step~\ref{step:flush-low} includes each such vertex in $\mis_\lo$
or gives it a low-MIS neighbor. All other low vertices either stay in the
MIS or remain adjacent to a vertex in $\mis_\lo\cup\mis_\mi$, since
the flush removes no vertex from these two MIS layers. Step~\ref{step:flush-middle}
then extends $\mis_\mi$ to a maximal independent set on the middle
vertices not dominated by $\mis_\lo$. Consequently, every low or middle
vertex $v\notin\mis$ satisfies $d_\lo(v)+d_\mi(v)>0$ after these
two passes. No later step removes a vertex from $\mis_\lo\cup\mis_\mi$.

Finally, Step~\ref{step:flush-high} constructs a maximal independent set
on the high vertices not dominated by $\mis_\lo\cup\mis_\mi$. Every remaining
high vertex outside $\mis$ therefore has a positive exact counter.
Thus every vertex outside $\mis$ has a domination certificate, and \textup{(IV)} holds
with $Q=\emptyset$ after the flush.
\end{proof}

The four invariants imply correctness at the end of each update.

\begin{lemma}\label{lem:correctness}
When the update algorithm terminates,
$\mis=\mis_\lo\cup\mis_\mi\cup\mis_\hi$ is a maximal independent set
of the current graph $G$.
\end{lemma}
\begin{proof}
There are two possible termination cases. In regular mode, the update
terminates only after all requests have been extracted. By
\Cref{lem:restore-independence,lem:data-correctness,lem:low-list-count,lem:request-coverage},
\textup{(I)}--\textup{(IV)} hold with $Q=\emptyset$. In flush mode,
the update terminates only after the low and middle passes and the final
high-layer reconstruction. The flush cases of the same four lemmas again
give \textup{(I)}--\textup{(IV)} with $Q=\emptyset$.
Thus, in every termination case, \textup{(I)} gives independence and
\textup{(IV)} ensures that every vertex outside $\mis$ has a domination certificate. By
\textup{(II)}--\textup{(III)}, each such vertex has an actual MIS neighbor.
Hence $\mis$ is maximal.
\end{proof}

\subsection{Probabilistic Properties against Adaptive Adversaries}

We use the two probabilistic games from the preliminaries to prove the
following lemmas. The running-time analysis requires three sampling guarantees:
reverse lists remain small, the stored high domination edge
lists of low and middle vertices remain small, and an empty high domination
edge list implies that its vertex has few neighbors in the high MIS. These
properties bound, respectively, the work of removing a vertex from $\mis_\hi$,
maintaining a low or middle vertex's high domination edge list, and removing
conflicting high-MIS neighbors from $\mis_\hi$ before a low or middle vertex enters the MIS. For
fixed sets, the guarantees follow from standard sampling bounds. The
difficulty here is that the reported MIS depends on the sampled lists,
and the adversary uses these outputs to choose later updates. Thus the
current graph and MIS cannot be treated as independent of the sampling
choices.

The membership-query game of \Cref{lem:adaptive-stopped-membership} gives
the reverse-list bound. The emptiness-query game of
\Cref{lem:adaptive-hidden-sample} bounds both the stored high domination
edge lists and the number of high-MIS neighbors when such a list is empty.
In each proof, a simulator acts as the player in the corresponding game.
We fix the adversary's internal randomness and all sampling coins except
those defining the sampled list under consideration. The excluded coins
include those shared with copies of the sampled edges in other lists.
The simulator uses the fixed coins to simulate the MIS algorithm and
the adaptive adversary.
The simulator must reproduce the MIS reported after each completed update to ensure that the adversary chooses the same subsequent updates. We verify
that it can do so using only the operations allowed by the game, and
then apply the game's probabilistic guarantees to the sampled list.

Throughout this analysis, a time interval is a consecutive portion of the
algorithm's execution, ordered by its individual operations. Its start and
end events may occur within a single edge update. Each game step is one
permitted update or query in the probabilistic game; one edge update of
the graph may generate several game steps.

\begin{lemma}\label{lem:adaptive-reverse-list-size}
Against any adaptive adversary generating a polynomial-length update
sequence, with high probability, simultaneously throughout the sequence,
\[
 |R_x^\hi|=\widetilde O(m^{3/8})\qquad(x\in\mis_\hi).
\]
On the same event, every reverse-list initialization samples
$\widetilde O(m^{3/8})$ vertices, including vertices rejected by the
adjacency test.
\end{lemma}

\begin{proof}
We use the simulation argument for the membership-query game of
\Cref{lem:adaptive-stopped-membership}. In this application, $U$
represents the adjacency list of one high-MIS vertex restricted to
neighbors whose high domination edge lists are stored, including both
sampled and unsampled edges. The set $S$ represents that vertex's reverse
high domination edge list. For each application, fix all sampling coins
except those defining this reverse list. Given these fixed coins, a simulator can
reproduce the reported MISs, and hence the adversary's next updates,
while consulting $S$ only through the operations allowed by the game.
We verify this claim below.

Fix a vertex $x$ and consider the maximal time interval from one entry
of $x$ into $\mis_\hi$ until its next departure. Entry or departure
through degree migration is included, and removal by a flush also ends
the time interval. A later reentry starts a new time interval, even within
the same edge update. Throughout this time interval, define
\[
 U:=\{(x,y)\in E(G):A_y^\hi\text{ is stored}\},
 \qquad S:=R_x^\hi.
\]
Thus $U$ is the adjacency list of $x$ restricted to middle vertices and
low candidates whose high domination edge lists have been initialized, and $S$ consists
of its sampled edges. An edge $(x,y)\in U$ belongs to $S$ exactly when
it is stored in $A_y^\hi$.

Condition on the history before $x$ enters $\mis_\hi$. The game starts with $U=S=\emptyset$. When $R_x^\hi$ is initialized, place every graph edge $(x,y)$ for which $A_y^\hi$ is stored into $U$. Assign each such edge an independent Bernoulli-$p_\hi$ bit, where $p_\hi=\tfrac12m^{-1/4}$, and include it in $S=R_x^\hi$ exactly when its bit is one. The simulator knows the graph, MIS, exact
counters, and all sampling bits not used to define $S$. For $(x,y)\in U$,
the same hidden bit determines whether the edge appears in $S$ and
in $A_y^\hi$, so the simulator cannot inspect that copy separately.
Fixing the adversary's internal randomness, the
simulator supplies each reported MIS to the adversary and obtains the
next graph update.

We first explain how updates to the adjacency list and reverse high
domination edge list correspond to the game's insertion and
reveal-and-delete operations. This correspondence specifies the lifetime
of each sampling bit and when the game ends.
Inserting an edge $(x,y)$ for which $A_y^\hi$ is stored, or
initializing a list at an existing neighbor $y$, inserts a fresh element
into $U$ with an independent sampling bit. This bit is kept as long as
the edge is present, $x$ stays in $\mis_\hi$, and the same list
$A_y^\hi$ remains stored. Deleting the edge or discarding $A_y^\hi$
removes the entry from $U$ and $S$ and discards its sampling bit.
Every later insertion of the same graph edge creates
a new game element with a fresh independent bit. In particular, a low vertex leaving
$C_1$ discards its high domination edge list; a later return initializes
that list using new independent samples. A middle vertex keeps its list
while it remains in the middle degree class. The departure of $x$ from $\mis_\hi$, including its removal by
a flush, ends the game.

We represent deletion of an entry from $U$ by the game's
reveal-and-delete operation. This gives the simulator the entry's sampling
bit so that it can reproduce the removal of any stored copies and the
resulting domination checks. This disclosure occurs only within the simulation; the graph adversary observes only the reported MISs. The game
operation simultaneously removes the entry from $U$ and $S$, so the
revealed bit can no longer contribute to the sample or be used in a
later insertion.

We now show that the data structure's decisions and reported MISs can be
simulated using only the query rules of
\Cref{lem:adaptive-stopped-membership}. It remains to verify that
whenever the simulator makes an ordinary membership query about an edge
still in $U$, a no answer ends the game. For an edge $(x,y)\in U$, put
\[
 B_y:=A_y^\hi\setminus\{(x,y)\}.
\]
The simulator knows $B_y$, since its sampling coins are among those fixed.
While $B_y\neq\emptyset$, the list $A_y^\hi$ is nonempty under either
value of the sampling bit for $(x,y)$. No query is needed if $y\in\mis$ or
$d_\lo(y)+d_\mi(y)>0$, either. We defer querying this bit until $y$ is extracted from the simulated queue with $d_\lo(y)=d_\mi(y)=0$ and $B_y=\emptyset$. At that point the
simulator asks whether $(x,y)\in S$. A yes answer establishes that $y$ is dominated
and lets the game continue. A no answer establishes that $A_y^\hi$ is empty. The algorithm then removes every high-MIS neighbor of $y$, including $x$, from $\mis_\hi$ and
includes $y$ in $\mis_\lo$ if $y\in V_\lo$, or in $\mis_\mi$ if
$y\in V_\mi$. The current game therefore terminates,
exactly as required by \Cref{lem:adaptive-stopped-membership}.

It remains to justify deferring the query, since the real algorithm tests list
emptiness before placing a vertex in $Q$. Whenever a change to a
vertex $y$'s counters or domination edge list requires such a test, the
simulator first checks whether $y\notin\mis$ and
$d_\lo(y)=d_\mi(y)=0$. If these conditions hold but the list test
depends on the hidden bit of $(x,y)$, the simulator puts $y$ in its queue
without querying that bit. The same rule applies
when $A_y^\hi$ is initialized. All other list tests use known coins.
Deleting an entry from $U$ uses reveal-and-delete, but the simulator does
not query the sampling bits of edges still in $U$ to decide whether to
issue a request.
Membership in $C_1$ and $C_2$, the flush test, and decisions to include vertices in the high MIS use
exact counters and are reproduced directly.

The simulated queue contains every real request and may contain additional
ones. By \Cref{lem:request-coverage}, every vertex outside the MIS with
neither a positive maintained MIS-neighbor counter nor a nonempty stored
high domination edge list is already in the real queue before an extraction.
A vertex $y$ extracted only from the simulated queue must therefore
belong to the MIS, satisfy $d_\lo(y)+d_\mi(y)>0$, or have a nonempty
stored list $A_y^\hi$. Its extraction changes only the simulated queue,
leaving MIS membership, counters, candidate sets, and domination edge
lists unchanged. Both queues use the same fixed priority order. After
omitting these additional extractions, both executions extract the same
next vertex and make the same changes to the MIS and data structures.
Induction therefore gives the same reported MISs and subsequent
adversarial updates.
In particular, a no answer to the membership query above occurs only
when the real algorithm also includes $y$ in the MIS and removes $x$ from it. Thus it ends
the game, and no other membership query about an edge still in $U$ is needed.

We can now apply the game lemma. By \Cref{lem:low-list-count}, fewer
than $B$ high domination edge lists are stored at low vertices at every
instant, even during preparation when $C_1$ itself may be larger. Hence
\[
 |U|\leq |V_\mi|+B=O(m^{5/8}+m^{1/2})=O(m^{5/8}),
\]
and the lemma gives
\[
 |R_x^\hi|=|S|=\widetilde O(p_\hi m^{5/8})
 =\widetilde O(m^{3/8}).
\]
This includes the state immediately before $x$ is removed from $\mis_\hi$; subsequent
cleanup only shortens its reverse list. There are polynomially many
operations and entries into $\mis_\hi$,
regardless of the sampling outcomes. Choosing a sufficiently small
inverse-polynomial failure probability and taking a union bound proves
the simultaneous reverse-list bound.

To bound the number of vertices sampled during initialization, condition on the history before a fresh
sampling call. The set $V_\mi$ and the indexed collection of high
domination edge lists stored at low vertices are then fixed, and the number
of sampled vertices is binomial with mean at most
$p_\hi(|V_\mi|+B)=O(m^{3/8})$. A Chernoff bound and a union bound over
all calls give $\widetilde O(m^{3/8})$ sampled vertices, including those that
fail the adjacency test.
\end{proof}

\begin{lemma}\label{lem:adaptive-high-list-size}
Against any polynomial-length adaptive update sequence, with high
probability, the following two bounds hold simultaneously for every stored
high domination edge list at a low or middle vertex, after initialization
and after every completed update to that list:
\[
 |A_v^\hi|=\widetilde O(m^{1/8}),
\]
and
\begin{equation}\label{eq:empty-high-list}
 A_v^\hi=\emptyset
 \quad\Longrightarrow\quad
 |N(v)\cap\mis_\hi|=\widetilde O(m^{1/4}).
\end{equation}
On the same event, every initialization of a list $A_v^\hi$ samples
$\widetilde O(m^{1/8})$ vertices, including vertices rejected by the
adjacency test.
\end{lemma}

\begin{proof}
We apply the emptiness-query game of \Cref{lem:adaptive-hidden-sample}
separately to each high domination edge list, with
$p=p_\hi=\tfrac12m^{-1/4}$. For each type of list, we specify the game, identify the corresponding list updates, and show how the simulator defers emptiness queries by queuing $v$ without testing the list. We then apply the game bounds. The
simulation must preserve the MIS reported after each completed update,
so the adaptive adversary chooses the same subsequent updates.

\medskip
\noindent\textbf{The high domination edge list of a low vertex.}\nopagebreak

\smallskip
\noindent\emph{1. Game specification.}
Fix a low vertex $v$ and follow one stored list $A_v^\hi$ over the time
interval from its initialization until it is discarded. Define
\[
 U:=\{(x,v)\in E(G):x\in\mis_\hi\},
 \qquad S:=A_v^\hi=\{e\in U:X_e=1\},
\]
where each edge $e$ newly placed in $U$ receives an independent
$X_e\sim\operatorname{Bernoulli}(p)$. Start with $U=S=\emptyset$
immediately before initialization. Place every current graph edge $(x,v)$ with $x\in\mis_\hi$ into $U$, and include it in $S$ exactly when $X_e=1$. Condition on the preceding history, fix the adversary's
internal randomness, and fix all sampling coins except those defining
$A_v^\hi$ in this game. The simulator acts as the game player: it
runs the MIS data structure and the adaptive adversary, supplying each
reported MIS to obtain the next graph update. It knows the graph, MIS,
exact counters, and all sampling bits not used to define $S$, but can
learn about $S$ only through the permitted emptiness queries. The game ends when
$A_v^\hi$ is discarded, including when $v$ leaves $C_1$, changes
degree class, or enters $\mis_\lo$. Any later initialization starts
a new game with fresh coins.
A flush also discards the list and ends the game.

\smallskip
\noindent\emph{2. Data-structure changes and game operations.}
Initialization inserts every edge from $v$ to $\mis_\hi$ into $U$.
Inserting $(x,v)$ with $x\in\mis_\hi$, or including a neighbor $x$ of
$v$ in $\mis_\hi$, inserts $(x,v)$ into $U$ with a fresh coin.
Deleting $(x,v)$ or removing $x$ from $\mis_\hi$ deletes the entry
from $U$ and $S$ without revealing its sampling bit. This bit is discarded;
a later insertion of the same graph edge uses a fresh independent bit.
For each $(x,v)\in U$, the simulator does not determine whether an entry
for that edge is stored in $R_x^\hi$. When $x$ leaves the high MIS,
the simulator deletes $(x,v)$ from the game's sets without learning its
sampling bit. Outside flush mode, it queues $v$ if $v\notin\mis$ and
$d_\lo(v)=d_\mi(v)=0$, regardless of whether the edge was sampled.

\smallskip
\noindent\emph{3. Deferring queries until $v$ is processed.}
The real execution checks whether $A_v^\hi$ is empty before issuing
certain requests for $v$, for example after initialization or removal
of a domination edge. The simulator instead puts $v$ in its queue
without querying $S$, provided $v\notin\mis$ and
$d_\lo(v)=d_\mi(v)=0$, suppressing duplicate requests. While the request
is pending, changes to the graph or MIS induce permitted updates to the
game sets $U$ and $S$. These updates reveal no sampling bits; any emptiness
query is deferred until $v$ is processed. Queuing $v$ does not end the game.
In any other vertex's reverse high domination edge list, the only potentially hidden entry is the edge to $v$. When that edge is deleted
from $U$, the simulator uses the same rule without asking whether the
edge was sampled. It also uses this rule when the graph or exact counters
at $v$ change. All other vertices' high domination edge lists use known
coins, and high-MIS decisions, membership in $C_1$ and $C_2$, and the flush test use
exact counters.

When $v$ is extracted, the simulator first checks whether $v\in\mis$
or $d_\lo(v)+d_\mi(v)>0$. If either condition holds, it discards the
request without changing MIS membership or any maintained list and
without querying $S$. Otherwise, it asks whether $S$ is empty. On a nonempty answer, the simulator discards the request without changing the MIS or any maintained list and continues the simulation.
An empty answer establishes that $A_v^\hi$ is empty. The algorithm then removes all high-MIS neighbors of $v$ from $\mis_\hi$, includes $v$ in $\mis_\lo$, and discards its list. This ends the game.
Thus every queried empty answer terminates the game, as required.

The additional requests preserve the reported MISs. By
\Cref{lem:request-coverage}, every vertex outside the MIS with neither a
positive maintained MIS-neighbor counter nor a nonempty stored high
domination edge list is already in the real queue before an extraction. Any additional
request is for $v$, so if $v$ is absent from the real queue, then
$v\in\mis$, $d_\lo(v)+d_\mi(v)>0$, or its stored list $A_v^\hi$
is nonempty. Its extraction changes only the simulated queue, leaving
MIS membership, counters, candidate sets, and domination edge lists
unchanged. Both queues use the same fixed priority order. After omitting
these additional extractions, their next extracted vertex is the same,
and its processing makes the same MIS changes and issues the same requests, except possibly
for additional requests for $v$. Induction over the extractions therefore
gives the same reported MISs and subsequent adversarial updates. In particular,
an empty answer after which the algorithm includes $v$ in the MIS can occur only when $v$ also belongs to
the real queue. The simulator may enumerate $U$ to issue these additional requests; the simulation's running time does not contribute to the algorithm's update bound.

During a flush no high domination edge list is tested for emptiness.
The simulator deletes each entry of $U$ and $S$ and discards its sampling
bit without revealing it. Discarding the high domination edge list
$A_v^\hi$ then ends the game.

At every state, $|U|\leq |V_\hi|=O(m^{3/8})$. The two conclusions of
\Cref{lem:adaptive-hidden-sample} therefore give
\[
 |A_v^\hi|=\widetilde O(p m^{3/8})=\widetilde O(m^{1/8}),
\]
and
\[
 A_v^\hi=\emptyset
 \quad\Longrightarrow\quad
 |N(v)\cap\mis_\hi|=\widetilde O(p^{-1})
 =\widetilde O(m^{1/4}).
\]
The second bound holds whether or not the list is queried at that state,
and includes the time immediately before $v$ joins $\mis_\lo$.

\medskip
\noindent\textbf{The high domination edge list of a middle vertex.}

\smallskip
\noindent\emph{1. Game specification.}
Fix $v\in V_\mi$ and define
\[
 U:=\{(x,v)\in E(G):x\in\mis_\hi\},
 \qquad S:=A_v^\hi=\{e\in U:X_e=1\}.
\]
Each new entry receives an independent
$X_e\sim\operatorname{Bernoulli}(p)$. Condition on the history before
each game, fix the adversary's internal randomness, and fix all sampling
coins except those defining $A_v^\hi$ in this game. As in the low-vertex case,
the simulator runs the MIS data structure and the adaptive adversary,
and accesses $S$ only through the game's emptiness queries. The first game
starts with $U=S=\emptyset$ immediately before list initialization.
Each game ends at the next of the following events: $v$ leaves $V_\mi$,
$v$ enters $\mis_\mi$, or a flush finishes clearing the high MIS.

A middle vertex keeps its list throughout its time in $V_\mi$, including
after joining the MIS. We start a new game immediately after $v$ joins $\mis_\mi$.
At that point $U=S=\emptyset$, since \emph{all} high-MIS
neighbors were removed from $\mis_\hi$ before $v$ joined $\mis_\mi$, including those whose edges
were not sampled. These deletions discard their sampling bits; every
later insertion uses a fresh independent bit. Likewise, a flush removes
all high-MIS neighbors of $v$ from $\mis_\hi$, deletes their entries from $U$ and $S$,
and discards their sampling bits without an emptiness query.
A new game starts with empty sets immediately after this
clearing, before the high MIS is reconstructed. Thus each game corresponds
to a time interval beginning at list initialization or one of these
restarts and ending at the next specified event. These successive game
time intervals cover the entire period during which the list is stored.

\smallskip
\noindent\emph{2. Data-structure changes and game operations.}
Initialization inserts every edge from $v$ to $\mis_\hi$ into $U$.
An edge insertion or entry of a neighbor into $\mis_\hi$ inserts the
corresponding entry with a fresh independent value of $X_e$.
An edge deletion or departure of the neighbor from $\mis_\hi$ deletes
that entry from $U$ and $S$ without revealing $X_e$. Its old bit is
discarded; any later insertion uses a fresh independent bit. As in the
low-vertex case, the simulator does not determine whether an entry for
$(x,v)$ is stored in $R_x^\hi$. When $x$ leaves the high MIS, it deletes
$(x,v)$ from the game's sets without learning the sampling bit. Outside
flush mode, it queues $v$ if $v\notin\mis$ and
$d_\lo(v)=d_\mi(v)=0$, regardless of whether the edge was sampled. Thus
$|U|=|N(v)\cap\mis_\hi|$ after every completed list update.

\smallskip
\noindent\emph{3. Deferring queries until $v$ is processed.}
Whenever the real execution would check whether $A_v^\hi$ is empty
to decide whether to queue $v$, the simulator puts $v$ in its queue
without querying $S$, provided $v\notin\mis$ and
$d_\lo(v)=d_\mi(v)=0$. This applies both at initialization and after
later changes to the counters or list. Duplicate requests are suppressed.
At extraction, if $v\in\mis_\mi$ or $d_\lo(v)+d_\mi(v)>0$,
the simulator discards the request without changing MIS membership or
any maintained list and without a query. Otherwise it
queries $S$. On a nonempty answer, the simulator discards the request without changing the MIS or any maintained list and continues the game. An empty answer establishes that $A_v^\hi$ is empty. The algorithm then removes all high-MIS neighbors of $v$ from $\mis_\hi$ and includes $v$ in $\mis_\mi$, ending the game.

The two executions can differ only by an extra request for $v$. Processing this request changes neither a maintained list nor MIS membership unless the real execution also needs to process $v$. The queue-coverage and fixed-order
argument from the low-vertex case therefore gives the same MIS changes.
Since $v$ never belongs to $C_1\cup C_2$ during this game, candidate
sets are identical as well. Both
executions make the same flush decisions and report the same MISs.
If a flush begins, the simulator deletes all entries of $U$ and $S$
without a query, as described above.

Applying \Cref{lem:adaptive-hidden-sample} with
$|U|\leq|V_\hi|=O(m^{3/8})$ gives
\[
 |A_v^\hi|=\widetilde O(p m^{3/8})=\widetilde O(m^{1/8}),
 \qquad
 A_v^\hi=\emptyset\ \Longrightarrow\
 |N(v)\cap\mis_\hi|=\widetilde O(p^{-1})
 =\widetilde O(m^{1/4}).
\]
As for low vertices, these estimates hold after initialization and every
completed list update, including immediately before $v$ joins $\mis_\mi$.

In both cases, there are polynomially many game operations and
initializations. Choosing sufficiently small failure
probabilities and taking a union bound over all vertices and all these
games proves the simultaneous list-size and empty-list assertions.

Finally, conditional on the history before an initialization, the number
of sampled vertices has distribution $\operatorname{Bin}(|\mis_\hi|,p)$,
with mean at most $p|V_\hi|=O(m^{1/8})$. A Chernoff bound and a union bound
over all calls give $\widetilde O(m^{1/8})$ sampled vertices per initialization,
including those rejected by the adjacency test.
\end{proof}

\subsection{Runtime Analysis}
\label{sec:runtime}

We prove the $\widetilde O(m^{5/8})$ amortized update bound using a potential function. For each completed edge update, we bound the actual running time plus the change in potential. The analysis follows the algorithm's
order: degree migration, edge insertion or deletion, and then maximality
restoration in regular mode or flush mode.
Throughout the cost analysis, we work on the event that the simultaneous guarantees of \Cref{lem:adaptive-reverse-list-size,lem:adaptive-high-list-size} hold.
Recall the parameters
\[
 b_0=m^{3/8},\qquad b_1=m^{5/8},\qquad
 p_\hi=\tfrac12m^{-1/4},\qquad B=m^{1/2}.
\]

\paragraph{The potential and the purpose of its terms.}
Define
\begin{equation}\label{eq:potential}
\begin{aligned}
 \Phi:={}&W_\lo|V_\lo\setminus\mis|
         +W_\mi|V_\mi\setminus\mis|
         +W_\hi|V_\hi\setminus\mis|\\
        &+\alpha|C_1\cup C_2|+\beta|C_1|+\Theta(\log n)|Q|,
\end{aligned}
\end{equation}
where
\[
 W_\lo=W_\mi=\widetilde\Theta(m^{5/8}),\qquad
 W_\hi=\widetilde\Theta(m^{3/8}),
\]
\[
 \alpha=\widetilde\Theta(m^{1/4}),\qquad
 \beta=\widetilde\Theta(m^{1/8}).
\]
The constants and logarithmic factors will be chosen to cover the costs
described below. Each term has a specific purpose:
\begin{itemize}[leftmargin=*]
 \item The first three terms pay for vertices joining the MIS in regular mode. Including
 a low, middle, or high vertex in the MIS decreases the corresponding
 term by $W_\lo$, $W_\mi$, or $W_\hi$, respectively. This decrease
 pays for its neighborhood scans and any high-MIS removals needed before
 the vertex joins the MIS.
 \item The term $\alpha|C_1\cup C_2|$ pays for flushes. We will show
 that at most $O(m^{3/8})$ vertices join this union per update. A vertex is
 counted again if it rejoins during a later update. Each flush removes
 at least $B$ vertices from the union. Transfers between
 $C_1$ and $C_2$ leave this term unchanged.
 \item The term $\beta|C_1|$ pays for deleting high domination edge
 lists when low vertices leave $C_1$ during regular mode.
 \item The term $\Theta(\log n)|Q|$ pays for extracting requests and
 checking whether the requested vertices should join the MIS.
\end{itemize}
Initially $\mis=V$ and $C_1=C_2=Q=\emptyset$, so $\Phi=0$.
The potential is always nonnegative. Thus the sum of the amortized costs over all updates upper-bounds their total actual running time.

\paragraph{Accounting for queue operations.}
We include the cost of issuing a request in the operation that examines
the vertex. Testing its counters and domination edge list takes $O(1)$
time. If the vertex is placed in $Q$, the $O(\log n)$ cost of placing it there and
the $\Theta(\log n)$ potential increase are both included. Each scanned
vertex or edge causes only $O(1)$ requests, so this adds at most an
$O(\log n)$ factor to the scan's cost. An attempt to issue a duplicate request changes neither $Q$ nor the potential.

Extracting or deleting a request decreases the queue term by
$\Theta(\log n)$, which pays for the queue operation and its
constant-time tests of MIS membership, exact counters, and the stored
high domination edge list, if present, when the coefficient is sufficiently large.
The same decrease pays for clearing $Q$. Updating a vertex's position
in $Q$ after degree migration does not change $|Q|$; its cost is included
in that migration.

\subsubsection{Preparation Phase}

For an update to $e=(s,t)$, we follow the preparation procedure: first
complete any degree migrations of $s$ and $t$ on the graph before the
edge update, then process the edge
insertion or deletion using the resulting degree classes. We first upper
bound the runtime of each degree migration together with the changes in
all potential terms other than $\beta|C_1|$.
The change in $\beta|C_1|$ is included when we combine preparation with
regular mode or flush mode below. During preparation, $|C_1|$ can
increase by $O(b_1)$, increasing this term by
$O(\beta b_1)=\widetilde O(m^{3/4})$. If the update enters regular
mode, its net increase is instead at most $\beta B$; a flush makes
$C_1$ empty and cancels any increase during preparation.

We first bound a list-deletion cost shared by the preparation cases.
By \Cref{lem:low-list-count}, fewer than $B$ high domination edge lists
are stored at low vertices, and no new ones are initialized during
preparation. Deleting each list $A_z^\hi$, the entry for every stored edge $(z,x)$ in $R_x^\hi$, and the list's entry in the
indexed collection takes $\widetilde O(m^{1/8})$ time by
\Cref{lem:adaptive-high-list-size}.

\begin{lemma}\label{lem:migration-time}
Each degree migration has cost $\widetilde O(m^{5/8})$, where this cost is the actual running time plus the changes in all terms of $\Phi$ other than $\beta|C_1|$.
\end{lemma}
\begin{proof}
The potential weights satisfy $W_\lo=W_\mi\geq W_\hi$ and
$b_0\alpha=\widetilde\Theta(m^{5/8})$. We analyze the four degree
migrations specified in
\Cref{case:migration-low-middle,case:migration-middle-low,case:migration-middle-high,case:migration-high-middle}.
Migration is performed before changing the graph. Since
$|\widehat{\deg}(w)-\deg(w)|=1$, a migration between the low and middle
degree classes scans $O(b_0)$ neighbors, and a migration between the
middle and high degree classes scans $O(b_1)$ neighbors. Updating the indexed class sets and issuing
requests are included in the bounds below. The second migration uses
the classes, counters, and lists produced by the first; each cost and
potential change is measured from the state immediately before that migration.
Throughout this proof, changes in $\beta|C_1|$ are deferred until the
flush test. Deleting high domination edge lists at low vertices costs
at most $\widetilde O(m^{5/8})$ in total, as established above.

\begin{enumerate}[(a),leftmargin=*]
 \item \emph{$V_\lo\to V_\mi$.}
 Deleting the old high domination edge list $A_w^\hi$, if stored,
 takes $\widetilde O(m^{1/8})$ time by
 \Cref{lem:adaptive-high-list-size}. Next, updating the adjacency lists
 and, if $w\in\mis$, decreasing $d_\lo(z)$ by one and increasing
 $d_\mi(z)$ by one at each neighbor $z$ takes $\widetilde O(b_0)$ time. Finally, initializing
 $A_w^\hi$ in the middle degree class takes $\widetilde O(m^{1/8})$ time,
 also within this bound.
 If $w\notin\mis$, the migration decreases
 $W_\lo|V_\lo\setminus\mis|$ by $W_\lo$ and increases
 $W_\mi|V_\mi\setminus\mis|$ by $W_\mi$. These changes cancel
 because $W_\lo=W_\mi$, and removing $w$ from $C_1\cup C_2$ can
 only decrease $\alpha|C_1\cup C_2|$. If $w\in\mis$, it remains
 in the MIS, so neither of these two potential terms changes. At most
 $O(b_0)$ low degree neighbors lose their last low-MIS neighbor and enter
 $C_1\cup C_2$, increasing $\alpha|C_1\cup C_2|$ by $O(b_0\alpha)$.
 These new candidates belong to $C_2$, since $w$ is now their
 middle-MIS neighbor. Thus $|C_1|$ does not increase in this migration.
 Together with the cost of deleting high domination edge lists at low
 vertices, the actual running time plus the changes in the potential
 terms considered here is at most $\widetilde O(m^{5/8})$.

 \item \emph{$V_\mi\to V_\lo$.}
 Deleting $A_w^\hi$, then scanning $N(w)$ to update the adjacency
 lists and, if $w\in\mis$, increase $d_\lo(z)$ by one and decrease
 $d_\mi(z)$ by one at each neighbor $z$ take
 $\widetilde O(b_0)$ time.
 If $w\in\mis$, its low degree neighbors gain a low-MIS neighbor and can
 only leave the candidate union. The cost of deleting their high
 domination edge lists is included in the bound above.
 If $w\notin\mis$, the migration decreases
 $W_\mi|V_\mi\setminus\mis|$ by $W_\mi$ and increases
 $W_\lo|V_\lo\setminus\mis|$ by $W_\lo$. These changes cancel
 because $W_\lo=W_\mi$. Placing $w$ in $C_1$ or $C_2$ increases
 $\alpha|C_1\cup C_2|$ by at most $\alpha$.
 Only $w$ itself can enter $C_1$, so its size increases by at most one
 and the increase in $\beta|C_1|$ is at most $\beta$.
 Including the list deletions, the actual running time plus the changes
 in the potential terms considered here is at most $\widetilde O(m^{5/8})$.

 \item \emph{$V_\mi\to V_\hi$.}
 Deleting $A_w^\hi$, updating the adjacency lists, and initializing
 $d_\hi(w)$ take $\widetilde O(b_1)$ time. If $w\in\mis$,
 decreasing $d_\mi(z)$ by one at every neighbor $z$ and increasing
 $d_\hi(z)$ by one at every high degree neighbor $z$ also takes
 $\widetilde O(b_1)$ time. In this case, initializing
 $R_w^\hi$ after $w$ enters $\mis_\hi$ by sampling middle vertices and
 high domination edge lists stored at low vertices takes
 $\widetilde O(m^{3/8})$ time by
 \Cref{lem:adaptive-reverse-list-size}, also within this bound.
 If $w\notin\mis$, the migration decreases
 $W_\mi|V_\mi\setminus\mis|$ by $W_\mi$ and increases
 $W_\hi|V_\hi\setminus\mis|$ by $W_\hi$, for a net change of
 $W_\hi-W_\mi\leq 0$. If $w\in\mis$, neither of these two
 potential terms changes. Its low degree neighbors may move from $C_2$ to $C_1$,
 but this leaves $\alpha|C_1\cup C_2|$ unchanged. At most
 $\deg(w)=O(b_1)$ vertices make this move, so the increase in
 $|C_1|$ is at most $O(b_1)$ and that in $\beta|C_1|$ is at most
 $O(\beta b_1)=\widetilde O(m^{3/4})$.
 Hence the actual running time plus the changes in the other potential
 terms is at most $\widetilde O(b_1)=\widetilde O(m^{5/8})$.

 \item \emph{$V_\hi\to V_\mi$.}
 Updating the adjacency lists takes $\widetilde O(b_1)$ time. If $w\in\mis$,
 deleting $R_w^\hi$, increasing $d_\mi(z)$ by one at every neighbor $z$,
 and decreasing $d_\hi(z)$ by one at every high degree neighbor $z$ also take
 $\widetilde O(b_1)$ time. Discarding $d_\hi(w)$ and then initializing
 $A_w^\hi$ also fit within this bound. If $w\in\mis$, a low degree neighbor
 that gains middle domination can move from $C_1$ to $C_2$. This move
 preserves the candidate union, and $|C_1|$ cannot increase in this
 migration. If its high domination edge list is
 stored, its deletion is included in the bound above; an uninitialized
 list requires no deletion.
 If $w\notin\mis$, the migration decreases
 $W_\hi|V_\hi\setminus\mis|$ by $W_\hi$ and increases
 $W_\mi|V_\mi\setminus\mis|$ by $W_\mi$, for a net increase of
 $W_\mi-W_\hi$; $w$ does not enter either candidate set.
 Including the list deletions, the runtime
 and changes in the potential terms considered here are at most
 $\widetilde O(b_1+m^{5/8})+W_\mi-W_\hi
 =\widetilde O(m^{5/8})$.
\end{enumerate}
These cases account for every operation and every potential term except
$\beta|C_1|$, proving the lemma.
\end{proof}

We now combine the migration costs with the costs of the subsequent edge
insertion or deletion and any removal needed to restore independence.

\begin{lemma}\label{lem:preparation-cost}
For every graph update, the preparation cost is $\widetilde O(m^{5/8})$, where this cost includes the actual running time and the changes in all terms of $\Phi$ other than $\beta|C_1|$.
This includes the degree
migrations, updates to the basic data structures and sampled domination
edge lists, all changes to $Q$, and removal of one MIS vertex when an
insertion violates independence.
\end{lemma}
\begin{proof}
Let $e=(s,t)$ be the inserted or deleted edge. We analyze the algorithm's steps in execution order.

\medskip
\noindent\emph{Degree migration.}
There is at most one degree migration for each of $s,t$. For each
migration, the actual running time plus the changes in the potential
terms other than $\beta|C_1|$ is at most $\widetilde O(m^{5/8})$ by
\Cref{lem:migration-time}. Both finish before the graph edge is changed.
All costs below use the resulting degree classes and data structures.

\medskip
\noindent\emph{Edge insertion.}
Inserting $e$ into $G$ changes two adjacency-list entries and at most one
exact counter at each of $s,t$. These first two steps take
$\widetilde O(1)$ time.

If both $s$ and $t$ belong to the MIS, the next step removes a vertex
$w\in\{s,t\}$ from $\mis$: $w=t$ if $s\prec t$, and $w=s$ otherwise.
Thus the earlier vertex under $\prec$ remains in the MIS. We
consider the degree class of $w$.
\begin{itemize}[leftmargin=*]
 \item If $w\in V_\lo$, both $s$ and $t$ belong to $V_\lo$. Scanning the neighbors of $w$ and updating their
 counters and membership in $C_1$ and $C_2$ take $\widetilde O(b_0)$ time, apart
 from the list deletions already bounded above. The removal increases
 $W_\lo|V_\lo\setminus\mis|$ by $W_\lo$ and can introduce
 $O(b_0)$ new candidates,
 increasing $\alpha|C_1\cup C_2|$ by $O(b_0\alpha)$. Including the update to
 the membership of $w$ in $C_1$ and $C_2$ gives $\widetilde O(m^{5/8})$ in total.
 The increase in $|C_1|$ is at most $O(b_0)$, contributing at most
 $O(\beta b_0)=\widetilde O(m^{1/2})$ to the potential.
 \item If $w\in V_\mi$, the neighborhood scan takes
 $\widetilde O(b_1)$ time and removing $w$ from the MIS increases
 $W_\mi|V_\mi\setminus\mis|$ by $W_\mi$.
 A low candidate losing its last middle-MIS neighbor moves
 from $C_2$ to $C_1$, without changing $\alpha|C_1\cup C_2|$.
 At most $\deg(w)=O(b_1)$ vertices make this move. Thus $|C_1|$
 increases by at most $O(b_1)$, and $\beta|C_1|$ increases by at
 most $O(\beta b_1)=\widetilde O(m^{3/4})$.
 The total for the other terms is again
 $\widetilde O(m^{5/8})$.
 \item If $w\in V_\hi$, scanning its high degree neighbors and traversing
 $R_w^\hi$ each take $\widetilde O(m^{3/8})$ time. Removing $w$
 from the MIS increases $W_\hi|V_\hi\setminus\mis|$ by $W_\hi$,
 while $d_\lo(z)$, $d_\mi(z)$,
 and candidate membership remain unchanged for every vertex $z$.
 Thus the actual time plus the potential
 increase is $O(W_\hi)$, including all requests placed in $Q$.
\end{itemize}

Finally, the algorithm recomputes the membership of each low degree endpoint in $C_1$ and $C_2$ from its updated counters and MIS-membership bit and samples
the new edge if it joins a current high-MIS vertex to a vertex with a
stored high domination edge list. Apart from the MIS removal already
counted, increasing the counters at $s$ or $t$ can only remove that
vertex from $C_1\cup C_2$ or move that vertex from $C_1$ to $C_2$.
The cost of deleting high domination edge lists at low vertices is
included in the $\widetilde O(m^{5/8})$ bound above.
Sampling the new edge stores at most two copies and takes
$\widetilde O(1)$ time. The sampling decision uses the endpoints' MIS memberships after independence has been restored.

\medskip
\noindent\emph{Edge deletion.}
If a sampled copy of $e$ is stored in $A_z^\hi$ for $z\in\{s,t\}$,
the algorithm first deletes both sampled copies and tests $z$ for a
request. It then removes $e$ from the adjacency lists and adjacency dictionaries and decrements the exact
counters at $s$ and $t$. Finally, it recomputes the membership of each low degree endpoint in $C_1$ and $C_2$ from its updated counters and MIS-membership bit and tests both vertices for a request. These steps take
$\widetilde O(1)$ time. One of $s,t$ may enter the candidate union,
increasing $\alpha|C_1\cup C_2|$ by at most $\alpha$.
A move from $C_2$ to $C_1$ leaves the union
unchanged. Only the vertex losing an MIS neighbor can enter $C_1$,
so $|C_1|$ increases by at most one and $\beta|C_1|$ by at most
$\beta$. High domination edge lists at vertices entering $C_1$ are
initialized only after the flush test, if the update enters regular mode.

Adding the migration costs and the subsequent insertion or deletion cost
proves the lemma. The explicit bounds above also show that preparation
increases $|C_1|$ by at most $O(b_1)=O(m^{5/8})$: there are at most
two migrations and one edge update.
\end{proof}

\subsubsection{Maximality Restoration}

We begin by bounding how much the candidate sets can grow during one edge update.

\begin{lemma}\label{lem:candidate-growth}
During any single edge update, the number of vertices that do not belong
to $C_1\cup C_2$ at the beginning of the update but subsequently
join $C_1\cup C_2$ is $O(b_0)=O(m^{3/8})$. Consequently,
throughout the update, the size of $C_1\cup C_2$ is at most its size at
the beginning of the update plus $O(m^{3/8})$.
\end{lemma}
\begin{proof}
Let $e=(s,t)$ be the updated edge. Moving a vertex from $C_1$ to $C_2$,
or from $C_2$ to $C_1$, does not
increase the size of their union. We therefore count only vertices
that join $C_1\cup C_2$ from outside the union. We consider all
places in the update procedure where a vertex can join the union.
\begin{enumerate}[(a),leftmargin=*]
 \item \emph{The preparation stage.}
 First consider degree migration. Only $s$ and $t$ can change
 degree class, so entry into the low degree class introduces at most two candidates.
 When a vertex in $\{s,t\}\cap\mis_\lo$ migrates to the middle degree class, its scan can
 additionally make $d_\lo(z)=0$ for $O(b_0)$ low degree neighbors. The other
 migrations do not introduce any further member of $C_1\cup C_2$.
 Next consider the edge update. An insertion can remove at most one
 vertex from $\mis_\lo$, causing at most $O(b_0)$ low degree neighbors to enter
 the union; the other insertion cases only move or remove candidates.
 A deletion can introduce a candidate only at $s$ or $t$.
 Thus preparation introduces $O(b_0)$ vertices into the union.
 \item \emph{A low vertex extracted from $Q$.}
 If the extracted vertex $v$ already belongs to $\mis$, satisfies
 $d_\lo(v)+d_\mi(v)>0$, or has a nonempty stored list $A_v^\hi$,
 the iteration ends without changing the candidate sets. Otherwise,
 the algorithm includes $v$ in $\mis_\lo$ and removes $v$ and its low degree
 neighbors from the candidate union. Thus a low iteration introduces no
 new member of $C_1\cup C_2$.
 \item \emph{A middle vertex extracted from $Q$.}
 A middle vertex never belongs to either candidate set. When processing its request, the algorithm may include it in $\mis_\mi$, which can only move low candidates from
 $C_1$ to $C_2$. Thus a middle iteration creates no candidate.
 \item \emph{A high vertex extracted from $Q$.}
 This case never causes a vertex to join either candidate set. The
 algorithm either discards the request or includes the vertex in $\mis_\hi$.
 \item \emph{Flush mode.}
 The low and middle passes remove candidates or move them between the
 two sets, and the final reconstruction of the high layer creates no
 candidate. Thus flush mode creates no additional source of candidates.
\end{enumerate}
Combining the contributions of $s,t$ and the $O(b_0)$ low vertices that
can lose their last low-MIS neighbor during preparation proves the lemma.
\end{proof}

The distinction between $C_1$ and $C_1\cup C_2$ is essential here.
Removing one vertex from $\mis_\mi$ can move $O(b_1)$ vertices from $C_2$ to
$C_1$, but these moves do not increase the union or its contribution
$\alpha|C_1\cup C_2|$ to the potential.

After preparation, testing $|C_1|\geq B$ takes $O(1)$ time. If the test
fails, then $|C_1|<B$. Thus the net increase in $\beta|C_1|$ from the
beginning of the update to the start of regular mode is at most $\beta B$,
regardless of how many vertices moved between $C_1$ and $C_2$ during
preparation. The algorithm initializes the missing high domination edge
lists at vertices in $C_1$. There are fewer than $B$ such vertices, and
each initialization takes $\widetilde O(m^{1/8})$ time without changing
either candidate set. Including the change in $\beta|C_1|$, enumeration
of $C_1$, and all requests gives
\begin{equation}\label{eq:low-list-initialization-cost}
 \widetilde O\bigl(B(m^{1/8}+\beta)\bigr)
 =\widetilde O(m^{5/8}).
\end{equation}
The algorithm makes no such initializations if the test succeeds.
During regular mode $C_1$ can only shrink, so there are no further
initializations of high domination edge lists at low vertices.

The next three lemmas analyze regular-mode extractions. In each case,
we use the following bound for candidate-set changes: a low vertex $z$'s
departure from $C_1$ decreases $\beta|C_1|$ by $\beta$, paying for
deleting its list $A_z^\hi$, the entry for every stored edge $(z,x)$ in $R_x^\hi$, and the list's entry in the indexed collection in $\widetilde O(m^{1/8})$ time by
\Cref{lem:adaptive-high-list-size}. Apart from these deletions,
updating membership in $C_1$ and $C_2$ takes $O(\log n)$ time.

\paragraph{Regular mode: extracting a low vertex.}
We next analyze the amortized runtime of processing a vertex of
$Q\cap V_\lo$.

\begin{lemma}\label{lem:regular-low-cost}
Extracting any $v\in V_\lo$ from $Q$ and completing its regular-mode iteration has nonpositive amortized cost: the actual running time plus the change in $\Phi$ is nonpositive.
\end{lemma}
\begin{proof}
We follow the low-vertex case of the restoration procedure. Every
iteration starts by extracting $v$, which decreases $|Q|$ by one and
pays for the queue operation and its constant-time MIS-membership,
counter, and list tests.

\medskip
\noindent\emph{No change to $\mis_\lo$.}
If $v\in\mis$, $d_\lo(v)+d_\mi(v)>0$, or
$A_v^\hi\neq\emptyset$, the algorithm stops processing this request without changing any candidate set or domination edge list and continues with the next request in $Q$, if any. Thus this
case creates no positive potential change.

\medskip
\noindent\emph{$v$ joins $\mis_\lo$.}
It remains to consider $d_\lo(v)=d_\mi(v)=0$ and
$A_v^\hi=\emptyset$. First, scanning $N(v)$ to find
$X_\hi=N(v)\cap\mis_\hi$ takes $\widetilde O(b_0)$ time. By
\eqref{eq:empty-high-list}, $|X_\hi|=\widetilde O(m^{1/4})$.
Next, remove each $x\in X_\hi$ from $\mis_\hi$. This scans its high degree neighbors and
traverses $R_x^\hi$. Each examined vertex or sampled edge can cause at
most one new request, so removing $x$ from $\mis_\hi$ places at most
$|N(x)\cap V_\hi|+|R_x^\hi|=\widetilde O(m^{3/8})$ vertices in $Q$.
Each new request costs $O(\log n)$ time and increases the queue term
of $\Phi$ by $\Theta(\log n)$; duplicate requests do not increase it.
Thus the total increase in the queue term is at most
\[
 O(\log n)\sum_{x\in X_\hi}
 \bigl(|N(x)\cap V_\hi|+|R_x^\hi|\bigr)
 =\widetilde O(|X_\hi|m^{3/8}).
\]
Including this increase, the queue-operation time, the scans, and the
increase of $W_\hi$ when $x$ leaves the MIS, each removal costs
$O(W_\hi)$.

Finally, including $v$ in $\mis_\lo$ decreases
$W_\lo|V_\lo\setminus\mis|$ by $W_\lo$, and the subsequent scan
of $N(v)$ to increment the exact low
counters takes $\widetilde O(b_0)$ time.
These counter increases and the subsequent candidate removals create no
new requests, since including $v$ in $\mis_\lo$ only increases domination.
Removing $v$ and its low degree neighbors from the candidate union can only
decrease $\alpha|C_1\cup C_2|$, and the separate decrease of
$\beta|C_1|$ pays for each discarded high domination edge list.
No vertex leaves $\mis_\lo\cup\mis_\mi$, so no new candidate enters the union.
Consequently,
\[
 \textnormal{actual time}+\Delta\Phi
 \leq -W_\lo+\widetilde O(b_0)
       +\widetilde O(m^{1/4})\,O(W_\hi)\leq0
\]
for a sufficiently large coefficient and polylogarithmic factor in
$W_\lo=\widetilde\Theta(m^{5/8})$.
\end{proof}

\paragraph{Regular mode: extracting a middle vertex.}
We now analyze the regular-mode iterations in which the extracted vertex
belongs to $V_\mi$.

\begin{lemma}\label{lem:regular-middle-cost}
Extracting any $v\in V_\mi$ from $Q$ and completing its regular-mode iteration has nonpositive amortized cost.
\end{lemma}
\begin{proof}
We follow the middle-vertex case of the restoration procedure. Every
iteration starts by extracting $v$, which decreases $|Q|$ by one and
pays for the queue operation.

\medskip
\noindent\emph{No change to $\mis_\mi$.}
The MIS-membership, counter, and list tests take $\widetilde O(1)$ time.
If $v\in\mis$, $d_\lo(v)+d_\mi(v)>0$, or $A_v^\hi\neq\emptyset$,
the algorithm stops processing this request without changing any candidate set or domination edge list and continues with the next request in $Q$, if any.
The decrease in $|Q|$ pays for all tests.

\medskip
\noindent\emph{$v$ joins $\mis_\mi$.}
We have $d_\lo(v)+d_\mi(v)=0$ and $A_v^\hi=\emptyset$.
First, scanning $N(v)$ to find $X_\hi=N(v)\cap\mis_\hi$ takes
$\widetilde O(b_1)$ time. Next, by \eqref{eq:empty-high-list}, the
actual runtime of removing every vertex of $X_\hi$ from $\mis_\hi$, scanning its vertices' high degree
neighbors, and deleting their reverse lists, plus the
resulting potential increase, is
\[
 O(|X_\hi|W_\hi)=\widetilde O(m^{1/4})\,O(W_\hi).
\]
This includes the queue-operation time and the potential increase from
every new request, as calculated in \Cref{lem:regular-low-cost}.
Finally, including $v$ in $\mis_\mi$ decreases
$W_\mi|V_\mi\setminus\mis|$ by $W_\mi$.
The algorithm then scans every neighbor $z\in N(v)$, including low degree
neighbors, and increments $d_\mi(z)$. This takes
$\widetilde O(b_1)$ time. For every low degree neighbor moved from $C_1$ to
$C_2$, the candidate union is unchanged and $\beta|C_1|$ decreases by
$\beta$. This decrease pays for deleting $A_z^\hi$ and the entry for every stored edge $(z,x)$ in $R_x^\hi$.
Choosing a sufficiently large coefficient and polylogarithmic factor in
$W_\mi$ gives
\[
 \textnormal{actual time}+\Delta\Phi
 \leq -W_\mi+\widetilde O(b_1)
       +\widetilde O(m^{1/4})\,O(W_\hi)\leq0.
\]
\end{proof}

\paragraph{Regular mode: extracting a high vertex.}
We finally analyze the regular-mode iterations in which the extracted
vertex belongs to $V_\hi$.

\begin{lemma}\label{lem:regular-high-cost}
Extracting any $v\in V_\hi$ from $Q$ and completing its regular-mode iteration has nonpositive amortized cost.
\end{lemma}
\begin{proof}
We again follow the cases in the restoration procedure. Every iteration
starts by extracting $v$, which decreases $|Q|$ by one and pays for the
queue operation and its constant-time MIS-membership and counter tests.

\medskip
\noindent\emph{Vertex $v$ is already in the MIS or dominated.}
The algorithm checks the membership bit and the three exact counters.
If $v\in\mis$ or $d_\lo(v)+d_\mi(v)+d_\hi(v)>0$, it performs no
adjacency scan and changes no counter, candidate set, MIS layer, or
domination edge list. The only potential change is the decrease of
$\Theta(\log n)|Q|$, which pays for the iteration.

\medskip
\noindent\emph{$v$ joins $\mis_\hi$.}
Otherwise all three exact counters equal zero, so $v$ can be included in
$\mis_\hi$. This decreases the potential by $W_\hi$. Scanning its
high degree neighbors and updating their counters take
$\widetilde O(m^{3/8})$ time. By
\Cref{lem:adaptive-reverse-list-size}, initializing $R_v^\hi$ takes
$\widetilde O(m^{3/8})$ time as well. No vertex is removed from the MIS, and
neither candidate term increases. With a sufficiently large coefficient
in $W_\hi$, its decrease pays for the entire iteration.
\end{proof}

Summing these three lemmas shows that regular queue processing has
nonpositive amortized cost. Together with \Cref{lem:preparation-cost}
and \eqref{eq:low-list-initialization-cost}, this bounds the entire
update, including the change in $\Phi$, by $\widetilde O(m^{5/8})$
whenever the update uses regular mode.

\paragraph{Flush mode.}
We first bound the decrease in the size of the candidate union. We then follow the
flush steps in order: copy $C_1$ and clear $Q$ and $\mis_\hi$,
process the low and middle vertices, and reconstruct $\mis_\hi$.

\begin{lemma}\label{lem:flush-candidate-decrease}
Let $C_1^0$ be the set $C_1$ at the start of a flush. Every vertex of $C_1^0$ leaves
$C_1\cup C_2$ during its low pass, and no new vertex enters this union
during the flush. Its size decreases by at least $|C_1^0|\geq B$.
\end{lemma}
\begin{proof}
Clearing $Q$ and $\mis_\hi$ changes neither candidate set.
Every vertex $v\in C_1^0$ has $d_\lo(v)=d_\mi(v)=0$ at the start of
the flush. The middle MIS remains unchanged during the low pass, so each vertex either already
has a low-MIS neighbor when visited or is included in the low MIS.
Both outcomes remove it from the candidate union. There are no class
changes or removals from $\mis_\lo\cup\mis_\mi$ during the flush, so these departures
persist and no new low vertex can lose its last low-MIS neighbor. The
later middle and high passes do not change this conclusion.
\end{proof}

\paragraph{Flush mode: copying $C_1$ and clearing $Q$ and $\mis_\hi$.}
\begin{lemma}\label{lem:flush-clear-time}
For Step~\ref{step:flush-clear}, the actual running time plus the change
in $\Phi$ is at most
\[
 \widetilde O(|C_1^0|)+O(|V_\hi|W_\hi)
 =\widetilde O(|C_1^0|+m^{3/4}),
\]
and the step does not change $C_1\cup C_2$.
\end{lemma}
\begin{proof}
Making the copy $C_1^0$ of $C_1$ takes
$\widetilde O(|C_1^0|)$ time. Clearing $Q$ is paid for by the decrease
in $\Theta(\log n)|Q|$.

The algorithm then removes every vertex of $\mis_\hi$ from the MIS. The degree
bound for $V_\hi$ gives $|V_\hi|=O(m^{3/8})$. Removing these vertices from $\mis_\hi$
increases $W_\hi|V_\hi\setminus\mis|$ by at most
$|V_\hi|W_\hi$. Removing each vertex $x$ from $\mis_\hi$ requires scanning its high degree neighbors
and traversing its reverse high domination edge list in
$\widetilde O(m^{3/8})$ time. Thus both stored entries of every sampled edge are removed from the domination edge lists
within $O(|V_\hi|W_\hi)$ time. Finally, discarding the empty high domination
edge lists at low vertices and their entries in the indexed collection takes
$\widetilde O(|C_1^0|)$ time, since each such list belongs to a vertex
of $C_1^0$. For every vertex $z$, this step
changes neither $d_\lo(z)$ nor $d_\mi(z)$ nor candidate membership.
\end{proof}

\paragraph{Flush mode: processing low and middle vertices.}
Every existing member of $\mis_\lo\cup\mis_\mi$ remains in the MIS during the low and middle passes.

\begin{lemma}\label{lem:flush-lower-time}
After Step~\ref{step:flush-clear}, the low and middle passes have amortized cost at most the following quantity, including their changes in $\Phi$:
\[
 -(\alpha+\beta)|C_1^0|+\widetilde O(|C_1^0|+m^{5/8}).
\]
\end{lemma}
\begin{proof}
At the start of these passes, the high MIS is empty and no high domination
edge list is stored at a low vertex. First, traversing $C_1^0$ in
Step~\ref{step:flush-low} takes $\widetilde O(|C_1^0|)$ time.
For each vertex that joins $\mis_\lo$, the algorithm scans $O(b_0)$ neighbors and decreases
$W_\lo|V_\lo\setminus\mis|$ by $W_\lo$. This decrease pays for
the scan and exact counter updates. Next, traversing $V_\mi$ in
Step~\ref{step:flush-middle} takes $\widetilde O(m^{5/8})$ time.
For each vertex that joins $\mis_\mi$, the algorithm scans $O(b_1)$ neighbors and decreases
$W_\mi|V_\mi\setminus\mis|$ by $W_\mi$. This decrease again pays
for the scan and counter updates. A vertex already in the MIS or dominated requires only
the constant-time tests included in these traversal costs. No vertex of
$\mis_\lo\cup\mis_\mi$ is removed during either pass.

By \Cref{lem:flush-candidate-decrease}, every vertex of $C_1^0$ leaves
$C_1\cup C_2$, and no new vertex enters the union. Thus
$\alpha|C_1\cup C_2|$ decreases by at least $\alpha|C_1^0|$.
Moreover, $C_1$ becomes empty, so $\beta|C_1|$ decreases by
$\beta|C_1^0|$. Combining these decreases with the traversal costs
proves the lemma.
\end{proof}

\paragraph{Flush mode: reconstructing the high layer.}
After the low and middle passes, $Q=C_1=\emptyset$. The degree classes
and every list $\mathcal E_z^\hi$ were already updated during preparation,
so the flush does not rebuild these adjacency lists.

\begin{lemma}\label{lem:flush-high-reconstruction}
For Step~\ref{step:flush-high}, the actual running time plus the change
in $\Phi$ is at most
\[
 O(|V_\hi|W_\hi)=\widetilde O(m^{3/4}).
\]
These operations do not change $C_1\cup C_2$.
\end{lemma}
\begin{proof}
After the low and middle passes, the algorithm scans
$\mathcal E_z^\hi$ for high vertices $z$ to construct a greedy MIS of
\[
 G[V_\hi\setminus N(\mis_\lo\cup\mis_\mi)].
\]
The same pass updates $d_\hi(w)$ for every high vertex $w$. Its runtime is
\[
 \widetilde O(|V_\hi|+|V_\hi|^2)
 =\widetilde O(m^{3/4}).
\]
Each vertex included in the rebuilt $\mis_\hi$ decreases
$W_\hi|V_\hi\setminus\mis|$ by $W_\hi$, so omitting these decreases only makes the
upper bound larger.

For every vertex $x$ selected for the rebuilt $\mis_\hi$, apply
\Cref{lem:output-sensitive-bernoulli} to $V_\mi$ with probability
$p_\hi$ and test adjacency for every returned vertex. By
\Cref{lem:adaptive-reverse-list-size}, this takes
$\widetilde O(m^{3/8})$ time per vertex. For each returned $z$ adjacent
to $x$, store an entry for $(z,x)$ in both $R_x^\hi$ and $A_z^\hi$.
Thus reconstruction of the sampled lists also takes
$\widetilde O(|V_\hi|m^{3/8})\subseteq O(|V_\hi|W_\hi)$ time.
Since $C_1=\emptyset$, no high domination edge list at a low vertex is
initialized. No scan of all graph edges is required.
\end{proof}

\begin{lemma}\label{lem:flush-cost}
For a flush, the actual running time plus the change in $\Phi$ is at most $-\beta|C_1^0|$.
\end{lemma}
\begin{proof}
The operations in
\Cref{lem:flush-clear-time,lem:flush-lower-time,lem:flush-high-reconstruction}
account for the entire flush in execution order. Adding their bounds gives
\[
 \textnormal{actual time}+\Delta\Phi
 \leq-(\alpha+\beta)|C_1^0|+\widetilde O(|C_1^0|+m^{3/4}).
\]
The flush starts only when $|C_1^0|\geq B$, and
\[
 \alpha B=\widetilde\Theta(m^{1/4+1/2})
          =\widetilde\Theta(m^{3/4}).
\]
A sufficiently large coefficient and polylogarithmic factor in $\alpha$
therefore make the displayed amortized cost at most $-\beta|C_1^0|$.
\end{proof}

The increase in $\beta|C_1|$ during preparation is at most
$\beta|C_1^0|$, since $C_1=C_1^0$ when the flush starts.
\Cref{lem:flush-cost} cancels this increase. Together with
\Cref{lem:preparation-cost}, it follows that, for an update using flush mode,
the actual running time plus the change in $\Phi$ is at most $\widetilde O(m^{5/8})$.

\paragraph{Coefficients of the potential.}
First choose a sufficiently large constant in the queue term
$\Theta(\log n)|Q|$. Then choose the constants and logarithmic factors
successively for $\beta$, $W_\hi$, $\alpha$, and $W_\lo=W_\mi$.
The queue coefficient pays for extractions and tests of MIS membership
and domination; $\beta$
pays for deleting a high domination edge list at a low vertex during regular mode; and
$W_\hi$ pays for including a vertex in the high MIS. The choice of $\alpha$ then pays for a flush, and
$W_\lo$ and $W_\mi$ pay for neighborhood scans and for removing high-MIS neighbors from $\mis_\hi$ before vertices join $\mis_\lo$ and $\mis_\mi$, respectively.
All these choices have the stated exponents.
Any additional polylogarithmic sampling cost can be absorbed by a common
factor in these weights.

\begin{proof}[Proof of \Cref{thm:main}]
Correctness follows from \Cref{lem:correctness}.
By \Cref{lem:adaptive-reverse-list-size,lem:adaptive-high-list-size},
the sampling bounds used in the running-time analysis hold with high
probability against adaptive adversaries. Combining
\Cref{lem:preparation-cost,lem:regular-low-cost,lem:regular-middle-cost,lem:regular-high-cost,lem:flush-cost} gives
$\widetilde O(m^{5/8})$ amortized update time.
\end{proof}

\clearpage

\section*{Acknowledgment}
This research is supported by the Fundamental and Interdisciplinary Disciplines Breakthrough Plan of the Ministry of Education of China (No. JYB2025XDXM118) and the ``111 Center'' (No. B26023).

\subsection*{Disclosure of AI Use}
\begin{flushleft}
\renewcommand{\arraystretch}{1.3}
\begin{tabular}{@{}l@{\qquad}c@{}}
\textbf{Activity} & \textbf{AI used} \\
Asking the research question & No \\
Coming up with the main approach & No \\
Developing the main proof ideas & No \\
Writing and exposition & Yes \\
Checking for bugs & Yes \\
Other supporting tasks & Yes \\
\end{tabular}
\end{flushleft}

\begin{sloppypar}
\paragraph{Details (elaborating on the points above).}
\begin{enumerate}
\item \emph{Asking the research question.}
The authors formulated the research question and set the objective of improving the update time for fully dynamic MIS against adaptive adversaries.
\item \emph{Coming up with the main approach.}
The authors developed the main algorithmic ideas, including the layered
candidate construction and adaptive sampling framework.
\item \emph{Developing the main proof ideas.}
The authors developed the main proof ideas, including the probabilistic
games for adaptivity analysis and the potential-based amortized analysis.
GPT assisted in refining and checking the details of these arguments.
\item \emph{Writing and exposition.}
GPT assisted with drafting and editing the manuscript, including the
overview, detailed algorithm, proofs, and figures.
\item \emph{Checking for bugs.}
GPT assisted in discovering bugs in earlier versions of the adaptivity analysis.
\item \emph{Other supporting tasks.}
The authors initially developed an algorithm with amortized update time
\(\widetilde O(m^{2/3-1/54})\). OpenAI GPT subsequently identified that many
subroutines were unnecessary. Removing these subroutines substantially
simplified the algorithm and yielded the current
\(\widetilde O(m^{5/8})\) amortized update bound.
\end{enumerate}
\end{sloppypar}

\bibliographystyle{alpha}
\bibliography{references}

@inproceedings{DBLP:conf/stoc/AssadiOSS18,
  author       = {Sepehr Assadi and Krzysztof Onak and Baruch Schieber and Shay Solomon},
  title        = {Fully Dynamic Maximal Independent Set with Sublinear Update Time},
  booktitle    = {Proceedings of the 50th Annual {ACM} {SIGACT} Symposium on Theory of Computing ({STOC} 2018)},
  pages        = {815--826},
  publisher    = {{ACM}},
  year         = {2018},
  url          = {https://doi.org/10.1145/3188745.3188922},
  doi          = {10.1145/3188745.3188922},
  bibsource    = {dblp computer science bibliography, https://dblp.org}
}

@inproceedings{DBLP:conf/stoc/BeimelKMNSS22,
  author       = {Amos Beimel and Haim Kaplan and Yishay Mansour and Kobbi Nissim and Thatchaphol Saranurak and Uri Stemmer},
  title        = {Dynamic Algorithms against an Adaptive Adversary: Generic Constructions and Lower Bounds},
  booktitle    = {Proceedings of the 54th Annual {ACM} {SIGACT} Symposium on Theory of Computing ({STOC} 2022)},
  pages        = {1671--1684},
  publisher    = {{ACM}},
  year         = {2022},
  url          = {https://doi.org/10.1145/3519935.3520064},
  doi          = {10.1145/3519935.3520064},
  eprinttype   = {arXiv},
  eprint       = {2111.03980},
  bibsource    = {dblp computer science bibliography, https://dblp.org}
}

@misc{DBLP:journals/corr/abs-1804-01823,
  author       = {Manoj Gupta and Shahbaz Khan},
  title        = {Simple Dynamic Algorithms for Maximal Independent Set and Other Problems},
  year         = {2018},
  url          = {https://arxiv.org/abs/1804.01823},
  eprinttype   = {arXiv},
  eprint       = {1804.01823},
  howpublished = {arXiv:1804.01823},
}

@inproceedings{DBLP:conf/sosa/GuptaK21,
  author       = {Manoj Gupta and Shahbaz Khan},
  title        = {Simple Dynamic Algorithms for Maximal Independent Set, Maximum Flow and Maximum Matching},
  booktitle    = {Proceedings of the 4th Symposium on Simplicity in Algorithms ({SOSA} 2021)},
  pages        = {86--91},
  publisher    = {{SIAM}},
  year         = {2021},
  url          = {https://doi.org/10.1137/1.9781611976496.10},
  doi          = {10.1137/1.9781611976496.10},
  bibsource    = {dblp computer science bibliography, https://dblp.org}
}

@misc{DBLP:journals/corr/abs-1804-08908,
  author       = {Yuhao Du and Hengjie Zhang},
  title        = {Improved Algorithms for Fully Dynamic Maximal Independent Set},
  year         = {2018},
  url          = {https://arxiv.org/abs/1804.08908},
  eprinttype   = {arXiv},
  eprint       = {1804.08908},
  bibsource    = {dblp computer science bibliography, https://dblp.org},
  howpublished = {arXiv:1804.08908},
}

@inproceedings{DBLP:conf/soda/AssadiOSS19,
  author       = {Sepehr Assadi and Krzysztof Onak and Baruch Schieber and Shay Solomon},
  title        = {Fully Dynamic Maximal Independent Set with Sublinear in $n$ Update Time},
  booktitle    = {Proceedings of the 30th Annual {ACM-SIAM} Symposium on Discrete Algorithms ({SODA} 2019)},
  pages        = {1919--1936},
  publisher    = {{SIAM}},
  year         = {2019},
  url          = {https://doi.org/10.1137/1.9781611975482.116},
  doi          = {10.1137/1.9781611975482.116},
  bibsource    = {dblp computer science bibliography, https://dblp.org}
}

@inproceedings{DBLP:conf/focs/BehnezhadDHSS19,
  author       = {Soheil Behnezhad and Mahsa Derakhshan and MohammadTaghi Hajiaghayi and Cliff Stein and Madhu Sudan},
  title        = {Fully Dynamic Maximal Independent Set with Polylogarithmic Update Time},
  booktitle    = {Proceedings of the 60th Annual {IEEE} Symposium on Foundations of Computer Science ({FOCS} 2019)},
  pages        = {382--405},
  publisher    = {{IEEE} Computer Society},
  year         = {2019},
  url          = {https://doi.org/10.1109/FOCS.2019.00032},
  doi          = {10.1109/FOCS.2019.00032},
  bibsource    = {dblp computer science bibliography, https://dblp.org}
}

@inproceedings{DBLP:conf/focs/ChechikZ19,
  author       = {Shiri Chechik and Tianyi Zhang},
  title        = {Fully Dynamic Maximal Independent Set in Expected Poly-Log Update Time},
  booktitle    = {Proceedings of the 60th Annual {IEEE} Symposium on Foundations of Computer Science ({FOCS} 2019)},
  pages        = {370--381},
  publisher    = {{IEEE} Computer Society},
  year         = {2019},
  url          = {https://doi.org/10.1109/FOCS.2019.00031},
  doi          = {10.1109/FOCS.2019.00031},
  bibsource    = {dblp computer science bibliography, https://dblp.org}
}

@article{DBLP:journals/talg/OnakSSW20,
  author       = {Krzysztof Onak and Baruch Schieber and Shay Solomon and Nicole Wein},
  title        = {Fully Dynamic {MIS} in Uniformly Sparse Graphs},
  journal      = {{ACM} Transactions on Algorithms},
  volume       = {16},
  number       = {2},
  pages        = {26:1--26:19},
  year         = {2020},
  url          = {https://doi.org/10.1145/3378025},
  doi          = {10.1145/3378025},
  bibsource    = {dblp computer science bibliography, https://dblp.org}
}

@inproceedings{DBLP:conf/soda/BernsteinBFKS26,
  author       = {Aaron Bernstein and Sayan Bhattacharya and Nick Fischer and Peter Kiss and Thatchaphol Saranurak},
  title        = {Separations between Oblivious and Adaptive Adversaries for Natural Dynamic Graph Problems},
  booktitle    = {Proceedings of the 37th Annual {ACM-SIAM} Symposium on Discrete Algorithms ({SODA} 2026)},
  pages        = {5669--5687},
  publisher    = {{SIAM}},
  year         = {2026},
  url          = {https://doi.org/10.1137/1.9781611978971.202},
  doi          = {10.1137/1.9781611978971.202},
  bibsource    = {dblp computer science bibliography, https://dblp.org}
}

@inproceedings{DBLP:conf/swat/GuptaKS26,
  author       = {Manoj Gupta and Shahbaz Khan and Madhu Surendra},
  title        = {Dynamic {MIS} Revisited: Incremental, Fault Tolerant and Fully Dynamic},
  booktitle    = {Proceedings of the 20th Scandinavian Symposium on Algorithm Theory ({SWAT} 2026)},
  series       = {LIPIcs},
  volume       = {370},
  pages        = {21:1--21:11},
  publisher    = {Schloss Dagstuhl -- Leibniz-Zentrum f{\"{u}}r Informatik},
  year         = {2026},
  url          = {https://doi.org/10.4230/LIPIcs.SWAT.2026.21},
  doi          = {10.4230/LIPIcs.SWAT.2026.21},
  bibsource    = {dblp computer science bibliography, https://dblp.org}
}

@inproceedings{DBLP:conf/stoc/HolmLT98,
  author       = {Jacob Holm and Kristian de Lichtenberg and Mikkel Thorup},
  title        = {Poly-Logarithmic Deterministic Fully-Dynamic Algorithms for Connectivity, Minimum Spanning Tree, 2-Edge, and Biconnectivity},
  booktitle    = {Proceedings of the 30th Annual {ACM} {SIGACT} Symposium on Theory of Computing ({STOC} 1998)},
  pages        = {79--89},
  publisher    = {{ACM}},
  year         = {1998},
  url          = {https://doi.org/10.1145/276698.276715},
  doi          = {10.1145/276698.276715},
  bibsource    = {dblp computer science bibliography, https://dblp.org}
}

@inproceedings{DBLP:conf/focs/Sankowski04,
  author       = {Piotr Sankowski},
  title        = {Dynamic Transitive Closure via Dynamic Matrix Inverse (Extended Abstract)},
  booktitle    = {Proceedings of the 45th Annual {IEEE} Symposium on Foundations of Computer Science ({FOCS} 2004)},
  pages        = {509--517},
  publisher    = {{IEEE} Computer Society},
  year         = {2004},
  url          = {https://doi.org/10.1109/FOCS.2004.25},
  doi          = {10.1109/FOCS.2004.25},
  bibsource    = {dblp computer science bibliography, https://dblp.org}
}

@article{DBLP:journals/jacm/DemetrescuI04,
  author       = {Camil Demetrescu and Giuseppe F. Italiano},
  title        = {A New Approach to Dynamic All Pairs Shortest Paths},
  journal      = {Journal of the {ACM}},
  volume       = {51},
  number       = {6},
  pages        = {968--992},
  year         = {2004},
  url          = {https://doi.org/10.1145/1039488.1039492},
  doi          = {10.1145/1039488.1039492},
  bibsource    = {dblp computer science bibliography, https://dblp.org}
}

@inproceedings{DBLP:conf/focs/BaswanaGS11,
  author       = {Surender Baswana and Manoj Gupta and Sandeep Sen},
  title        = {Fully Dynamic Maximal Matching in {$O(\log n)$} Update Time},
  booktitle    = {Proceedings of the 52nd Annual {IEEE} Symposium on Foundations of Computer Science ({FOCS} 2011)},
  pages        = {383--392},
  publisher    = {{IEEE} Computer Society},
  year         = {2011},
  url          = {https://doi.org/10.1109/FOCS.2011.89},
  doi          = {10.1109/FOCS.2011.89},
  bibsource    = {dblp computer science bibliography, https://dblp.org}
}

@inproceedings{DBLP:conf/focs/NanongkaiSW17,
  author       = {Danupon Nanongkai and Thatchaphol Saranurak and Christian Wulff{-}Nilsen},
  title        = {Dynamic Minimum Spanning Forest with Subpolynomial Worst-Case Update Time},
  booktitle    = {Proceedings of the 58th Annual {IEEE} Symposium on Foundations of Computer Science ({FOCS} 2017)},
  pages        = {950--961},
  publisher    = {{IEEE} Computer Society},
  year         = {2017},
  url          = {https://doi.org/10.1109/FOCS.2017.92},
  doi          = {10.1109/FOCS.2017.92},
  bibsource    = {dblp computer science bibliography, https://dblp.org}
}

@inproceedings{DBLP:conf/stoc/ChuzhoyZ23,
  author       = {Julia Chuzhoy and Ruimin Zhang},
  title        = {A New Deterministic Algorithm for Fully Dynamic All-Pairs Shortest Paths},
  booktitle    = {Proceedings of the 55th Annual {ACM} {SIGACT} Symposium on Theory of Computing ({STOC} 2023)},
  pages        = {1159--1172},
  publisher    = {{ACM}},
  year         = {2023},
  url          = {https://doi.org/10.1145/3564246.3585196},
  doi          = {10.1145/3564246.3585196},
  bibsource    = {dblp computer science bibliography, https://dblp.org}
}

@inproceedings{DBLP:conf/focs/Liu24,
  author       = {Yang P. Liu},
  title        = {On Approximate Fully-Dynamic Matching and Online Matrix-Vector Multiplication},
  booktitle    = {Proceedings of the 65th Annual {IEEE} Symposium on Foundations of Computer Science ({FOCS} 2024)},
  pages        = {228--243},
  publisher    = {{IEEE} Computer Society},
  year         = {2024},
  url          = {https://doi.org/10.1109/FOCS61266.2024.00006},
  doi          = {10.1109/FOCS61266.2024.00006},
  bibsource    = {dblp computer science bibliography, https://dblp.org}
}

@inproceedings{DBLP:conf/stoc/BrandLS23,
  author       = {Jan van den Brand and Yang P. Liu and Aaron Sidford},
  title        = {Dynamic Maxflow via Dynamic Interior Point Methods},
  booktitle    = {Proceedings of the 55th Annual {ACM} {SIGACT} Symposium on Theory of Computing ({STOC} 2023)},
  pages        = {1215--1228},
  publisher    = {{ACM}},
  year         = {2023},
  url          = {https://doi.org/10.1145/3564246.3585135},
  doi          = {10.1145/3564246.3585135},
  bibsource    = {dblp computer science bibliography, https://dblp.org}
}

@inproceedings{DBLP:conf/focs/0001S21,
  author       = {Wenyu Jin and Xiaorui Sun},
  title        = {Fully Dynamic $s$-$t$ Edge Connectivity in Subpolynomial Time (Extended Abstract)},
  booktitle    = {Proceedings of the 62nd Annual {IEEE} Symposium on Foundations of Computer Science ({FOCS} 2021)},
  pages        = {861--872},
  publisher    = {{IEEE}},
  year         = {2021},
  url          = {https://doi.org/10.1109/FOCS52979.2021.00088},
  doi          = {10.1109/FOCS52979.2021.00088},
  bibsource    = {dblp computer science bibliography, https://dblp.org}
}

@inproceedings{DBLP:conf/stoc/HolmNR025,
  author       = {Jacob Holm and Wojciech Nadara and Eva Rotenberg and Marek Soko{\l}owski},
  title        = {Fully Dynamic Biconnectivity in {$\widetilde O(\log^2 n)$} Time},
  booktitle    = {Proceedings of the 57th Annual {ACM} {SIGACT} Symposium on Theory of Computing ({STOC} 2025)},
  pages        = {156--165},
  publisher    = {{ACM}},
  year         = {2025},
  url          = {https://doi.org/10.1145/3717823.3718270},
  doi          = {10.1145/3717823.3718270},
  bibsource    = {dblp computer science bibliography, https://dblp.org}
}

@article{DBLP:journals/iandc/HenzingerP21,
  author       = {Monika Henzinger and Pan Peng},
  title        = {Constant-Time Dynamic Weight Approximation for Minimum Spanning Forest},
  journal      = {Information and Computation},
  volume       = {281},
  pages        = {104805},
  year         = {2021},
  url          = {https://doi.org/10.1016/j.ic.2021.104805},
  doi          = {10.1016/j.ic.2021.104805},
  bibsource    = {dblp computer science bibliography, https://dblp.org}
}

@inproceedings{DBLP:conf/wea/HanauerH020,
  author       = {Kathrin Hanauer and Monika Henzinger and Christian Schulz},
  title        = {Faster Fully Dynamic Transitive Closure in Practice},
  booktitle    = {Proceedings of the 18th International Symposium on Experimental Algorithms ({SEA} 2020)},
  series       = {LIPIcs},
  volume       = {160},
  pages        = {14:1--14:14},
  publisher    = {Schloss Dagstuhl -- Leibniz-Zentrum f{\"{u}}r Informatik},
  year         = {2020},
  url          = {https://doi.org/10.4230/LIPIcs.SEA.2020.14},
  doi          = {10.4230/LIPIcs.SEA.2020.14},
  bibsource    = {dblp computer science bibliography, https://dblp.org}
}

@inproceedings{DBLP:conf/focs/BrandK23,
  author       = {Jan van den Brand and Adam Karczmarz},
  title        = {Deterministic Fully Dynamic {SSSP} and More},
  booktitle    = {Proceedings of the 64th Annual {IEEE} Symposium on Foundations of Computer Science ({FOCS} 2023)},
  pages        = {2312--2321},
  publisher    = {{IEEE}},
  year         = {2023},
  url          = {https://doi.org/10.1109/FOCS57990.2023.00142},
  doi          = {10.1109/FOCS57990.2023.00142},
  bibsource    = {dblp computer science bibliography, https://dblp.org}
}

@inproceedings{DBLP:conf/stoc/Wajc20,
  author       = {David Wajc},
  title        = {Rounding Dynamic Matchings Against an Adaptive Adversary},
  booktitle    = {Proceedings of the 52nd Annual {ACM} {SIGACT} Symposium on Theory of Computing ({STOC} 2020)},
  pages        = {194--207},
  publisher    = {{ACM}},
  year         = {2020},
  url          = {https://doi.org/10.1145/3357713.3384258},
  doi          = {10.1145/3357713.3384258},
  bibsource    = {dblp computer science bibliography, https://dblp.org}
}

@inproceedings{DBLP:conf/focs/BehnezhadG24,
  author       = {Soheil Behnezhad and Alma Ghafari},
  title        = {Fully Dynamic Matching and Ordered {Ruzsa-Szemer{\'{e}}di} Graphs},
  booktitle    = {Proceedings of the 65th Annual {IEEE} Symposium on Foundations of Computer Science ({FOCS} 2024)},
  pages        = {314--327},
  publisher    = {{IEEE} Computer Society},
  year         = {2024},
  url          = {https://doi.org/10.1109/FOCS61266.2024.00027},
  doi          = {10.1109/FOCS61266.2024.00027},
  bibsource    = {dblp computer science bibliography, https://dblp.org}
}

@inproceedings{DBLP:conf/focs/GaoLP21,
  author       = {Yu Gao and Yang P. Liu and Richard Peng},
  title        = {Fully Dynamic Electrical Flows: Sparse Maxflow Faster Than {Goldberg--Rao}},
  booktitle    = {Proceedings of the 62nd Annual {IEEE} Symposium on Foundations of Computer Science ({FOCS} 2021)},
  pages        = {516--527},
  publisher    = {{IEEE}},
  year         = {2021},
  bibsource    = {dblp computer science bibliography, https://dblp.org}
}

@inproceedings{DBLP:conf/focs/Solomon16,
  author       = {Shay Solomon},
  title        = {Fully Dynamic Maximal Matching in Constant Update Time},
  booktitle    = {Proceedings of the 57th Annual {IEEE} Symposium on Foundations of Computer Science ({FOCS} 2016)},
  pages        = {325--334},
  publisher    = {{IEEE} Computer Society},
  year         = {2016},
  url          = {https://doi.org/10.1109/FOCS.2016.43},
  doi          = {10.1109/FOCS.2016.43}
}

@inproceedings{DBLP:conf/spaa/0001T24,
  author       = {Mohsen Ghaffari and Anton Trygub},
  title        = {Parallel Dynamic Maximal Matching},
  booktitle    = {Proceedings of the 36th {ACM} Symposium on Parallelism in Algorithms and Architectures ({SPAA} 2024)},
  pages        = {427--437},
  publisher    = {{ACM}},
  year         = {2024},
  url          = {https://doi.org/10.1145/3626183.3659982},
  doi          = {10.1145/3626183.3659982}
}

@inproceedings{DBLP:conf/stoc/NeimanS13,
  author       = {Ofer Neiman and Shay Solomon},
  title        = {Simple Deterministic Algorithms for Fully Dynamic Maximal Matching},
  booktitle    = {Proceedings of the 45th Annual {ACM} {SIGACT} Symposium on Theory of Computing ({STOC} 2013)},
  publisher    = {ACM},
  year         = {2013},
  doi          = {10.1145/2488608.2488703},
  url          = {https://doi.org/10.1145/2488608.2488703},
  pages        = {745--754},
}

@inproceedings{DBLP:conf/stoc/BernsteinBKS25,
  author       = {Aaron Bernstein and Sayan Bhattacharya and Peter Kiss and Thatchaphol Saranurak},
  title        = {Deterministic Dynamic Maximal Matching in Sublinear Update Time},
  booktitle    = {Proceedings of the 57th Annual {ACM} {SIGACT} Symposium on Theory of Computing ({STOC} 2025)},
  pages        = {132--143},
  publisher    = {{ACM}},
  year         = {2025},
  url          = {https://doi.org/10.1145/3717823.3718153},
  doi          = {10.1145/3717823.3718153}
}

@inproceedings{DBLP:conf/stoc/ChuzhoyKS26,
  author       = {Julia Chuzhoy and Sanjeev Khanna and Junkai Song},
  title        = {A Faster Deterministic Algorithm for Fully Dynamic Maximal Matching},
  booktitle    = {Proceedings of the 58th Annual {ACM} {SIGACT} Symposium on Theory of Computing ({STOC} 2026)},
  pages        = {1604--1615},
  publisher    = {{ACM}},
  year         = {2026},
  url          = {https://doi.org/10.1145/3798129.3800868},
  doi          = {10.1145/3798129.3800868}
}

@article{DBLP:journals/talg/BhattacharyaGKL22,
  author       = {Sayan Bhattacharya and Fabrizio Grandoni and Janardhan Kulkarni and Quanquan C. Liu and Shay Solomon},
  title        = {Fully Dynamic {$(\Delta+1)$}-Coloring in {$O(1)$} Update Time},
  journal      = {{ACM} Transactions on Algorithms},
  volume       = {18},
  number       = {2},
  pages        = {10:1--10:25},
  year         = {2022},
  url          = {https://doi.org/10.1145/3494539},
  doi          = {10.1145/3494539}
}

@article{DBLP:journals/talg/HenzingerP22,
  author       = {Monika Henzinger and Pan Peng},
  title        = {Constant-Time Dynamic {$(\Delta+1)$}-Coloring},
  journal      = {{ACM} Transactions on Algorithms},
  volume       = {18},
  number       = {2},
  pages        = {16:1--16:21},
  year         = {2022},
  url          = {https://doi.org/10.1145/3501403},
  doi          = {10.1145/3501403}
}

@inproceedings{DBLP:conf/stoc/0001G24,
  author       = {Mohsen Ghaffari and Christoph Grunau},
  title        = {Dynamic {O(Arboricity)} Coloring in Polylogarithmic Worst-Case Time},
  booktitle    = {Proceedings of the 56th Annual {ACM} {SIGACT} Symposium on Theory of Computing ({STOC} 2024)},
  pages        = {1184--1191},
  publisher    = {{ACM}},
  year         = {2024},
  url          = {https://doi.org/10.1145/3618260.3649782},
  doi          = {10.1145/3618260.3649782}
}

@inproceedings{DBLP:conf/soda/BehnezhadRW25,
  author       = {Soheil Behnezhad and Rajmohan Rajaraman and Omer Wasim},
  title        = {Fully Dynamic {$(\Delta+1)$}-Coloring Against Adaptive Adversaries},
  booktitle    = {Proceedings of the 36th Annual {ACM-SIAM} Symposium on Discrete Algorithms ({SODA} 2025)},
  pages        = {4983--5026},
  publisher    = {{SIAM}},
  year         = {2025},
  url          = {https://doi.org/10.1137/1.9781611978322.169},
  doi          = {10.1137/1.9781611978322.169}
}

@inproceedings{DBLP:conf/icalp/FlinH25,
  author       = {Maxime Flin and Magn{\'u}s M. Halld{\'o}rsson},
  title        = {Faster Dynamic {$(\Delta+1)$}-Coloring Against Adaptive Adversaries},
  booktitle    = {Proceedings of the 52nd International Colloquium on Automata, Languages, and Programming ({ICALP} 2025)},
  series       = {LIPIcs},
  volume       = {334},
  pages        = {79:1--79:21},
  publisher    = {Schloss Dagstuhl -- Leibniz-Zentrum f{\"u}r Informatik},
  year         = {2025},
  url          = {https://doi.org/10.4230/LIPIcs.ICALP.2025.79},
  doi          = {10.4230/LIPIcs.ICALP.2025.79}
}

@inproceedings{GilHalldorsson27Coloring,
  author       = {Yuval Gil and Magn{\'u}s M. Halld{\'o}rsson},
  title        = {Dynamic {$(\Delta+1)$}-Coloring Against an Adaptive Adversary: Faster, Simpler, Worst-Case, and Bounded Recourse},
  booktitle    = {Proceedings of the 38th Annual {ACM-SIAM} Symposium on Discrete Algorithms ({SODA} 2027)},
  year         = {2027},
  publisher    = {{SIAM}},
  note         = {To appear. Full version: arXiv:2610.26182},
  url          = {https://arxiv.org/abs/2610.26182},
  eprinttype   = {arXiv},
  eprint       = {2610.26182}
}

@inproceedings{AssadiYazdanyar27Coloring,
  author       = {Sepehr Assadi and Helia Yazdanyar},
  title        = {Dynamic {$(\Delta+1)$} Vertex Coloring with Adaptive Adversaries: Entropy Compression to the Rescue},
  booktitle    = {Proceedings of the 38th Annual {ACM-SIAM} Symposium on Discrete Algorithms ({SODA} 2027)},
  year         = {2027},
  publisher    = {{SIAM}},
  note         = {To appear. Full version: arXiv:2610.38787},
  url          = {https://arxiv.org/abs/2610.38787},
  eprinttype   = {arXiv},
  eprint       = {2610.38787},
}

@misc{KhanKumar26Predictions,
  author       = {Shahbaz Khan and Anshul Kumar},
  title        = {Dynamic Maximal Independent Set using Predictions},
  howpublished = {arXiv:2610.16637},
  year         = {2026},
  doi          = {10.48550/arXiv.2610.16637},
  url          = {https://arxiv.org/abs/2610.16637}
}

@inproceedings{BhattacharyaCHN18Coloring,
  author       = {Sayan Bhattacharya and Deeparnab Chakrabarty and Monika Henzinger and Danupon Nanongkai},
  title        = {Dynamic Algorithms for Graph Coloring},
  booktitle    = {Proceedings of the 29th Annual {ACM-SIAM} Symposium on Discrete Algorithms ({SODA} 2018)},
  pages        = {1--20},
  publisher    = {{SIAM}},
  year         = {2018},
  url          = {https://doi.org/10.1137/1.9781611975031.1},
  doi          = {10.1137/1.9781611975031.1}
}

@inproceedings{DuanHeZhang19EdgeColoring,
  author       = {Ran Duan and Haoqing He and Tianyi Zhang},
  title        = {Dynamic Edge Coloring with Improved Approximation},
  booktitle    = {Proceedings of the 30th Annual {ACM-SIAM} Symposium on Discrete Algorithms ({SODA} 2019)},
  pages        = {1937--1945},
  publisher    = {{SIAM}},
  year         = {2019},
  url          = {https://doi.org/10.1137/1.9781611975482.117},
  doi          = {10.1137/1.9781611975482.117}
}

@inproceedings{Christiansen23VizingChains,
  author       = {Aleksander B. G. Christiansen},
  title        = {The Power of Multi-step {Vizing} Chains},
  booktitle    = {Proceedings of the 55th Annual {ACM} {SIGACT} Symposium on Theory of Computing ({STOC} 2023)},
  pages        = {1013--1026},
  publisher    = {{ACM}},
  year         = {2023},
  url          = {https://doi.org/10.1145/3564246.3585105},
  doi          = {10.1145/3564246.3585105}
}

@inproceedings{BhattacharyaCPS24Nibbling,
  author       = {Sayan Bhattacharya and Mart{\'i}n Costa and Nadav Panski and Shay Solomon},
  title        = {Nibbling at Long Cycles: Dynamic (and Static) Edge Coloring in Optimal Time},
  booktitle    = {Proceedings of the 35th Annual {ACM-SIAM} Symposium on Discrete Algorithms ({SODA} 2024)},
  pages        = {3393--3440},
  publisher    = {{SIAM}},
  year         = {2024},
  url          = {https://doi.org/10.1137/1.9781611977912.122},
  doi          = {10.1137/1.9781611977912.122}
}

@inproceedings{Christiansen26EdgeColoring,
  author       = {Aleksander B. G. Christiansen},
  title        = {Deterministic Dynamic Edge Colouring},
  booktitle    = {Proceedings of the 37th Annual {ACM-SIAM} Symposium on Discrete Algorithms ({SODA} 2026)},
  pages        = {1047--1096},
  publisher    = {{SIAM}},
  year         = {2026},
  url          = {https://doi.org/10.1137/1.9781611978971.43},
  doi          = {10.1137/1.9781611978971.43}
}

@inproceedings{BhattacharyaCPS24Arboricity,
  author       = {Sayan Bhattacharya and Mart{\'i}n Costa and Nadav Panski and Shay Solomon},
  title        = {Arboricity-Dependent Algorithms for Edge Coloring},
  booktitle    = {Proceedings of the 19th Scandinavian Symposium and Workshops on Algorithm Theory ({SWAT} 2024)},
  series       = {LIPIcs},
  volume       = {294},
  pages        = {12:1--12:15},
  publisher    = {Schloss Dagstuhl -- Leibniz-Zentrum f{\"u}r Informatik},
  year         = {2024},
  url          = {https://doi.org/10.4230/LIPIcs.SWAT.2024.12},
  doi          = {10.4230/LIPIcs.SWAT.2024.12}
}

@inproceedings{ChristiansenRV24EdgeColoring,
  author       = {Aleksander B. G. Christiansen and Eva Rotenberg and Juliette Vlieghe},
  title        = {Sparsity-Parameterised Dynamic Edge Colouring},
  booktitle    = {Proceedings of the 19th Scandinavian Symposium and Workshops on Algorithm Theory ({SWAT} 2024)},
  series       = {LIPIcs},
  volume       = {294},
  pages        = {20:1--20:18},
  publisher    = {Schloss Dagstuhl -- Leibniz-Zentrum f{\"u}r Informatik},
  year         = {2024},
  url          = {https://doi.org/10.4230/LIPIcs.SWAT.2024.20},
  doi          = {10.4230/LIPIcs.SWAT.2024.20}
}

@misc{HaeuplerMRST26DynamicLLL,
  author       = {Bernhard Haeupler and Slobodan Mitrovi{\'c} and Srikkanth Ramachandran and Wen-Horng Sheu and Robert Tarjan},
  title        = {Dynamic Construction of the {Lov{\'a}sz} Local Lemma},
  howpublished = {arXiv:2604.20836},
  year         = {2026},
  url          = {https://arxiv.org/abs/2604.20836},
  doi          = {10.48550/arXiv.2604.20836}
}

\end{document}